%% file: liner_20180504_arxiv.tex
\documentclass[twocolumn]{aastex62}		         % better single-column preprint formatting

\usepackage{float}
\usepackage[]{natbib}
\usepackage{graphicx}	
\usepackage{amsmath}
\shorttitle{Power Sources of LINERs}
\newcounter{species} 
\def\ion#1#2{\hbox{\setcounter{species}{#2}#1\,{\scriptsize\Roman{species}}\relax}}

\def\lsim{\lower0.3em\hbox{$\,\buildrel <\over\sim\,$}}
\def\gsim{\lower0.3em\hbox{$\,\buildrel >\over\sim\,$}}

\usepackage[normalem]{ulem}
\def\crossout#1{}
\def\crossout#1{\sout{#1}}

\accepted{\today}
\submitjournal{ApJ}
\begin{document}		
\title{\large The Shocking Power Sources of LINERs\footnote{Based on observations made with the NASA/ESA {\it Hubble Space Telescope}, obtained at the Space Telescope Science Institute, which is operated by the Association of Universities for Research in Astronomy, Inc., under NASA contract NAS 5-26555. These observations are associated with program \# HST-GO-12595.}}

\correspondingauthor{Mallory Molina}
\email{mem468@psu.edu}

\author[0000-0001-8440-3613]{Mallory Molina} 
\affil{Department of Astronomy and Astrophysics and Institute for Gravitation and the Cosmos, The Pennsylvania State University, 525 Davey Lab, University Park, PA 16803, USA}

\author[0000-0002-3719-940X]{Michael Eracleous}
\affil{Department of Astronomy and Astrophysics and Institute for Gravitation and the Cosmos, The Pennsylvania State University, 525 Davey Lab, University Park, PA 16803, USA}

\author[0000-0002-3026-0562]{Aaron J. Barth}
\affiliation{Department of Physics and Astronomy, 4129 Frederick Reines Hall, University of California, Irvine, CA, 92697-4575, USA}

\author{Dan Maoz}
\affiliation{School of Physics and Astronomy, Tel-Aviv University, Tel-Aviv 69978, Israel}

\author[0000-0001-8557-2822]{Jessie C. Runnoe}
\affil{Department of Astronomy and Astrophysics and Institute for Gravitation and the Cosmos, The Pennsylvania State University, 525 Davey Lab, University Park, PA 16803, USA}
 \affiliation{Department of Astronomy, University of Michigan, 1085 S. University Avenue, Ann Arbor, MI 48109}
 
 \author[0000-0001-6947-5846]{Luis C. Ho}
\affiliation{Kavli Institute for Astronomy and Astrophysics, Peking University, 5 Yiheyuan Road, Haidian District, Beijing 100871, P. R. China}
 
\author{Joseph C. Shields}
\affiliation{Department of Physics and Astronomy, Ohio University, Clippinger Labs 251, Athens, OH 45701}

\author[0000-0002-1881-5908]{Jonelle L. Walsh}
\affiliation{Mitchell Institute for Fundamental Physics and Astronomy, Department of Physics and Astronomy, Texas A\&M University, 4242 TAMU, College Station, TX 77845}


\begin{abstract}
  The majority of low-ionization nuclear emission-line regions (LINERs) harbor supermassive black holes with very low accretion rates. However, the accretion flows do not produce enough ionizing photons to power the emission lines emitted on scales of $\sim100$~pc, and therefore additional sources of power are required. We present and analyze \textit{Hubble Space Telescope} spectra of three nearby luminous LINERs that are spatially resolved on scales of $\lsim 9$~pc. The targets have multiple indicators of an accreting black hole, as well as a deficient ionizing photon budget. We measure diagnostic emission line ratios as a function of distance from the nucleus and compare them to models for different excitation mechanisms: shocks, photoionization by the accreting black hole, and photoionization by young or old hot stars. We also consider the kinematics of the line-emitting gas, as revealed by the widths and shifts of the emission lines. We conclude that, in LINERs with low-luminosity active nuclei, shocks by jets or other outflows are crucial in exciting the gas in and around the nucleus, as suggested by other authors. The physical model that best describes our targets comprises a low-luminosity, accretion-powered active nucleus that photoionizes the gas within $\sim20$~pc of the galaxy center, and shock excitation of the gas at larger distances.

% and shocks that excite the gas at larger distances. We conclude that, in LINERs with low-luminosity active nuclei, shocks by jets or other outflows are crucial in exciting the gas in and around the nucleus, as suggested by other authors.
\end{abstract}

%INTRODUCTION
%%%%%%%%%%%%%%%%%%%%%%%%%%%%%%%%%%%%%%%%%%%%%%%%%%%%%%%%%%%%%%%%%%%%%%%%%%%%%%%%%
\section{Introduction}
\label{sec:intro}
Low-ionization nuclear emission-line regions (LINERs) were first identified by \citet{Heckman1980} based on the relative strength of their low ionization lines and classified according to the ratios of their oxygen lines ($\textrm{[\ion{O}{2}]}\lambda3727/\textrm{[\ion{O}{3}]}\lambda5007 >1$ and $\textrm{[\ion{O}{1}]}\lambda6300/\textrm{[\ion{O}{3}]}\lambda5007 \\>1/3$). They are now known to be very common; they are found in approximately 50\% of nearby galaxies \citep{HoFS97V}. A detailed discussion on the properties of LINERs is given in \citet{Ho2008}. The power source behind their emission lines has been under debate since their discovery and the following explanations have been considered with varying degrees of success in accurately describing their observational signatures: (1) photoionization by low-luminosity active galactic nuclei (LLAGN) \citep{Halpern1983,Ferland1983}, (2) photoionization by young, hot stars such as Wolf-Rayet stars \citep{Terlevich1985}, hot O stars \citep{Filippenko1992,Shields1992} and compact starbursts \citep{Barth2000}, (3) photoionization by exposed cores of evolved stars such as post-asymptotic giant branch (pAGB) stars \citep{Binette94,Yan2012,Belfiore2016,Papaderos2013,Singh2013} and planetary nebulae \citep{Taniguchi2000}, and (4) collisional or photo-excitation either by slow, or fast, photoionizing shocks \citep{Fosbury78,Heckman1980,Dopita1996,Dopita1997}. Models invoking a combination of the above power sources have also been considered \citep[e.g.,][]{contini97,contini01,Martins04,Sabra03}.

The LLAGN interpretation of LINERs was initially motivated by their significant X-ray emission \citep{Ferland1983,Halpern1983}. Although LLAGNs are found via radio and X-ray observations in the majority of LINERs \citep{Nagar2005,Dudik2005,Dudik2009,Flohic2006,Filho06,Gonzalez2009}, they are not powerful enough to photoionize the gas in their vicinity on $\sim~100$~pc (i.e. a few arcseconds in nearby galaxies) scales on which the characteristic emission lines are detected \citep[][and references therein]{Flohic2006,Eracleous2010a}. Imaging studies of LINERs have found extended line emitting regions and complex circumnuclear dust morphologies that might obscure and further prevent the LLAGN from fully ionizing the surrounding gas \citep{Barth1999,Pogge00,Simoeslopes07,Gonzalez08,Gonzalez2009,Masegosa11}. Wolf-Rayet stars \citep[i.e.,``warmers'';][]{Terlevich1985} could successfully mimic the X-ray emission produced by a LLAGN, as well as provide the hard ionizing photons necessary to explain the relative intensities of the observed optical emission lines. If Wolf-Rayet stars were the primary source of ionizing photons for LINER-like emission lines, then most LINERs would be in the immediate post-starburst phase, which is unlikely given their high occurrence rate \citep{HoFS97V} and stellar populations \citep{Gonzalez04,Cid04}. Compact starbursts containing hot O stars offer an alternative explanation for the relative intensities of LINER emission lines \citep{Filippenko1992,Shields1992}, but have difficulty explaining the broad Balmer emission wings often seen in LINER spectra \citep{Filippenko1996}, since this would require long lived supernova remnants. Moreover, the census of stellar populations in LINERs by \citet{Gonzalez04} and \citet{Cid04} does not show a high incidence of compact, nuclear starbursts. pAGB stars and planetary nebulae \citep{Binette94,Taniguchi2000,Yan2012,Belfiore2016} are more plausible stellar based models, but are applicable only to a subset of LINERs due to their inability to explain large H$\alpha$ equivalent widths \citep{HoFS03}. Shock models can adequately describe the off-nuclear optical and ultraviolet (UV) emission-line spectra of some LINERs, such as M87, obtained with the \textit{Hubble Space Telescope} \citep[\textit{HST}; see ][]{Dopita1996,Dopita1997,Sabra03}. Meanwhile the UV spectra of some LINER nuclei, such as NGC~4579, do not have emission line ratios that are well described by shock excitation, and yet show high ionization lines that would require a hard extreme UV source, such as the continuum from an AGN, or fast shocks \citep[][]{Barth1996,Dopita2015}. Successful shock models could have a wide range of shock velocities \citep{Filippenko1996}, but the gas must be continuously shocked to maintain LINER-like ratios. The shocks could potentially be driven by radio jets, which are fairly common in LINERs and whose kinetic power is considerably higher than the electromagnetic luminosity of the LLAGN \citep{Nagar2005,Filho02,Maoz2007}. Alternatively, the shocks could result from supernovae or winds from either young or evolved stars.

LINERs are found in a large fraction of nearby galaxies and the LLAGNs they host define the low luminosity end of the population of active galactic nuclei (AGNs). Yet, none of the models considered above can provide a complete and universal explanation for their energetics. Furthermore, extended LINER-like emission is seen in inactive galaxies observed in the Spectrographic Areal Unit for Research on Optical Nebulae (SAURON), Calar Alto Legacy Integral Field spectroscopy Area (CALIFA) and Mapping Nearby Galaxies at Apache Point Observatory (MaNGA) integral field unit surveys \citep{Sarzi10,Papaderos2013,Singh2013,Belfiore2016}, in red, inactive galaxies \citep{Yan06,Yan2012}, and in post-starburst galaxies \citep{Yan06,Graves07}. In these cases the line emission arises over a substantial fraction of the volume of the galaxy and has been attributed to photoionization by pAGB or other old, hot stars. \citet{Yan2012} have made the distinction between the original ``nuclear'' LINERs and these ``extended'' LINERs while \citet{Belfiore2016} have used the term LIER (low-ionization emission region; {\it not} nuclear) to describe such galaxies. Thus, we use here the term LINER to denote {\it compact}, nuclear emission line regions, a few hundred parsec in size, found in nearby galaxies and the term LIER to refer to large, galaxy size emission line regions with LINER-like spectra.

Understanding the excitation mechanism of the gas in nearby LINERs is also important for understanding more distant galaxies. Since the LLAGNs in the majority of nearby LINERs do not produce enough ionizing photons to power the emission lines seen on $\sim 100$~pc scales, we set out to find indications of other power sources that excite the gas on these large scales. To this end, we obtained spatially resolved spectra of three nearby LINERs with the Space Telescope Imaging Spectrograph (STIS) on \textit{HST}, spanning the spectral range from the near-ultraviolet (NUV) to the red (H$\alpha$). Our goal is to identify and track the energy sources powering the emission lines within the central $\sim 100$~pc and determine if different sources dominate on different spatial scales. The STIS spectra have spatial resolution better than 9 pc in three wavelength bands, thus we use them to measure emission line ratios as a function of distance from the nucleus. We compare these measurements to physical models for the excitation mechanisms suggested for LINERs, including \ion{H}{2} region models \citep{Kewley2006, Dopita2006}, fast and slow shock models \citep{Shull79,Dopita1995,Allen2008}, pAGB models \citep{Binette94}, and AGN photoionization models \citep{Groves2004,Nagao02}. In addition to the emission line ratios, we consider the velocity field of the gas, as described by the full width at half maximum (FWHM) and centroid of strong, isolated emission lines. While this work is motivated by the assessment of the photon budget by \citet{Eracleous2010a}, we cannot carry out an analogous assessment for mechanical processes, such as shocks, because the models only provide the relative strengths of the emission lines. Therefore our tests here must rely on emission line ratios.

In Section~\ref{sec:obsdata} we describe the selection of our targets and summarize their properties. We describe the data and their basic reduction in Section~\ref{sec:data}, the method for separating  the unresolved nuclear source from the extended emission in Section~\ref{sec:2dspec}, and the subtraction of starlight from the spectra in Section~\ref{sec:stars}. In Section~\ref{sec:lines} we detail the measurements of the emission lines and in Section~\ref{sec:tempdens} we present the inferred temperature and density as a function of distance from the center of each galaxy. In Section~\ref{sec:ddresult} we make a detailed comparison of the measured line ratios with physical models. We consider those results and discuss implications for the LINER population in Section~\ref{sec:discussion} and summarize our findings and conclusions in Section~\ref{sec:summary}.

%Experimental Design and Target Selection
%%%%%%%%%%%%%%%%%%%%%%%%%%%%%%%%%%%%%%%%%%%%%%%%%%%%%%%%%%%%%%%%%%%%%%%%%%%%%%%%%
\section{Experimental Design and Target Selection}
\label{sec:obsdata}
To diagnose the excitation mechanisms of the emission line gas and their relative importance on small scales around the nuclei of our target galaxies, we obtained long-slit spectra with the \textit{HST}/STIS. By covering the spectral range from the NUV ($\sim$1900~\AA) to the red ($\sim$6800~\AA), we measured both well known diagnostic line ratios such as $\textrm{[\ion{O}{3}]}/\textrm{H}\beta$, $\textrm{[\ion{N}{2}]}/\textrm{H}\alpha$ and $\textrm{[\ion{O}{1}]}/\textrm{H}\alpha$, as well as a number of optical lines that are good probes of shocks, such as [\ion{Fe}{14}]$~\lambda$5303, and [\ion{Fe}{10}]$~\lambda$6374. We also looked for [\ion{Ne}{4}]$~\lambda$2423, a NUV line that is also a good probe of shocks. This line was detected in NGC~1052 by \cite{Dopita2015}, but we were not able to detect it in our NUV spectra because of their lower signal-to-noise ratio (S/N).

We selected as our targets the three galaxies listed in Table~\ref{table: redshift}, based on the following criteria: (1) a deficit of ionizing photons from the LLAGN as described in \citet{Eracleous2010a}, (2) multiple indicators for the presence of an AGN, and (3) existing archival H$\alpha$ spectra from \textit{HST}/STIS \citep{Walsh2008} that we could use to make our observing program economical. Moreover, \citet{Walsh2008} studied the velocity field of the line-emitting gas in our three target galaxies using these spectra, and we take advantage of their results in our interpretation of all the data.

\input{tgt_prop.tex}

Our targets include NGC~4278, which has a severe ionizing photon deficit, and NGC~4579, for which photoionization from the LLAGN is plausible only if all the ionizing photons are absorbed by the line emitting gas \citep{Eracleous2010a}. We also targeted NGC~1052, for which emission line imaging \citep{Pogge00} and X-ray spectroscopy \citep{Brenneman2009} show the nucleus is likely obscured along some lines of sight, suggesting that the AGN cannot be the only power source. Furthermore, previous observations of NGC~1052 with the Faint Object Spectrograph on \textit{HST} by \citet{Gabel2000} show shock excited NUV emission lines.

All three targets have radio emission. NGC~4278 has compact jets, with a length of $0.\!\!^{\prime\prime}$015 (1.2~pc) in the 6~cm band \citep{Giroletti2005,Helboldt07}. NGC~4579 has an unresolved radio core that is smaller than $0.\!\!^{\prime\prime}0025$ ($< 0.2$~pc) in the 5~GHz band \citep{Falcke2000}. NGC~1052 has a pair of large jets, each of length $15^{\prime\prime}$ (1.3~kpc) in the 20~cm band \citep{Wrobel84,jones84}, as well as resolved jets on smaller scales of length $0.\!\!^{\prime\prime}008$ (0.7~pc) in the 2~cm band \citep{claussen1998,Kellermann98}. The small and large scale jets in NGC 1052 show a change in position angle, from approximately $60^{\circ}$ to $95^{\circ}$ as described in \cite{claussen1998}. Because of the spatial scales probed by our observations, we consider the small-scale jet to be the most relevant to our discussion hence we will refer to the small-scale radio jet of NGC~1052 as ``the jet'' hereafter.

In Figure~\ref{fig:slits} we show \textit{HST} images of the targets covering a region of a few arcseconds in size around the nucleus. The directions of the resolved, smaller scale radio jet of NGC~1052 and the radio jet in NGC~4278 are also marked in these images.

%Observations and Basic Data Reduction
%%%%%%%%%%%%%%%%%%%%%%%%%%%%%%%%%%%%%%%%%%%%%%%%%%%%%%%%%%%%%%%%%%%%%%%%%%%%%%%%%
\section{Observations and Basic Data Reduction}
\label{sec:data}
\input{obs_sum.tex} We obtained new H$\beta$-region optical (hereafter, ``blue'') and NUV long-slit spectra of the targets in May 2012 using the charge-coupled device (CCD) and NUV-Multi-Anode Microchannel Array (NUV-MAMA) detectors of STIS, respectively, employing the $52^{\prime\prime}\times0.\!\!^{\prime\prime}2$ slit. We supplemented these data with archival H$\alpha$ (hereafter, ``red'') spectra obtained in 1999 and 2000 through the same slit with the STIS CCD and G750M grating, and presented by \citet{Walsh2008}. The physical separation between the spectra along the slit for each galaxy and detector are listed in Table~\ref{table: redshift}, and the log of observations is given in Table~\ref{table: dates}. We used the same slit position angle for the new observations taken with the CCD and NUV-MAMA detectors. 

\begin{figure*}
\centering
	\includegraphics[width=0.7\textwidth]{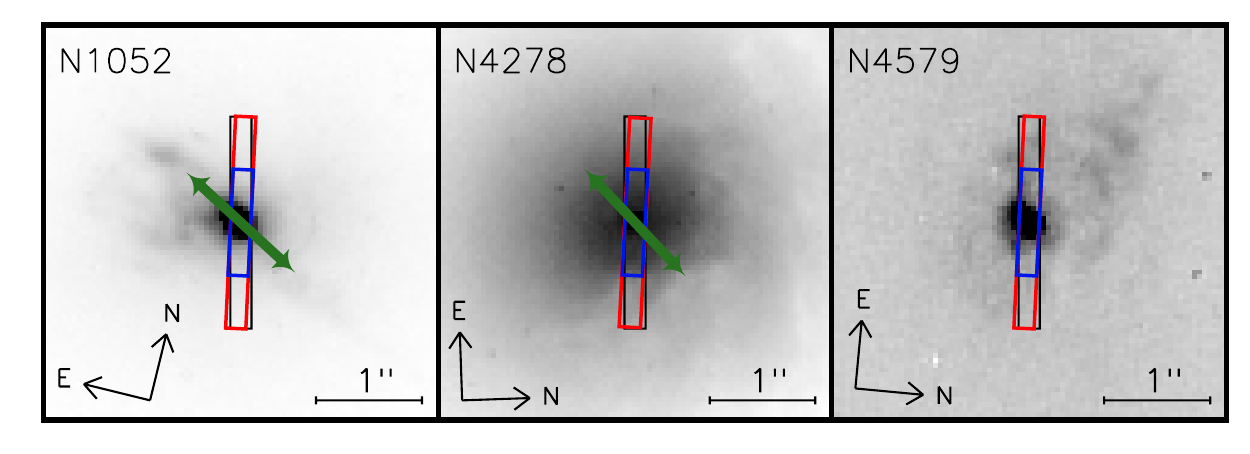}
	\caption{Images with slit and approximate jet orientation superposed for the three objects studied in this paper. The black (vertical) rectangle shows the slit placement of the original red (H$\alpha$) observation from \citet{Walsh2008}. The red (slightly rotated, full length) and blue (slightly rotated, half length) rectangles show the slit placement for the blue (H$\beta$) and NUV observations, respectively. See Section~\ref{sec:data} for a more detailed discussion on the observations. The length of each rectangle represents the physical extent to which we could detect emission lines in each wavelength band. The green arrows show the direction of the radio jets in each object (length not to scale). The radio morphologies of all three objects are described in Section~\ref{sec:obsdata}. \textit{Left: }NGC~1052 in a narrow band around the H$\alpha+$[\ion{N}{2}]$~\lambda\lambda$6548,6583 lines with the WFPC2 F658N with the small-scale jet orientation shown by the green arrows, see Section~\ref{sec:obsdata} for details. \textit{Center: }NGC~4278 in a wide continuum band with the WFPC2 PC F555W. \textit{Right: }NGC~4579 in the same band as NGC~1052.}
\label{fig:slits}
\end{figure*}

The archival spectra cover the wavelength range 6300--6860~\AA\,with a scale of $0.\!\!^{\prime\prime}05$~pixel\textsuperscript{$-1$} along the slit and a dispersion of 0.56~\AA~pixel\textsuperscript{$-1$}. The blue spectra were taken with the CCD detector and the G430L grating, covering a wavelength range of  2900--5700~\AA\, with a scale of $0.\!\!^{\prime\prime}05$~pixel\textsuperscript{$-1$} along the slit and a dispersion of 2.73~\AA~pixel\textsuperscript{$-1$}. The NUV spectra were taken with the NUV-MAMA detector and the G230L grating, covering a wavelength range of 1570--3180~\AA\, with a scale of $0.\!\!^{\prime\prime}025$~pixel\textsuperscript{$-1$} along the slit and a dispersion of 1.58~\AA~pixel\textsuperscript{$-1$}.  The slit alignment of the new data compared to the archival observations, as well as the jet orientations for NGC~1052 and NGC~4278, are shown in Figure~\ref{fig:slits}. 

In the new CCD observations we placed the object near the readout amplifier and took multiple exposures for the blue spectra to reduce the effects of charge transfer inefficiency on the resulting images. To further mitigate these effects, as well as the effect of cosmic rays, we used the STIS-ALONG-SLIT dither pattern where each exposure was linearly offset by three pixels in the spatial direction, corresponding to a $0.\!\!^{\prime\prime}15$ shift between CCD exposures and a  $0.\!\!^{\prime\prime}075$ shift between NUV-MAMA exposures.

Each raw 2--D spectral image, including the archival red spectra, was processed through the \texttt{calstis} pipeline. Before the final step in the \texttt{calstis} pipeline, we removed most of the hot pixels and cosmic rays in our blue and red spectral images using the routine \texttt{L.A.\ Cosmic} \citep{Dokkum2001}. The NUV spectral images were not processed through \texttt{L.A. Cosmic} as the NUV-MAMA detector is not affected by cosmic rays. To account for the dither pattern, we shifted the images to realign them using the \texttt{IRAF}\footnote{IRAF is distributed by the National Optical Astronomy Observatories, which are operated by the Association of Universities for Research in Astronomy, Inc., under cooperative agreement with the National Science Foundation.} task \texttt{imshift}, and used the \texttt{IRAF} task \texttt{imcombine} to median combine the aligned images, adopting the standard deviation from individual exposures as the error bar for each pixel in the resulting 2--D spectrum. The archival red spectra consist of a single exposure for each galaxy, so this last step was not carried out. 

We found that the rectification of the curvature of the 2--D spectrum, carried out by \texttt{calstis}, left an unacceptably large residual curvature, so we applied an improved geometric rectification. We fitted a low order polynomial to the spatial peak position as a function of location along the dispersion direction in all the final 2--D spectra. We chose the lowest order polynomial that ensured that, after the fit was subtracted from the data, the position of the spatial profile peak was both constrained within a tenth of a pixel along the slit, and showed no systematic trend in the dispersion direction. Following this correction, we proceeded to the scientific analysis of the 2--D spectra, as we describe in the following sections.
%2-d spectra
%%%%%%%%%%%%%%%%%%%%%%%%%%%%%%%%%%%%%%%%%%%%%%%%%%%%%%%%%%%%%%%%%%%%%%%%%%%%%%%%%
\section{Separating Resolved and Unresolved Light in the 2--D Spectra}
\label{sec:2dspec}
All three of our objects have multiple indicators of the presence of an AGN, including broad emission lines in their nuclear spectra obtained with STIS. Thus, we searched for and removed any contribution from the bright unresolved nuclear source in the rows of the 2--D spectra near the nucleus. By taking this measure, we separated the spatially unresolved nuclear source in each band from the extended, resolved source.
\subsection{Fitting Method}
\label{ssec:2dresfit}
To separate the two spatial components, we modeled the light profiles of each wavelength bin in each 2--D spectrum as a broad base from the extended, resolved light with a strongly peaked unresolved core. We adopted the Nuker Law to describe the resolved starlight \citep{Lauer95}, under the assumption that a substantial portion (but not necessarily all) of the emission line flux follows the distribution of the starlight. For the strong, unresolved core (i.e., the AGN), we parameterized the point spread function (PSF) of the instrument by fitting the spatial profiles of stars observed with the same instrument configuration as the 2--D spectrum with a combination of two Gaussians for the red and NUV spectra and three Gaussians for the blue spectra. A log of the archival observations of the stars we used is given in Table~\ref{table: stars}. The final spatial model thus consists of a linear combination of a Nuker model, which describes both the stellar continuum and light from any other source that follows the spatial distribution of the starlight, and a pointlike central component modeled using the PSF, which describes the light from the unresolved nuclear source. Fits to the stellar continuum of individual spectra extracted from different distances from the center of a galaxy are carried out at a later stage in the analysis,  as detailed in Section \ref{sec:stars}. We allowed the fitting routine to under fit but not over fit each spatial profile in spectral pixels with emission lines. 

We fitted the spatial model to the spatial profile along each column of a 2-D spectrum (i.e., at each wavelength). Our procedure involved binning the model to match the detector pixels before comparing it to the flux in the 39 central pixels of the observed spatial profile. The total model is described by a total of eight free parameters; six parameters are needed for the Nuker model \citep[the five parameters in equation 3 of][plus a shift]{Lauer95} and two for the PSF (amplitude and shift). We did not convolve the Nuker model with the PSF of the instrument since we were able to achieve our primary goal of fitting the inner parts of the spatial profile equally well with and without such a convolution. 

The rationale behind our spatial modeling was that the extended {\it line} emission could be described by a Nuker law plus positive deviations that represent clumps of emission line gas and is justified by the emission line morphologies revealed by narrow band images \citep[e.g.,][]{Pogge00}. Moreover, our primary goal is to identify the contribution of the unresolved nuclear source to the light profile and remove it and then ascribe what is left in the light profile to extended emission.

We applied this fitting method to each 2--D spectrum separately. The exact centroids of the two components of the model were allowed to vary within two pixels of each other (the uncertainty associated with the location of the PSF in each wavelength bin). 

\begin{figure}[t]
\includegraphics[width=0.49\textwidth]{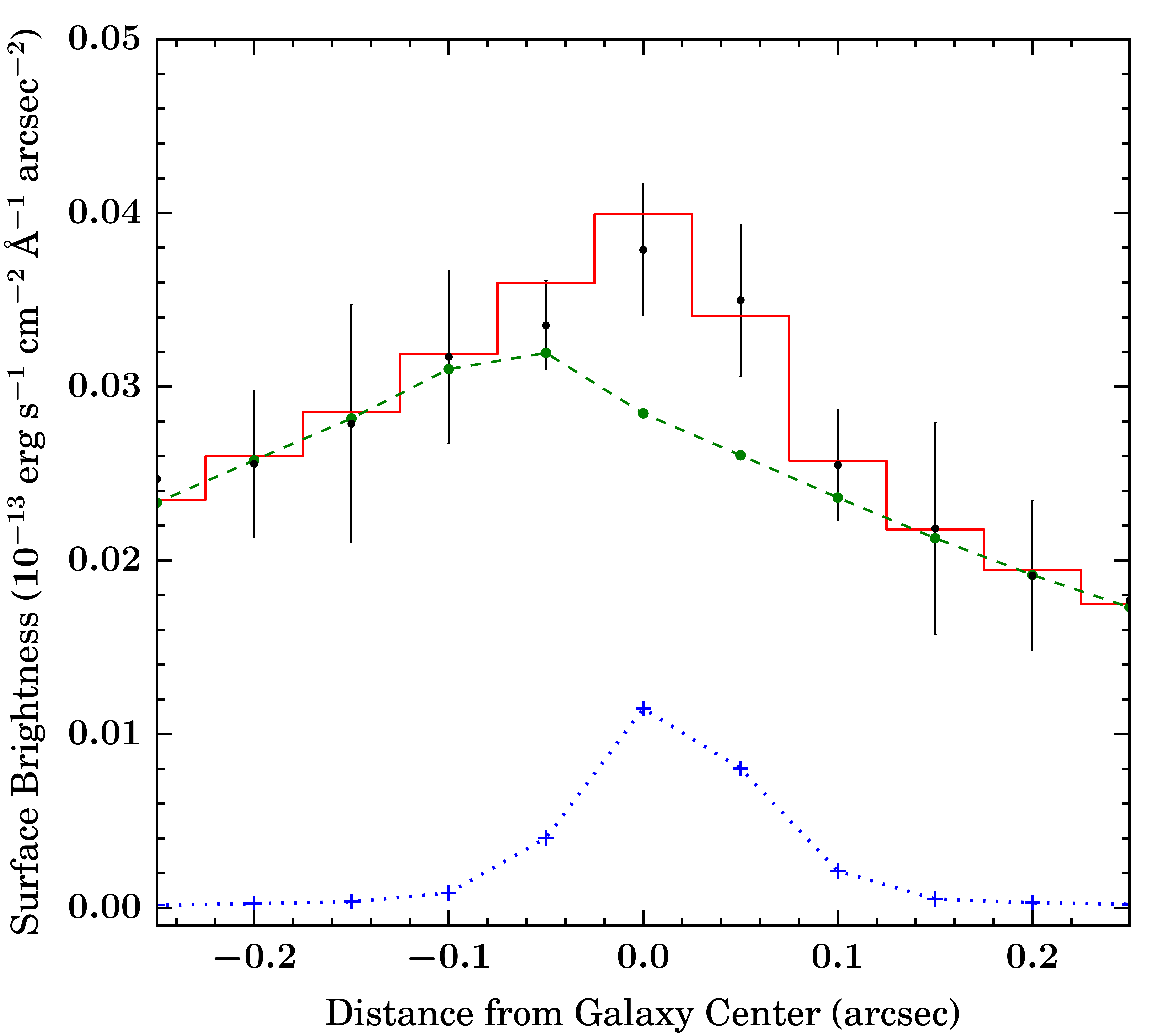}
\caption{An example of a fit to the spatial profile at a wavelength of 5106.5~\AA\ in the blue spectrum of NGC~1052, where the light is dominated by the continuum. The data are shown as black points with error bars, the blue `+' signs connected by the dotted blue line represents the binned PSF model, and the binned Nuker model is shown as green dots connected by a dashed line. These are both intrinsically symmetric but may appear asymmetric after binning. The two model components are added together to get the total model, represented by the red histogram. Our requirement for a good fit is that the model was either within or below the error bars at all points. The Nuker model and PSF model do not have the same peak position, but it is within the accepted separation as described in Section~\ref{ssec:2dresfit} of the text.}
\label{fig:cmods}
\end{figure}

\begin{figure}[t]
\includegraphics[width=0.49\textwidth]{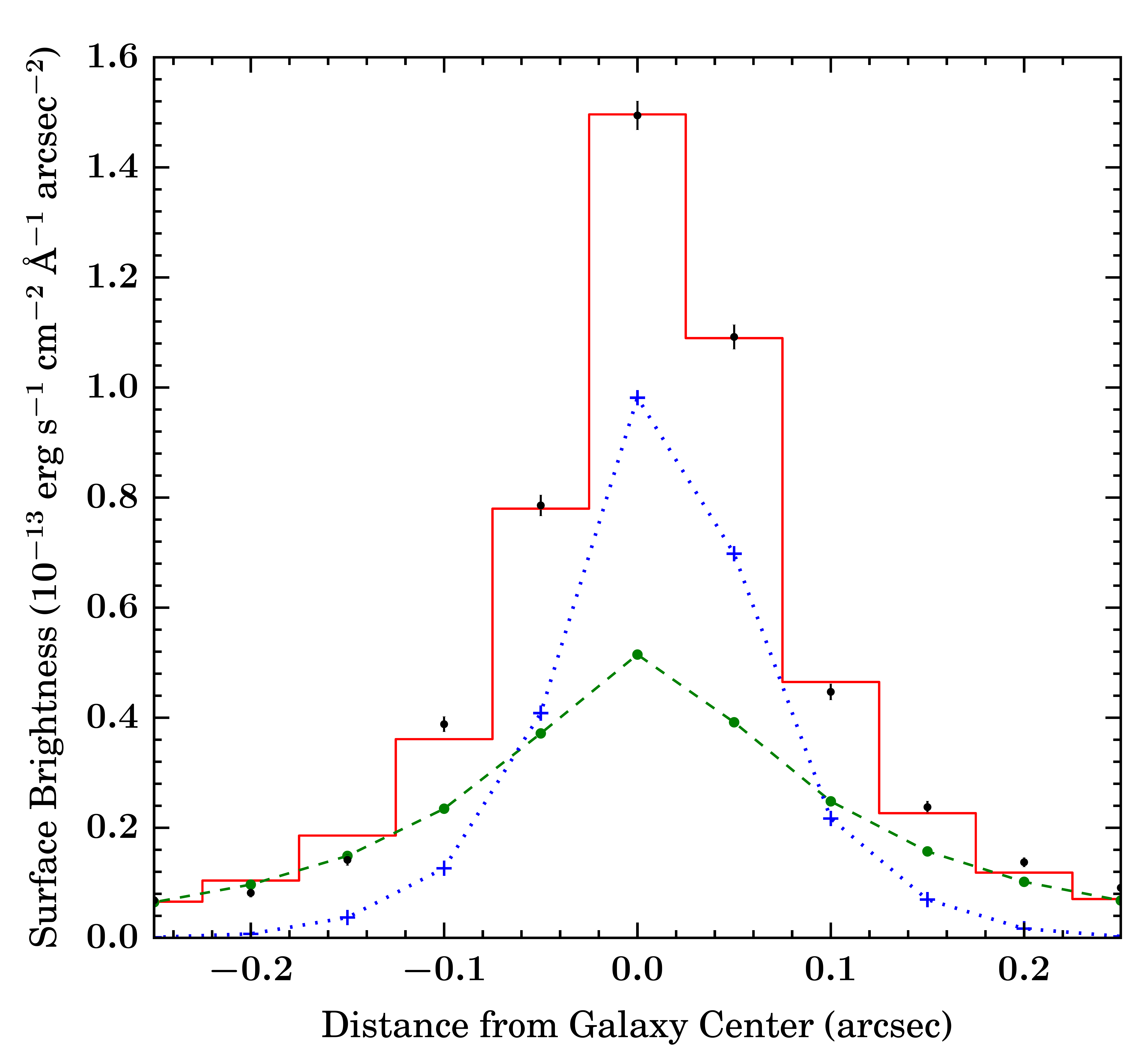}
\caption{An example of a fit to the spatial profile at a rest wavelength of 6559.3~\AA\ in the red spectrum of NGC~1052, where the light is dominated by the H$\alpha$ emission line. Here, the S/N is considerably higher than the continuum column shown in Figure~\ref{fig:cmods} and the contribution from the unresolved nuclear source is comparable to that of the resolved emission at the center of the galaxy. The data are shown as black points with error bars, the blue `+' signs connected by the dotted line represents the binned PSF model, and the binned Nuker model is shown as green dots connected by a dashed line. These are both intrinsically symmetric but may appear asymmetric after binning. The two model components are added together to get the total model, represented by the red histogram.  See detailed discussion in Section~\ref{ssec:2dresfit} of the text.}
\label{fig:mods}
\end{figure}

In Figures~\ref{fig:cmods} and~\ref{fig:mods} we show examples of fits to the spatial profiles at a continuum and an emission line wavelength, respectively. In each figure we compare the binned model to the data and we also show the components that make up the model, after binning them. In Figure~\ref{fig:cmods}, the best fitting Nuker component appears asymmetric as a consequence of binning it before comparing it with the data. The intrinsic Nuker component is symmetric but has a sharp core that falls near the boundary between two spectral bins. Once the best fit was obtained, we used the resulting PSF model to subtract the contribution of the unresolved source from each spectral pixel and isolate the extended emission. We also integrated the PSF flux in each spectral pixel to obtain the spectrum of the unresolved nuclear source.  At the end of the fitting process, we examined the spectra to ensure that broad emission lines were detected only in the spectrum of the unresolved source. Below we give specific notes for each object and present an example of the outcome. 

\input{stellar_obs.tex}

\begin{figure}[t]
\includegraphics[width=0.5\textwidth]{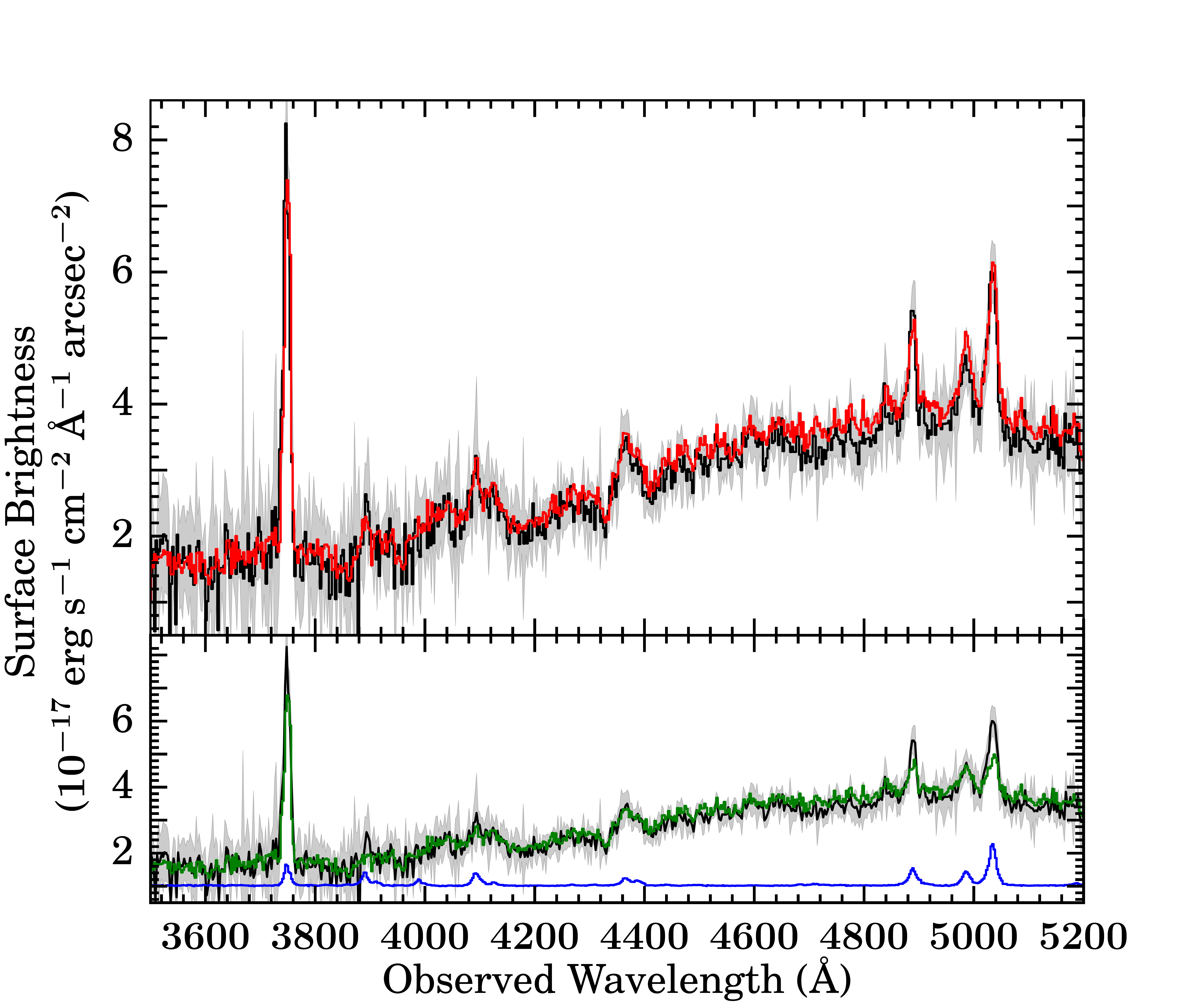}
\caption{An example of the decomposition of a blue spectrum for NGC~1052 into spectra of the resolved and unresolved light. The spectrum was extracted from a region four rows away from the nucleus. The data and errors are represented by the black solid line and gray shaded region, respectively. The red line in the top panel shows the total model. The blue (emission lines with no continuum) and green (emission lines with continuum) lines in the bottom panel show the contribution of the unresolved light and the Nuker model, respectively. The Nuker model accurately captures the starlight continuum and some of the emission line flux, while the unresolved light only contributes significantly in the emission lines. See detailed discussion of the fitting process in Section~\ref{ssec:2dresfit}.}	\label{fig:spec_mods}
\end{figure}

\begin{figure}[t]
\includegraphics[width=0.5\textwidth]{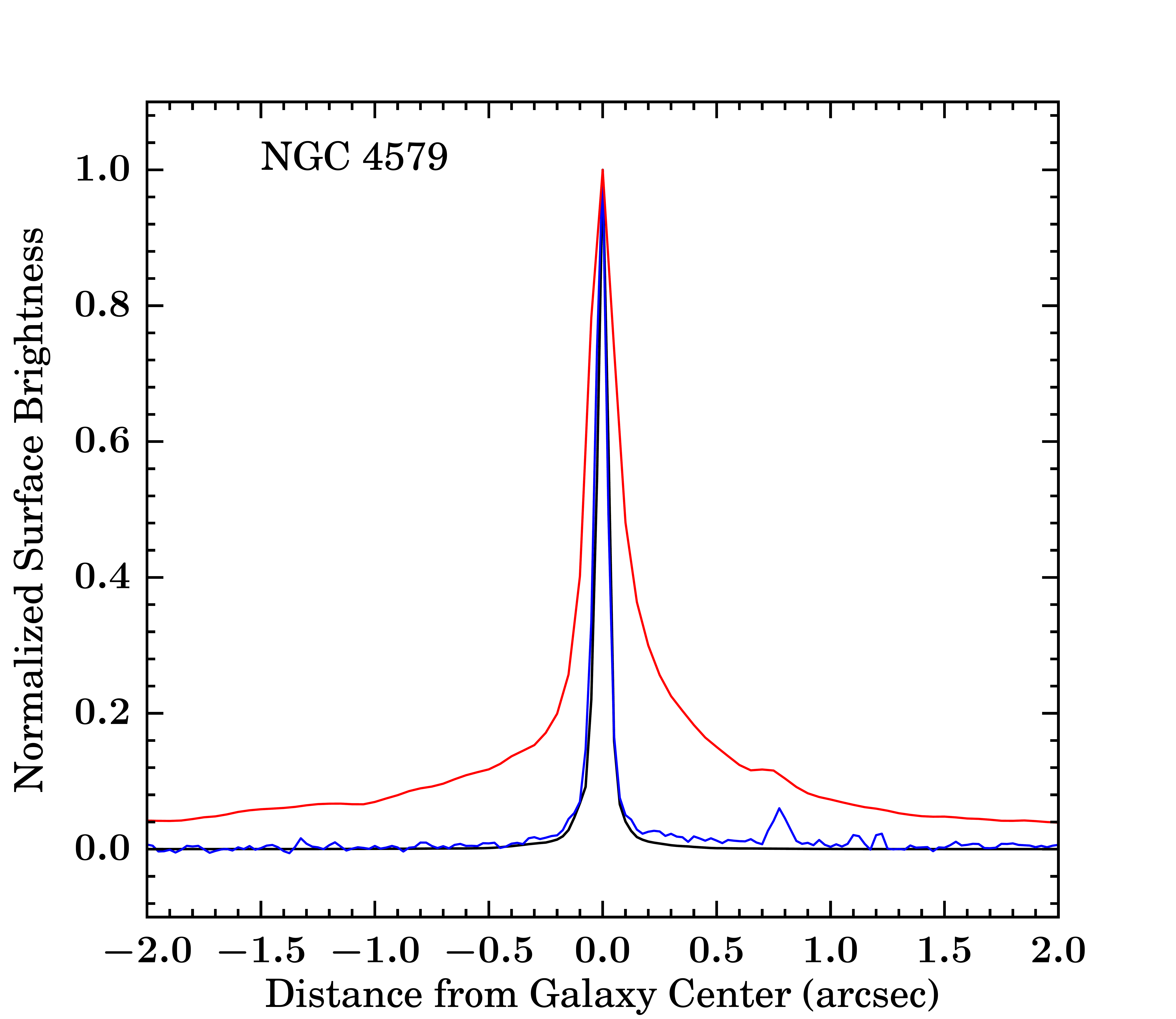}
\caption{Comparison of the spatial profiles of NGC~4579. We show the NUV spectrum (at 2962.6~\AA) in blue (noisy narrow profile), the blue spectrum (at 5340.7~\AA) in red (smooth broader profile) and the NUV stellar PSF template (integrated from 1570 to 3180~\AA) in black (smooth narrow profile) respectively. We choose the blue spectrum for this comparison as it has clearly detected resolved emission. The spatial profiles are in units of erg~s\textsuperscript{$-1$}~cm\textsuperscript{$-2$}~\AA\textsuperscript{$-1$}~arcsec\textsuperscript{$-2$} and normalized to unit maximum. The NUV spatial profile does not have the broad wings that appear in the blue spatial profile, but instead looks almost identical to the stellar PSF, implying the NUV spectrum is dominated by the light from the unresolved nuclear source. See Section~\ref{ssec:2dresfit} for more details.}\label{fig:n4579_hb_uv}
\end{figure}

\begin{description}

\item[NGC~1052]-- We successfully decomposed the spatially unresolved nuclear source and the extended, resolved sources in all three spectral bands for NGC~1052. Figure~\ref{fig:spec_mods} shows the outcome of this decomposition for a spectrum located four rows from the nucleus. As expected, the Nuker Law spectrum follows the starlight continuum, while the PSF model spectrum contributes only to the emission lines. The broad lines are successfully isolated in the NUV and blue spectra of the spatially unresolved nuclear source. 

The H$\alpha$ emission line in the spectrum of the central region (central row) of the extended emission has broad wings. It is possible that there is some residual contamination of this spectrum from light from the unresolved nuclear source (the measurements from this row are labeled with a ``c'' in the diagnostic diagrams presented in Section~\ref{sec:ddresult}). However, it is also possible that the broad emission we detect here represents scattered light within the galaxy (not the telescope or instrument optics) from the spatially unresolved nuclear source. The latter interpretation is supported by the fact that \citet{Barth1999} found that the broad H$\alpha$ wings in NGC~1052 are preferentially polarized, suggesting a substantial contribution from scattered light. We also note that we further constrained the Nuker profile to vary smoothly near the peak of the emission lines, i.e., a range of $\sim5$--10 \AA\, bracketing the peaks of the emission lines, in the red spectrum of NGC~1052 in order to obtain a good fit at the line peak. Figure \ref{fig:mods} shows an example of such a spatial profile fit.

\item[NGC~4278]-- We found only tentative evidence of spatially unresolved nuclear light in the blue and red spectra. A good fit was also possible with a model that included no unresolved source. Moreover, the only viable fit to the NUV spectrum did not include an unresolved source. Therefore, in the rest of the analysis we assume that there is no detectable unresolved source in NGC~4278.

\item[NGC~4579]-- The spatially unresolved nuclear source in NGC~4579 is so bright that we were not able to completely remove its contribution to the resolved light in the red and blue bands. The central blue spectrum of the resolved source exhibits wings on the H$\beta$ emission line, presumably a result of residual light from the unresolved nuclear source. Similarly, in the red band we were not able to separate the resolved and unresolved sources in the central pixels.  The contamination of the resolved light by the unresolved source is confined to two rows around the nucleus. We found that the NUV spectrum is dominated by light from the unresolved nuclear source, as illustrated in Figure~\ref{fig:n4579_hb_uv}: the blue spatial profile is sharply peaked with broad wings from the resolved light, but the NUV spatial profile only has a sharp peak and almost exactly matches the PSF template.
\end{description}

\subsection{Assessment of the Spatial Decompositions and Discussion of Uncertainties}\label{ssec:2derr}

We re-iterate the cautionary notes about the results of the spatial decomposition at $\pm2$ pixel rows from the nucleus of NGC~4579. As we describe in later sections, we add together pairs of pixel rows to obtain spatially resolved 1--D spectra, which implies that the spectra of the resolved emission of NGC~4579 in the two spatial bins closest to the nucleus ($\pm 8$ pc from the center) are not useful for our later analysis. A similar caution applies to the red spectrum of the innermost spatial bins in NGC~1052 ($\pm 4.3$ pc from the center). We consider the results of the spatial decomposition at larger distances from the nuclei of these two galaxies reliable because the PSF contribution drops by nearly an order of magnitude at three pixels from the peak and and becomes lower than the contribution of the resolved light \citep[see][and Figures~\ref{fig:cmods} and~\ref{fig:mods} of this paper]{bowers97}.

To assess further the robustness of the spatial decomposition, we checked whether the spectra of the unresolved sources in NGC~1052 and NGC~4579 in the three different bands connect smoothly to each other. Thus, we fitted a simple power law model \citep[a very common description of AGN optical continua, e.g.,][]{Vanden2001} to continuum windows in the three spectra of each object. This approach is further justified by the fact that we observe broad emission lines and no stellar absorption lines in the spectrum of the unresolved sources of both galaxies (we quantify the contribution of starlight in Section~\ref{sec:stars}, below).  We found that the continua in all three bands are indeed described by a common power law model and that the post-fit residuals have a Gaussian distribution about zero. To estimate the uncertainty in the relative normalization of the three individual spectra of each object, we explored how much the normalization of each of the three spectra can be changed while still keeping the distribution of residuals from that spectrum within one standard deviation (``$1\sigma$'') from zero. Thus we found normalization uncertainties of $\pm74$\%, $\pm41$\%, $\pm37$\% for the NUV, blue, and red spectra of NGC~1052 and  $\pm12$\%, $\pm12$\%, and $\pm10$\% for the corresponding spectra of NGC~4579. The magnitude of the uncertainties depends directly on the strength of the continuum in each band; it is highest for the NUV spectrum of NGC~1052 where the continuum is fairly weak. This uncertainty affects only ratios of lines that are detected in different spectra taken with different gratings and only when the spectra are within $\pm2$ pixel rows from the nucleus where the contribution of the PSF is substantial. It is taken into account in later error calculations for the emission line strengths measured from the spectra of the unresolved and resolved sources of these galaxies. We return to this issue in Section~\ref{ssec:reddening} where we discuss how this uncertainty affects our estimates of extinction.

%Starlight
%%%%%%%%%%%%%%%%%%%%%%%%%%%%%%%%%%%%%%%%%%%%%%%%%%%%%%%%%%%%%%%%%%%%%%%%%%%%%%%%%

\begin{figure}[t]
\includegraphics[width=0.5\textwidth]{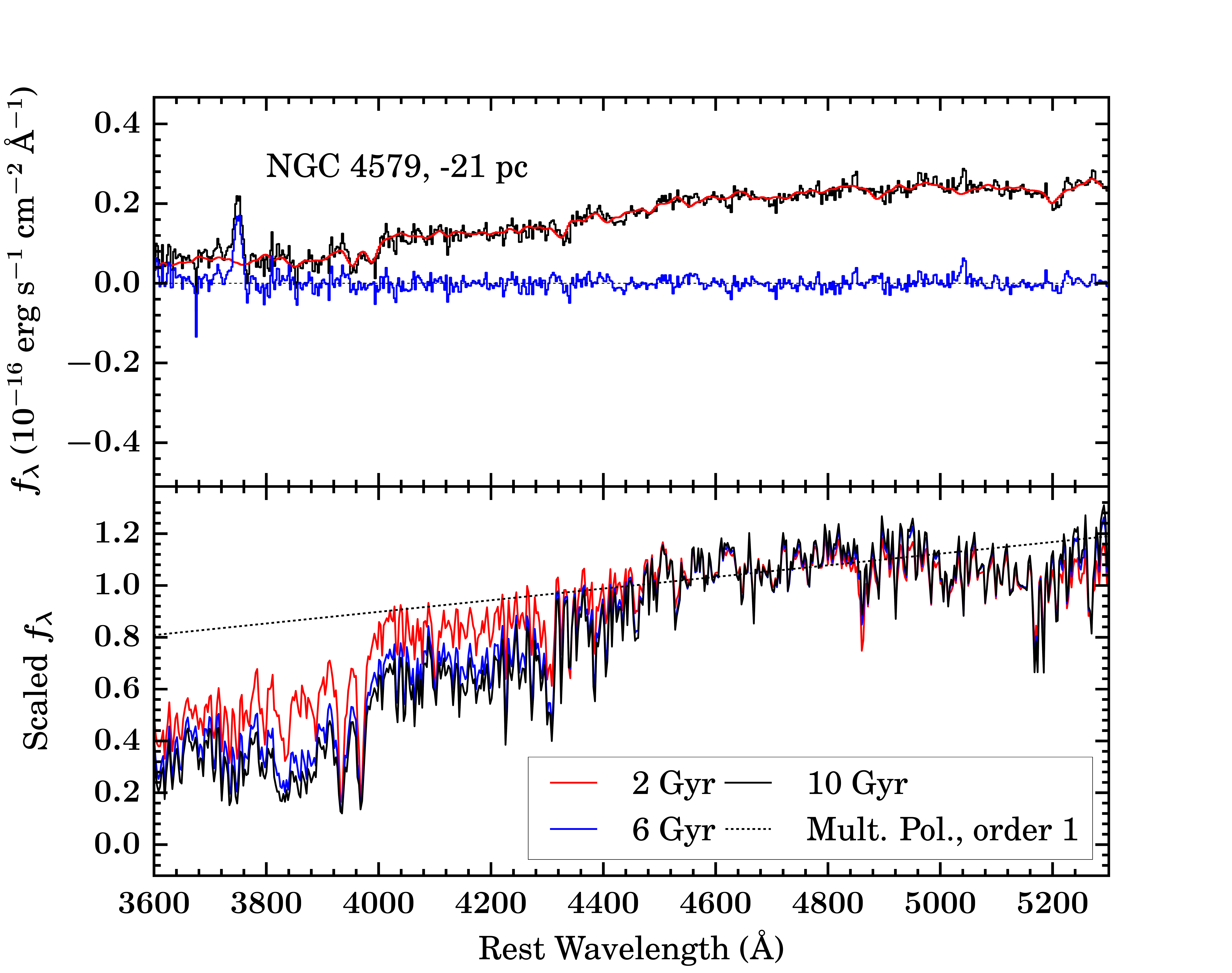}
\caption{Example of the stellar continuum fit to a spectrum extracted from a distance of approximately 21~pc from the nucleus of NGC~4579. The top panel shows the observed spectrum in black, the continuum fit in red, and the residual spectrum in blue. The bottom panel shows the components of the model continuum: a multiplicative polynomial of order one (dotted line) and the 2, 6, and 10~Gyr stellar populations (red, blue and black lines; listed in order of continuum strength around the \ion{Ca}{2} H \& K $\lambda\lambda 3969,3934$ lines), scaled to match at 4700~\AA. See detailed discussion in Section~\ref{sec:stars} of the text.}	\label{fig:starlight}
\end{figure}

\begin{figure}[t]
\includegraphics[width=0.5\textwidth]{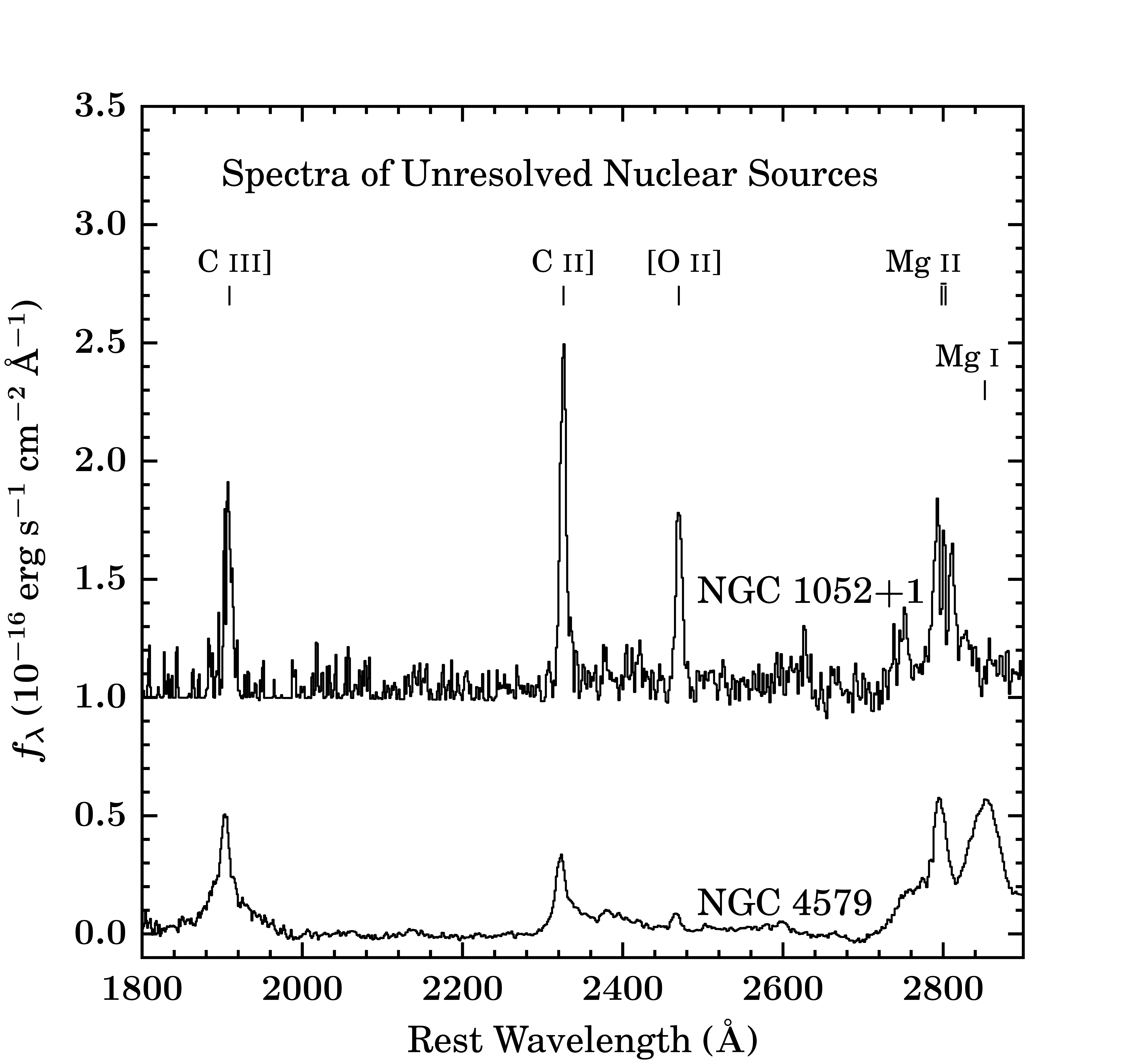}
\caption{NUV spectra of the spatially unresolved nuclear sources in NGC~1052 and NGC~4579. The spectra are offset vertically from each other for clarity, by the offsets given in the figure. The prominent emission lines are marked and labeled. NGC~4579 shows no resolved nuclear emission in this band, so the spectrum presented here is the spatially integrated spectrum. Both objects show broad \ion{Mg}{2} emission lines, indicative of an AGN. See Section~\ref{sec:lines} for a detailed discussion of Figures~\ref{fig:unres_stack_uv} through~\ref{fig:int_stack_ha}.}	\label{fig:unres_stack_uv}
\end{figure}        

\begin{figure}[t]
\includegraphics[width=0.5\textwidth]{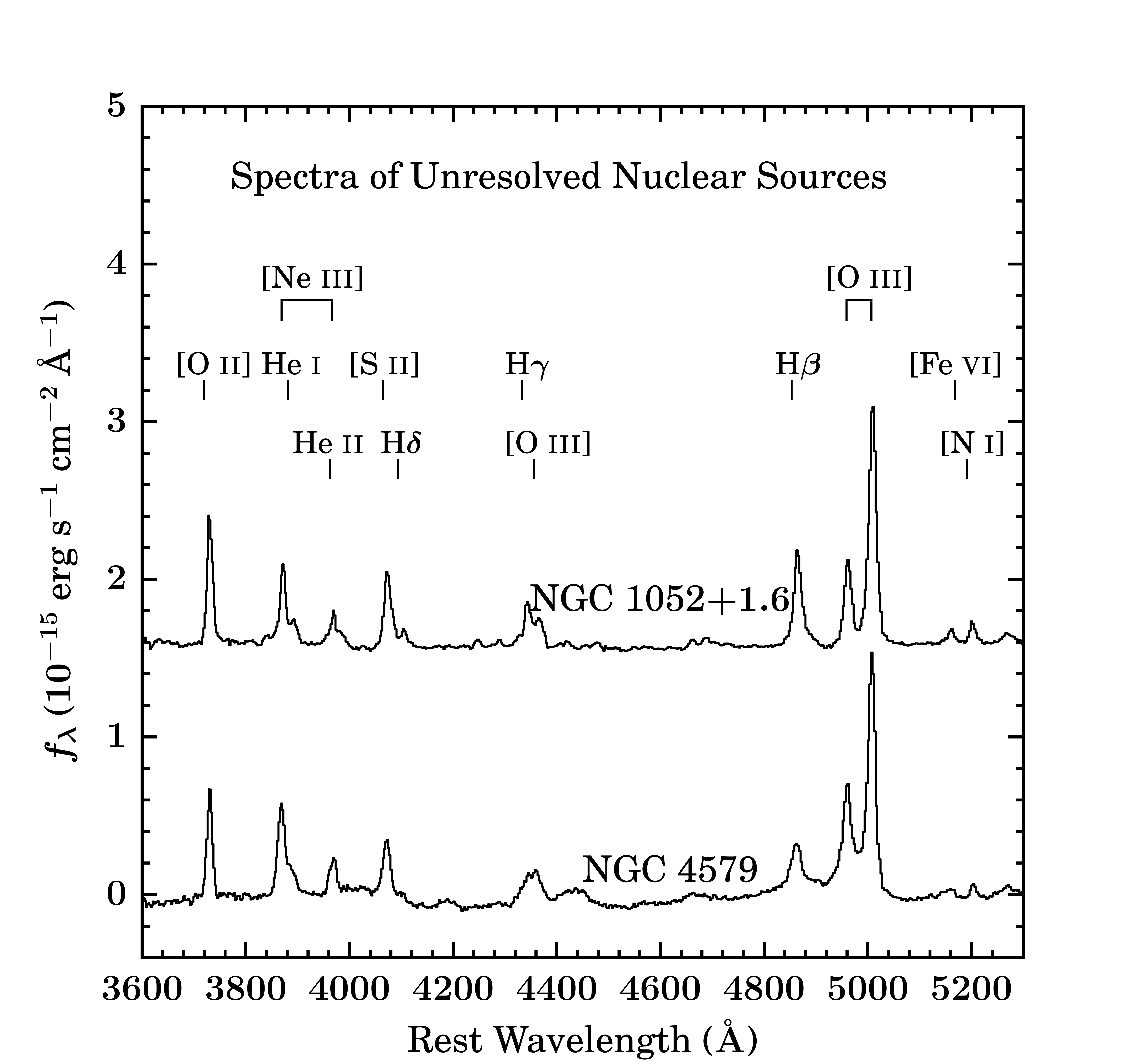}
\caption{Same as Figure~\ref{fig:unres_stack_uv}, for the blue spectra. NGC~4579 has broad Balmer emission lines indicative of an AGN.} \label{fig:unres_stack_hb}
%\end{figure}
\bigskip
%\begin{figure}[H]
\includegraphics[width=0.5\textwidth]{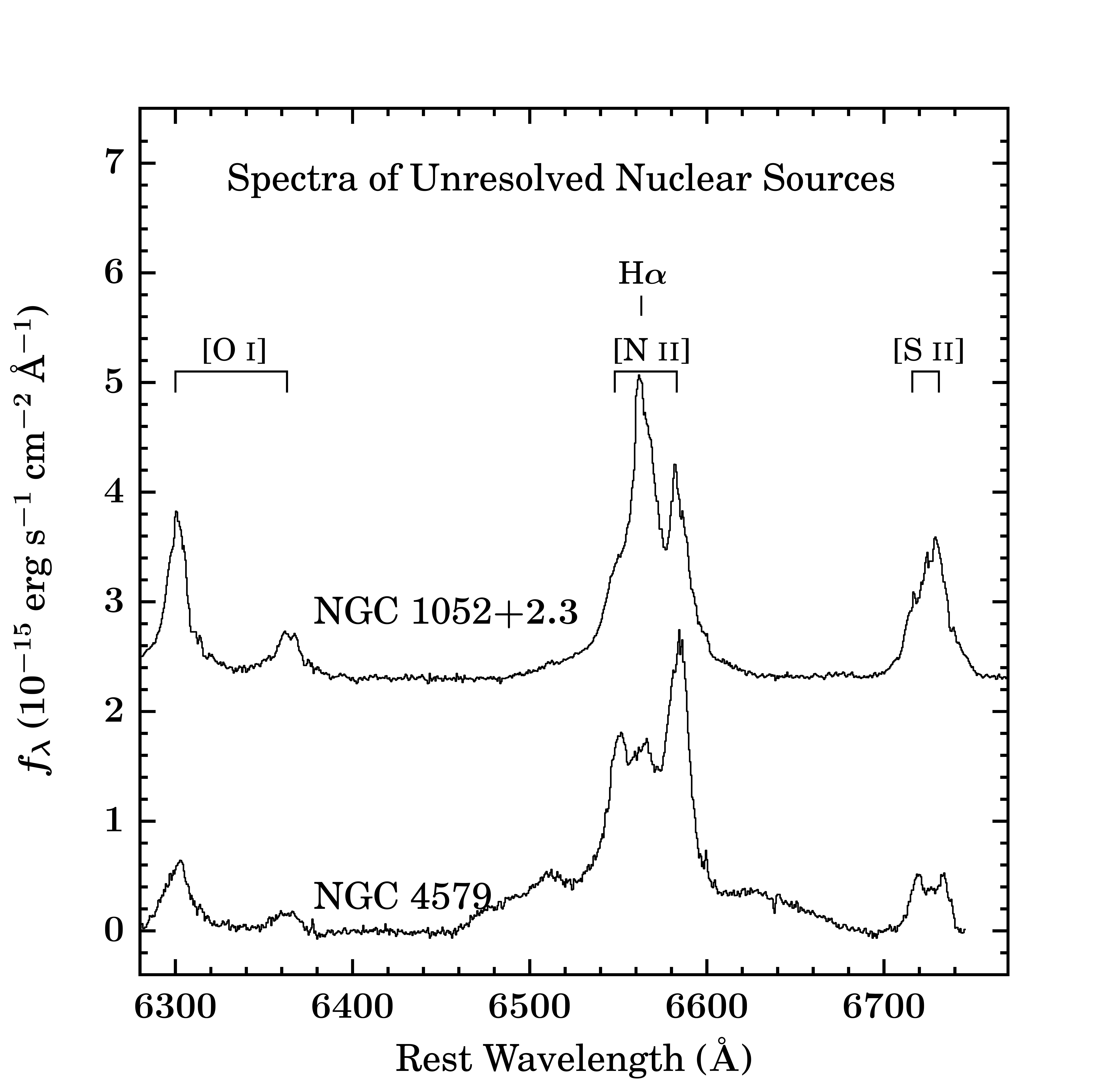}
\caption{Same as Figure~\ref{fig:unres_stack_uv}, for the red spectra. Both objects show broad Balmer emission lines indicative of an AGN.}
\label{fig:unres_stack_ha}
\end{figure}

\section{Subtraction of Starlight from the Extracted 1--D Spectra}
\label{sec:stars}

\subsection{Fitting Method}
\label{ssec:starfit}

After removing the contribution of the spatially unresolved nuclear source, we extracted 1--D spectra from each row of the extended (and resolved) 2--D spectrum of each galaxy as a function of distance from the nucleus along the slit. The spectra of the resolved light include starlight, while the spectra of the unresolved nuclear source in NGC~1052 and NGC~4579 have continua dominated by non-stellar light from the AGN and a contribution from starlight. We used the Penalized Pixel-Fitting code \texttt{pPXF} \citep{Cappellari2004} to fit the total continuum in each 1--D spectrum and isolate the emission lines. Our primary goal was to isolate the emission lines and not to determine the properties of the continuum. Therefore, we sought good fits to portions of the continuum where absorption lines are present but did not thoroughly explore the parameter space of stellar continuum models.

We adopted the stellar population synthesis models (known as simple stellar population, or SSP, models) from the MILES database \citep{Sanchez2006} for both the blue and red spectra, and the \citet[][BC03]{Bruzual2003} UV stellar models for the NUV spectra. \texttt{pPXF} requires the same wavelength sampling for the input spectra and stellar templates. The MILES library allows for user defined wavelength sampling, but the spectra it includes only cover a wavelength range of $\sim3500$--7400\AA. Therefore, we created the NUV stellar continuum templates by rebinning the BC03 models to match the wavelength scale of the observed NUV spectra. We assumed a Salpeter initial mass function for both sets of models, the Padova 2000 isochrones \citep{Girardi00} for the MILES SSP models, and the Padova 1994 evolutionary tracks \citep{padova94} for the BC03 models. The free parameters of both the MILES SSP models and BC03 models are  the metallicity $Z$ (0.02--$1.6\;Z_{\odot}$ for MILES, 0.005--$2.6\;Z_{\odot}$ for BC03) and population age (0.06--17.8 Gyr for MILES, 0--20~Gyr for BC03). We used a power law to model the continuum of the AGN in the 1--D unresolved nuclear spectra of NGC~1052 and NGC~4579. The starlight and power law continuum models were combined after weighting each by a low order multiplicative polynomial to account for any change to the overall shape of the continuum caused by extinction. All required components were fitted simultaneously by \texttt{pPXF}.

\begin{figure}[t]
\includegraphics[width=0.5\textwidth]{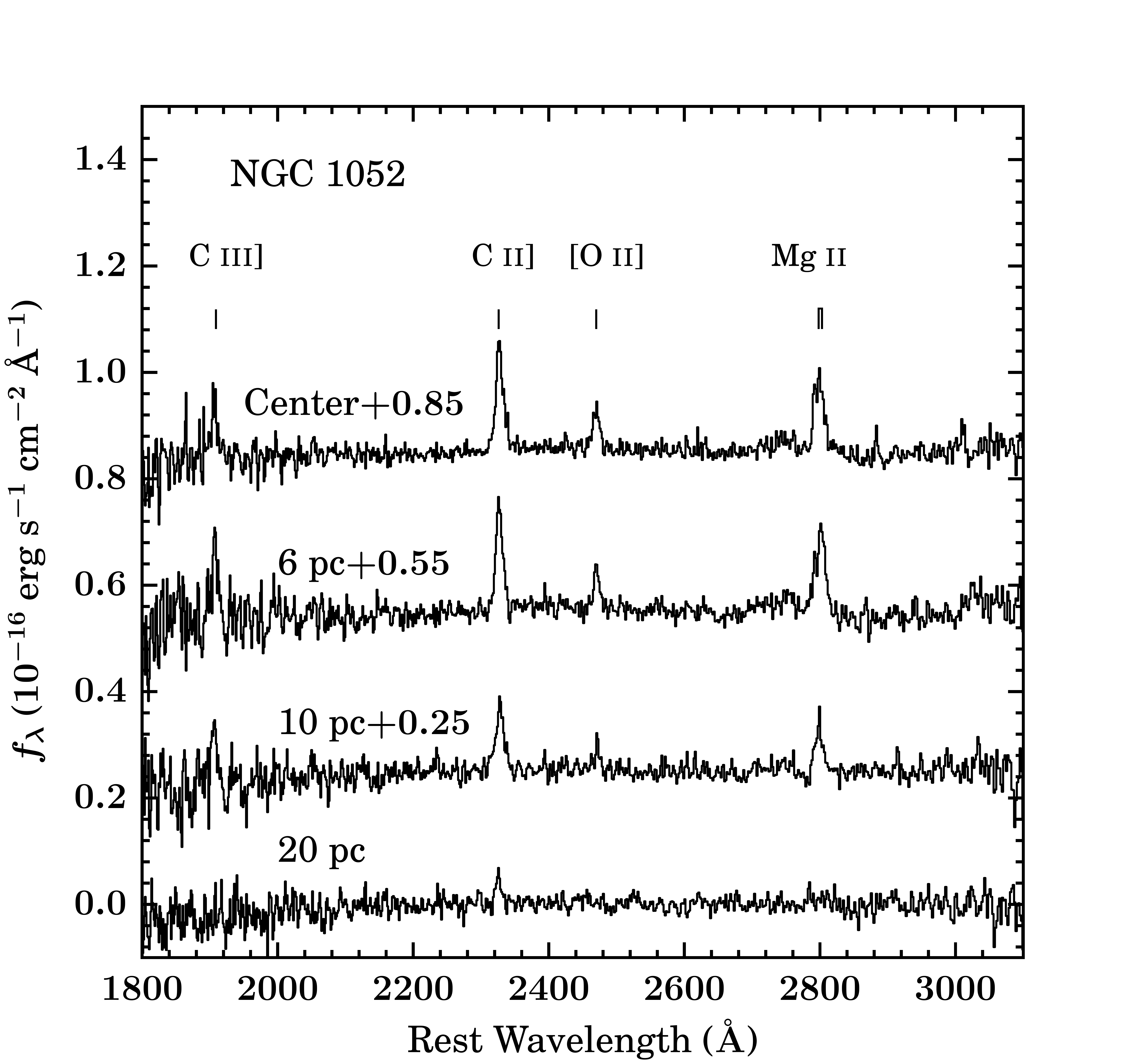}
\caption{The NUV spectra of the extended, resolved source in NGC 1052 at various distances moving outwards from the nucleus. The spectra are offset vertically from each other for clarity, by the offsets given in the figure. The prominent emission lines are marked and labeled. The \ion{C}{3}]$~\lambda1909$ line can be difficult to detect and measure because of the noise at the blue end of the spectra.}\label{fig:n1052_stack_uv}
\end{figure}
  
\begin{figure}[t]
\includegraphics[width=0.5\textwidth]{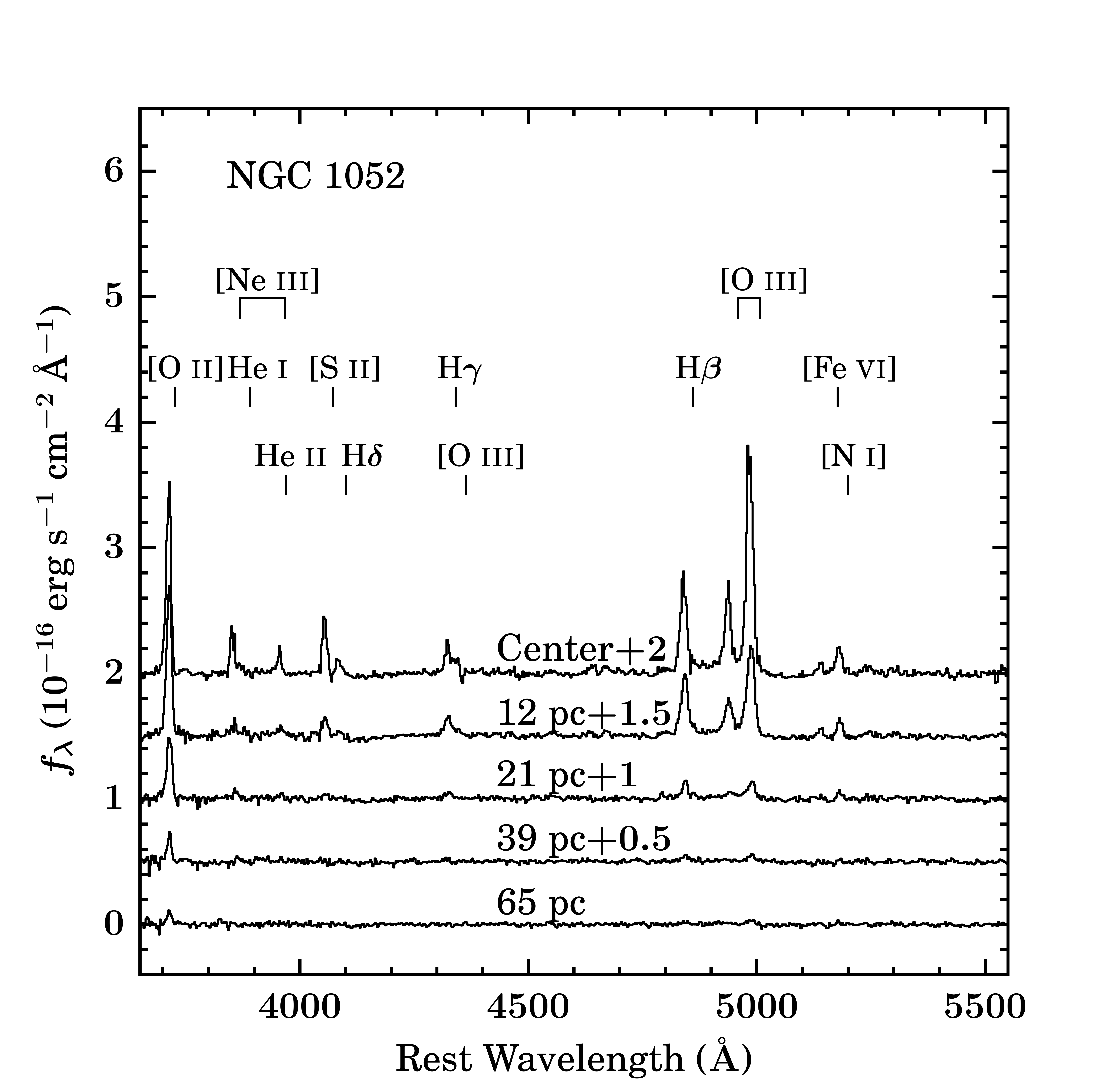}
\caption{Same as Figure~\ref{fig:n1052_stack_uv}, for NGC~1052, blue spectra. The [\ion{O}{2}]$~\lambda$3727 line is seen over the greatest spatial extent; it becomes stronger than the [\ion{O}{3}]$~\lambda$5007 line at 12~pc from the nucleus.}
\label{fig:n1052_stack_hb}
%\end{figure}        
%\bigskip
%\begin{figure}[H]
\includegraphics[width=0.5\textwidth]{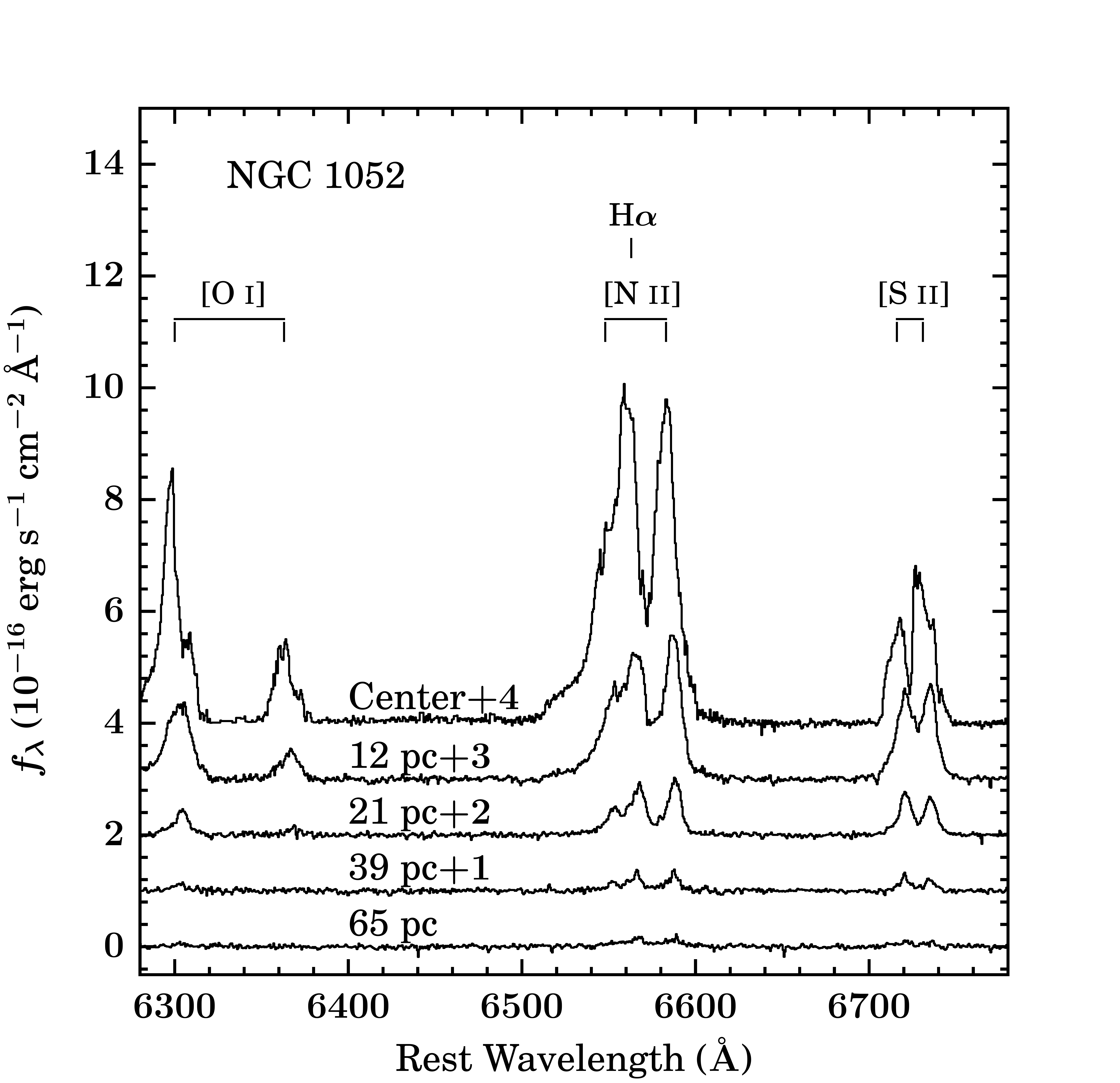}
\caption{Same as Figure~\ref{fig:n1052_stack_uv}, for NGC~1052, red spectra.}
\label{fig:n1052_stack_ha}
\end{figure}
 
\subsection{Fitting Results and Evaluation of Uncertainties}
\label{ssec:starres}

Figure~\ref{fig:starlight} shows an example of a fit to a single spectrum extracted from NGC~4579. The upper panel shows the fit and the residual spectrum, and the lower panel shows scaled versions of all the templates used in the fit. While each galaxy required a different combination of stellar population ages, all three galaxies were best fitted with the largest value of metallicity for the given family of models as judged from the slope of the underlying stellar continuum.

The spectra of NGC~1052 were well fitted by combining 2 and 10~Gyr stellar population models with a low order multiplicative polynomial. While all spectra required the 10 Gyr population to accurately describe the starlight, the 2~Gyr population was included only when necessary, and with no discernible trend as a function of distance from the nucleus. We also used a power law continuum in the NUV and blue spectra of the unresolved nuclear source in NGC~1052. 

The spectra of NGC~4278 were also well fitted by combining 2~Gyr and 10~Gyr stellar population models with a low order multiplicative polynomial. The 10~Gyr population was required at all distances from the nucleus, while the 2~Gyr population was used most heavily near the nucleus. There was no detected unresolved nuclear source, and no power law required for the fit. The spectra of NGC~4579 required 2, 3, 6, and 10 Gyr stellar populations and a low order multiplicative polynomial, with no trend in population age with distance from the nucleus. A power law was included in the model for the blue spectrum of the unresolved nuclear source.
%\subsection{Uncertainties in the Starlight Subtraction}
%\label{ssec:starerr}

We bootstrapped our best fitting continuum model to calculate the error  spectrum associated with the best fitting model.  Specifically, we simulated 1000 spectra by perturbing each data point according to its associated error bar, and repeated the fitting process using the parameter values found in the best fitted continuum model as input parameters. The distribution of resulting model parameters was Gaussian and we took the standard deviation of this distribution as the error on the parameters of the continuum fit. This process was carried out for each extracted 1--D spectrum, as well as the unresolved 1--D spectrum, for all three galaxies. We incorporated the error bars from this exercise in our error propagation calculations. 

\begin{figure}[t]
\includegraphics[width=0.5\textwidth]{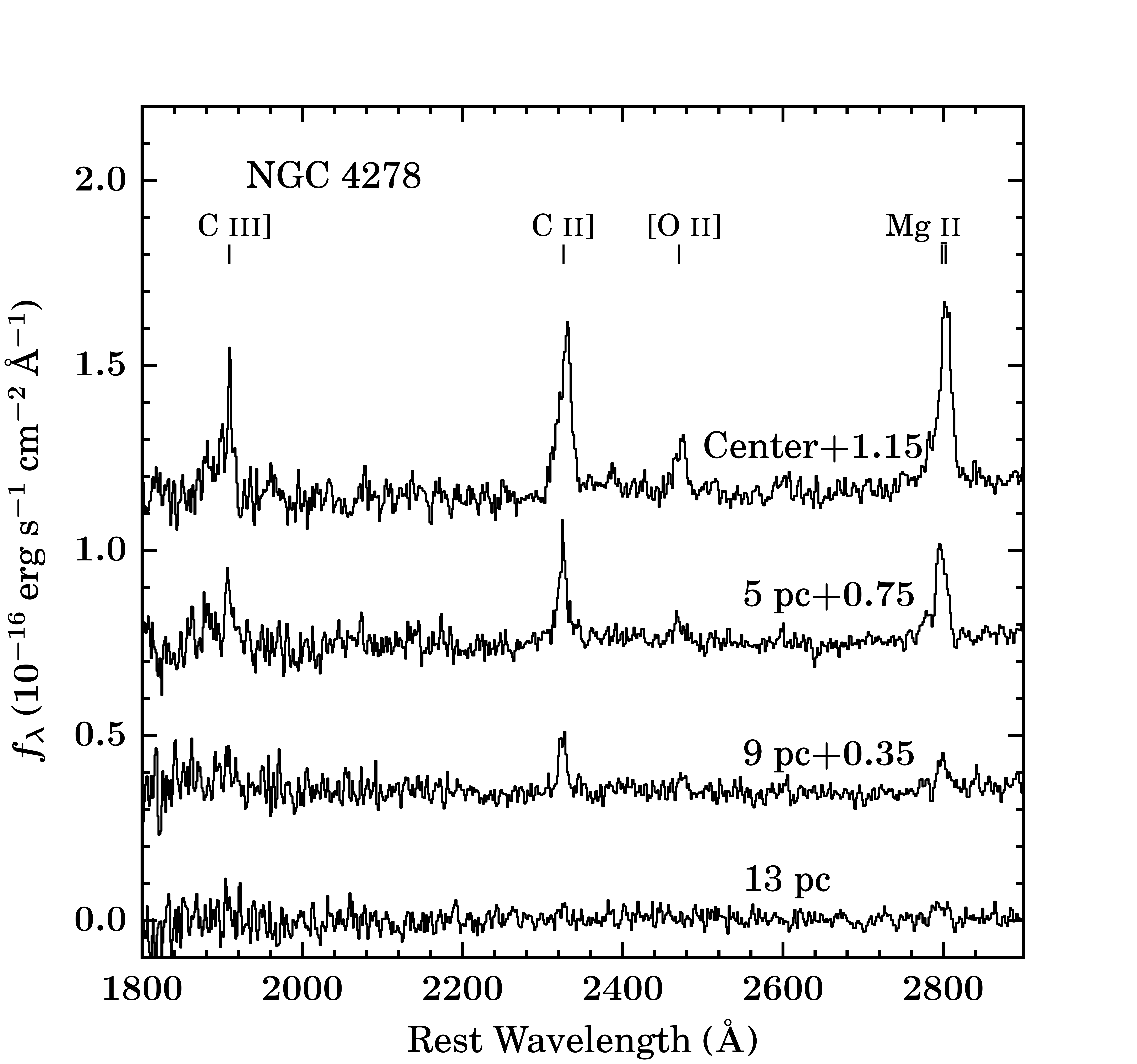}
\caption{Same as Figure~\ref{fig:n1052_stack_uv}, for NGC~4278, NUV spectra. The excess noise at the blue end of the spectrum makes \ion{C}{3}]$~\lambda1909$ difficult to detect past 5~pc from the center.}
\label{fig:n4278_stack_uv}
\end{figure}        

\begin{figure}[t]
\includegraphics[width=0.5\textwidth]{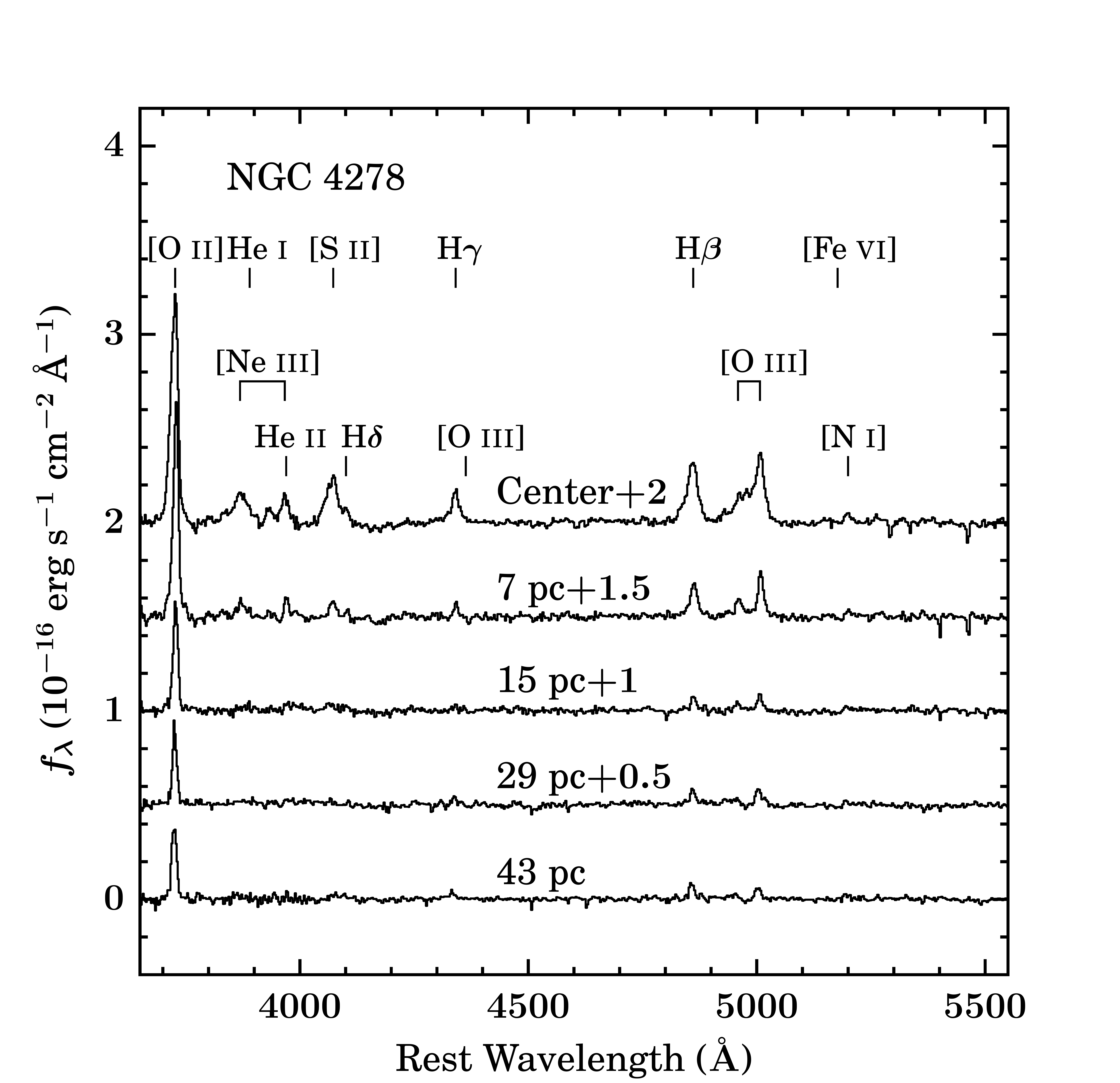}
\caption{Same as Figure~\ref{fig:n1052_stack_uv}, for NGC~4278, blue spectra. The [\ion{O}{2}]$~\lambda$3727 line is stronger than the [\ion{O}{3}]$~\lambda$5007 line at all distances.} \label{fig:n4278_stack_hb}
%\end{figure}
%  
%\begin{figure}[H]
\includegraphics[width=0.5\textwidth]{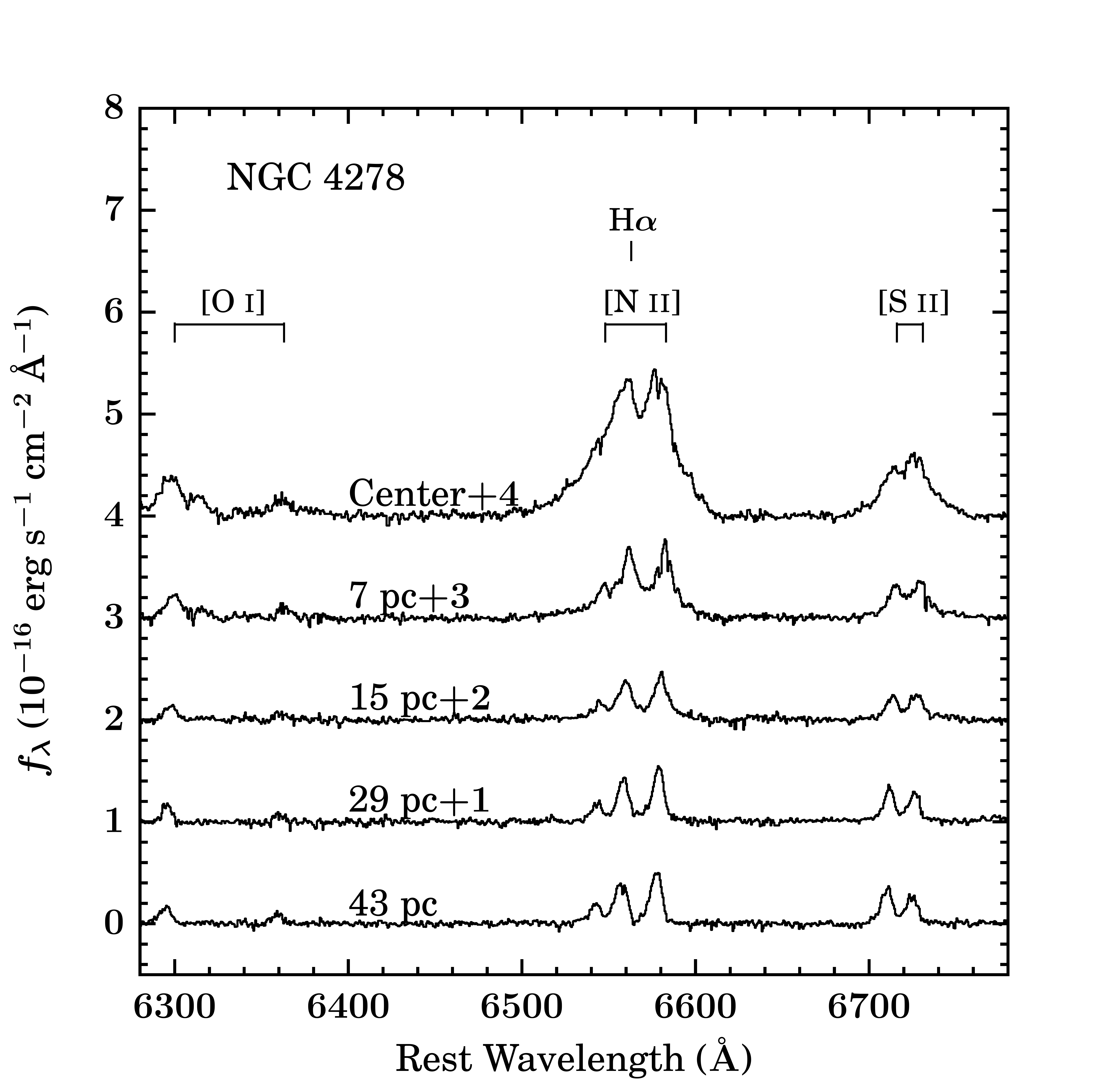}
\caption{Same as Figure~\ref{fig:n1052_stack_uv}, for NGC~4278, red spectra.} 
\label{fig:n4278_stack_ha}
\end{figure}

%Emission Lines
%%%%%%%%%%%%%%%%%%%%%%%%%%%%%%%%%%%%%%%%%%%%%%%%%%%%%%%%%%%%%%%%%%%%%%%%%%%%%%%%%
\section{Emission Line Spectra and Measurements}
\label{sec:lines}
\subsection{Presentation and Qualitative Discussion of Spectra}
\label{ssec:sdresult}
\begin{figure}[t]
\includegraphics[width=0.5\textwidth]{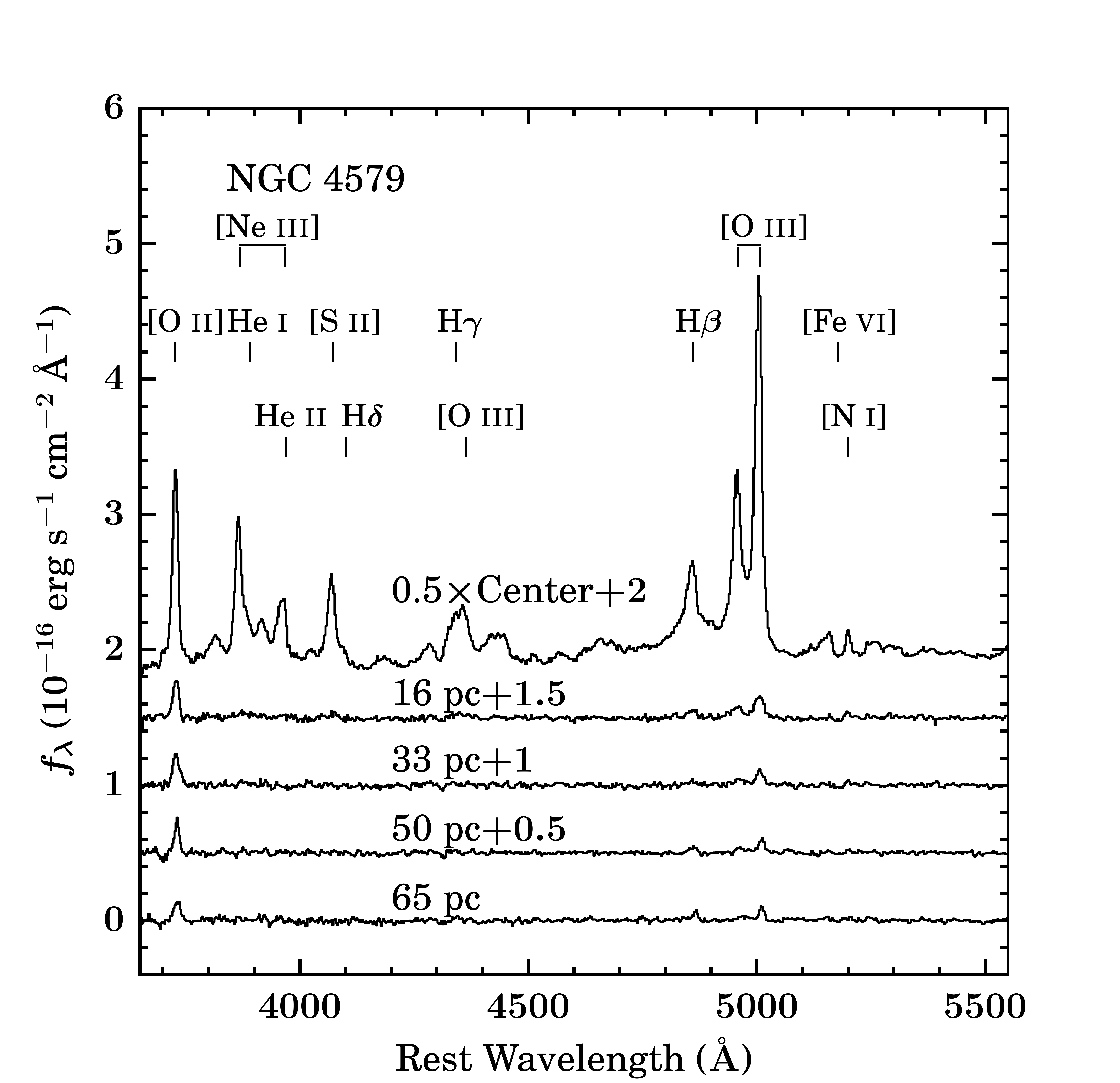}
\caption{Same as Figure~\ref{fig:n1052_stack_uv}, for NGC~4579, blue spectra. The central spectrum is scaled by 0.5 for easier comparison with the other spectra. [\ion{O}{2}]$~\lambda$3727 is seen at the greatest spatial extent, and becomes stronger relative to [\ion{O}{3}]$~\lambda$5007 at 16~pc from the nucleus.} \label{fig:n4579_stack_hb}
%\end{figure}
%
%\begin{figure}[t]
\includegraphics[width=0.5\textwidth]{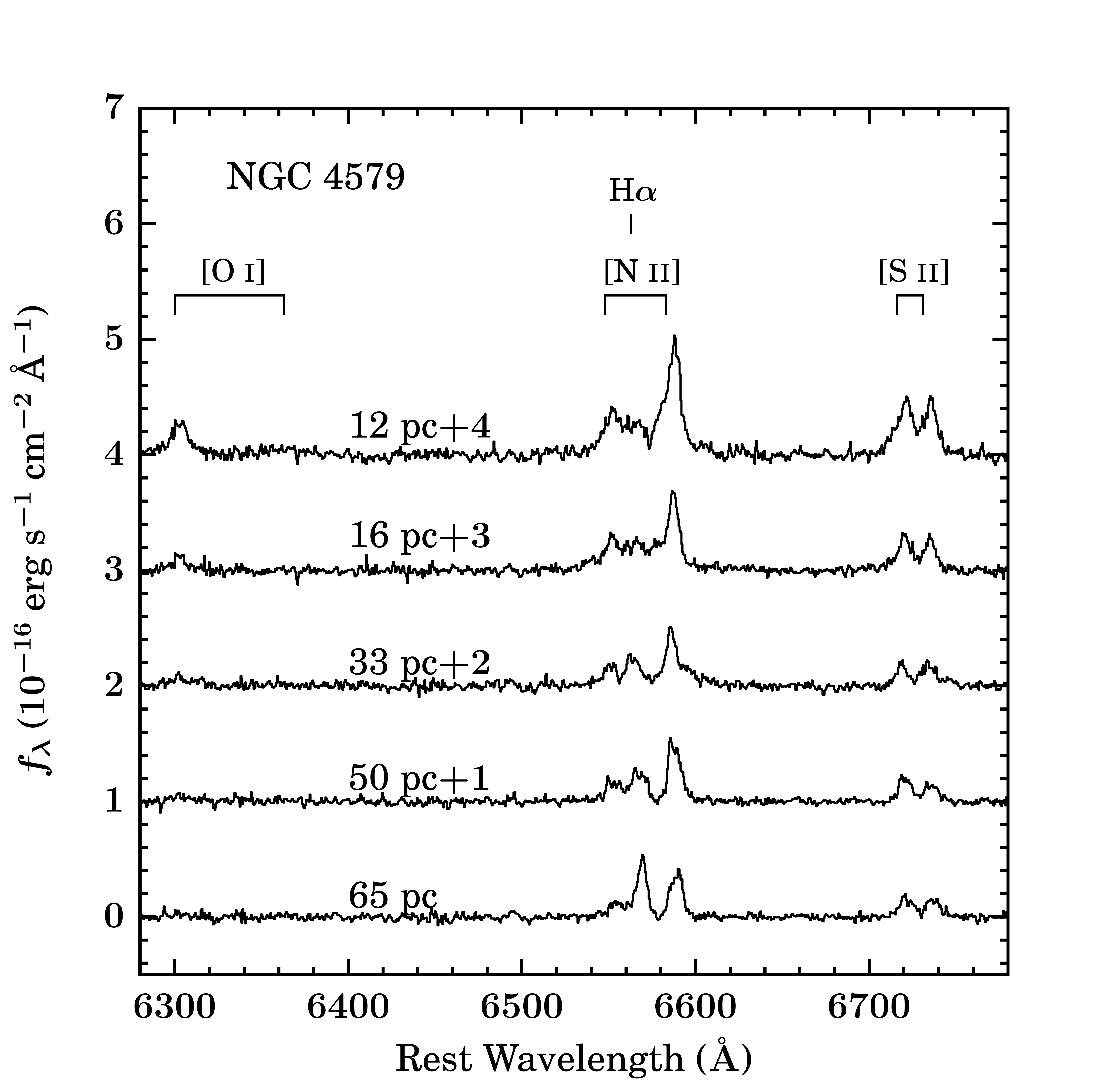}
\caption{Same as Figure~\ref{fig:n1052_stack_uv}, for NGC~4579, red spectra. We do not detect resolved emission in the nucleus of the galaxy because of the strength of the unresolved emission.}
\label{fig:n4579_stack_ha}
\end{figure}

\begin{figure}[t]
\includegraphics[width=0.5\textwidth]{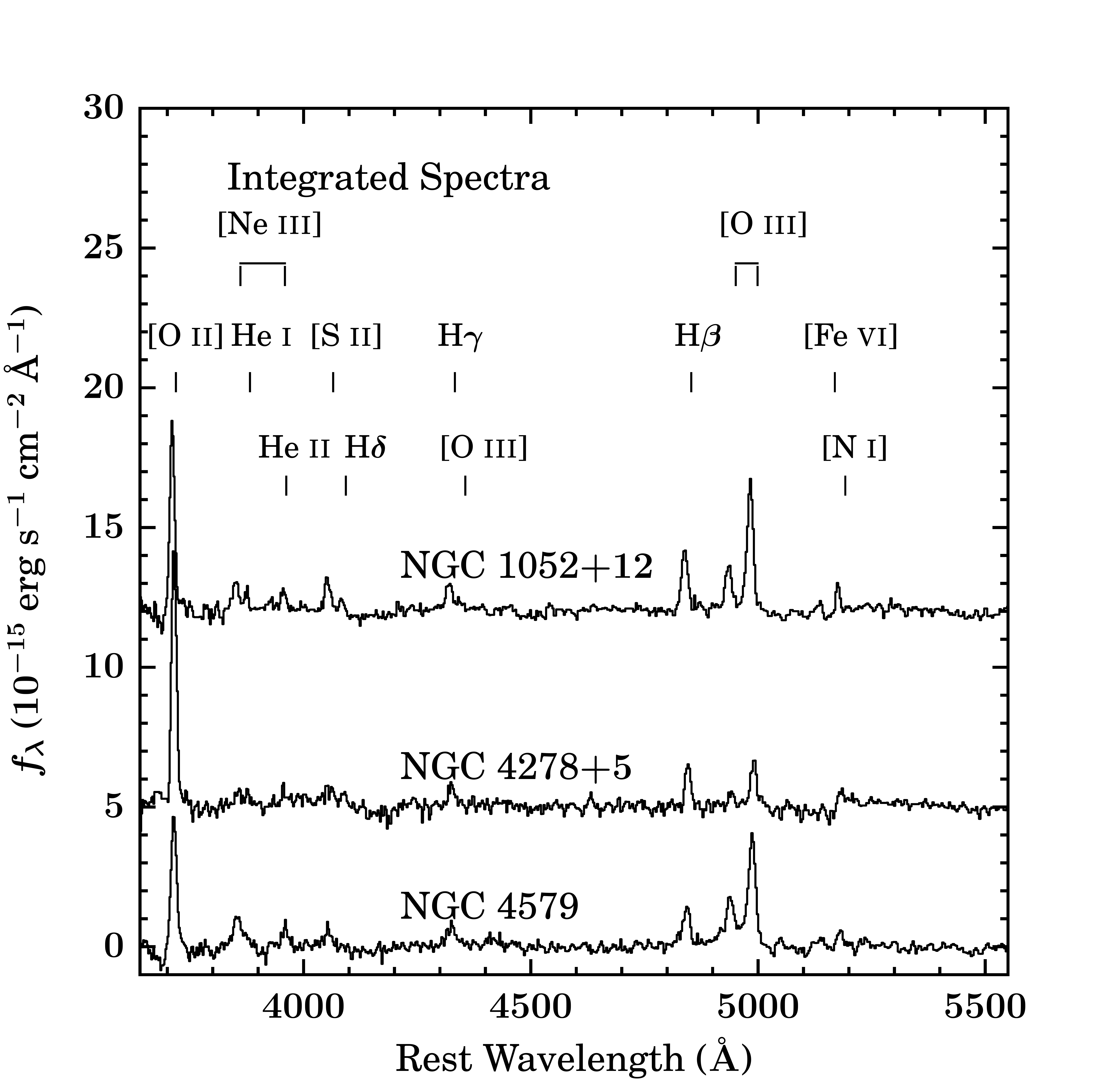}
\caption{Same as Figure~\ref{fig:unres_stack_uv}, for integrated blue spectra, constructed as described in Section~\ref{ssec:sdresult}. The [\ion{O}{2}]$~\lambda$3727 line is significantly stronger than the [\ion{O}{3}]$~\lambda$5007 line in all objects.}\label{fig:int_stack_hb}
%\end{figure}
% 
%\begin{figure}[t]
\includegraphics[width=0.5\textwidth]{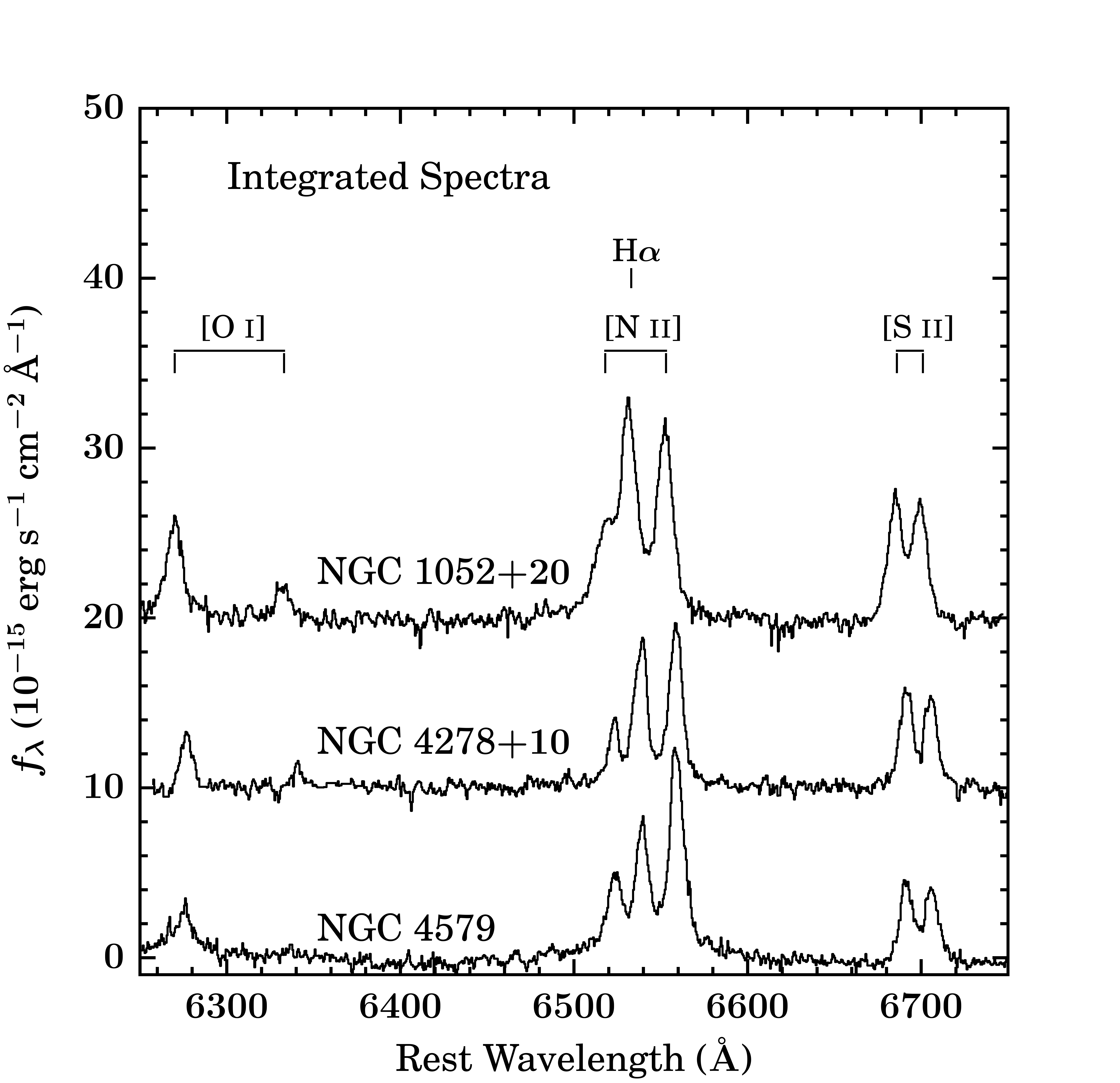}
\caption{Same as Figure~\ref{fig:unres_stack_uv}, for integrated red spectra, constructed as described in Section~\ref{ssec:sdresult}. The broad wings on the H$\alpha$ line are discernible in NGC~1052 and NGC~4579.}
\label{fig:int_stack_ha}
\end{figure}

A subset of the final spectra that we used in our analysis are shown in Figures~\ref{fig:unres_stack_uv} through~\ref{fig:int_stack_ha}, where they are presented without extinction corrections. We combined the spectra of the resolved light in pairs of two for our analysis because the PSF of each instrument is approximately two pixels wide, which means that adjacent spectra are not independent. We obtained the nuclear spectrum of the extended, resolved light by combining the two 1--D spectra that bracket the peak of the spatial profile.

The spectra of the three objects exhibit different qualitative behaviors as a function of distance from the nucleus, as described below.

\begin{description}
  
\item[NGC~1052]-- The spectrum of the unresolved nuclear source shows broad H$\alpha$, H$\beta$ and \ion{Mg}{2}\;$\lambda\lambda$2798,2803 lines as well as optical \ion{Fe}{2} blends in the vicinity of H$\beta$. The NUV spectra of the extended resolved light have detectable emission lines only within 40~pc. In the blue spectra, the strength of the [\ion{O}{3}]$~\lambda$5007 line declines faster with distance from the nucleus than that of the [\ion{O}{2}]$~\lambda$3727 line. H$\alpha$ and [\ion{N}{2}]$~\lambda\lambda$6548,6583 are the strongest lines in the red spectra of the extended resolved light, and are detected to distances of up to 80~pc.

\item[NGC~4278]-- The NUV spectra of the extended resolved light have detectable emission lines only within the central 18~pc.  The relative intensities of the lines in the blue spectra of the extended, resolved light do not vary significantly as a function of distance from the nucleus. The H$\alpha$ and [\ion{N}{2}]$~\lambda\lambda$6548,6583 lines are stronger than the [\ion{S}{2}]$~\lambda\lambda6716,6731$ doublet at and near the nucleus but their relative strengths change with distance from the nucleus and become comparable by 29~pc from the nucleus.
 
\item[NGC~4579]-- The NUV spectrum is dominated by the spatially unresolved nuclear source with no discernible contribution from extended emission, hence Figure~\ref{fig:unres_stack_uv} shows the integrated NUV spectrum. The [\ion{O}{3}]$~\lambda$5007 line declines in strength faster than the [\ion{O}{2}]$~\lambda$3727 line with distance from the nucleus. The unresolved nuclear source dominates and overwhelms any resolved emission in the red spectra in the vicinity of the nucleus. Therefore, the red spectra of the extended, resolved light shown in Figure~\ref{fig:n4579_stack_ha} start at 12~pc from the nucleus. H$\alpha$ and [\ion{N}{2}]$~\lambda\lambda$6548,6583  are the strongest lines in the red spectra of the extended, resolved light, detected out to 80~pc. The NUV and blue spectra of the unresolved nuclear source also show \ion{Fe}{2} and \ion{Fe}{3} complexes that are characteristic of AGN spectra. These complexes are also evident in the central blue spectrum of the resolved emission, which is contaminated by light from the unresolved nuclear source.

\end{description}

%\clearpage

\begin{figure}
\includegraphics[width=0.5\textwidth]{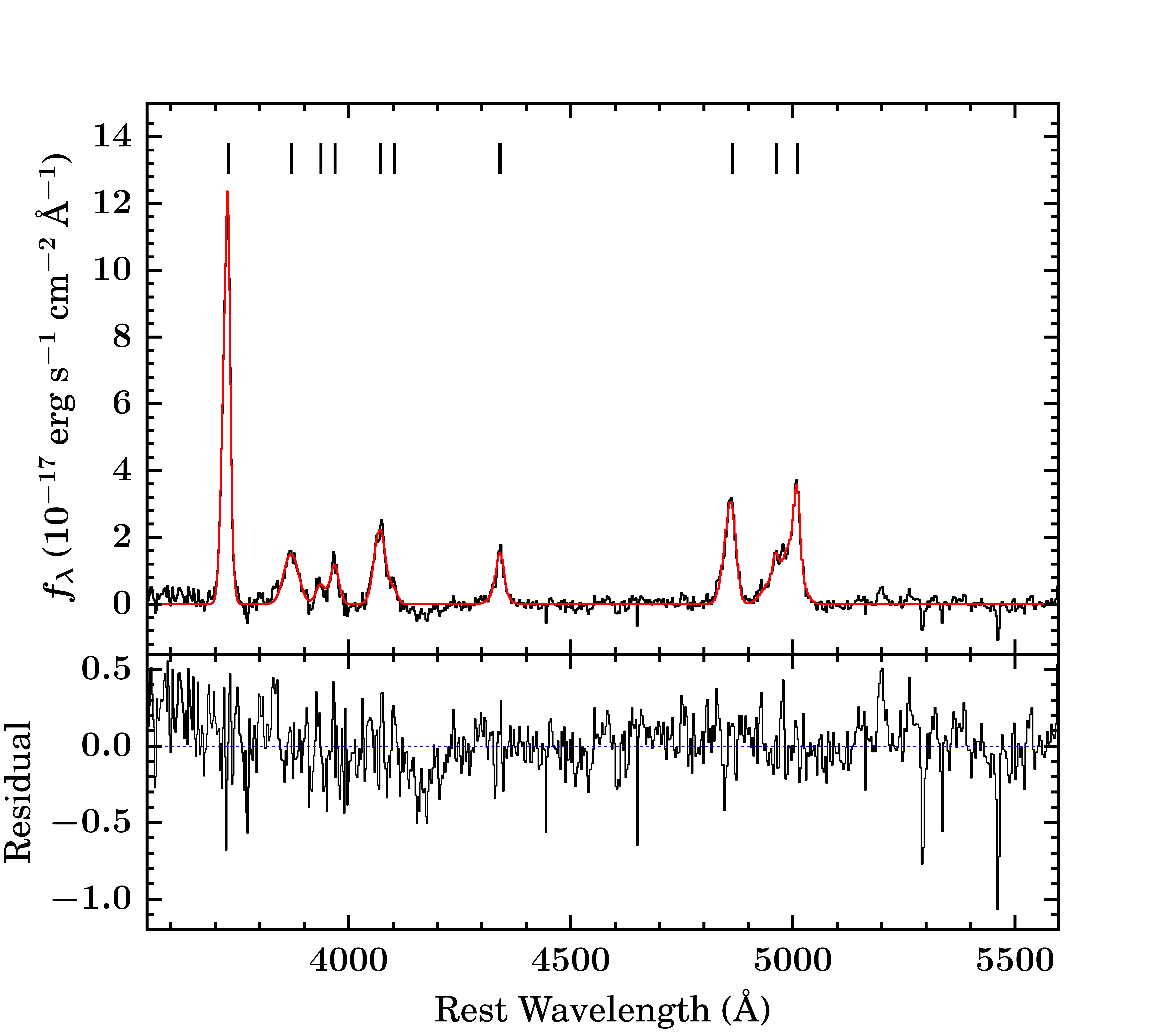}
\caption{The emission line decomposition for the central blue spectrum of the extended resolved source in NGC~4278. The top panel has the data shown in black and the fit from \texttt{specfit} shown in red (smooth, solid line). Each emission line is marked with a vertical hash. The bottom panel shows the residual when the model is subtracted from the data, with a blue dashed line at zero to guide the eye.  See Section~\ref{ssec:elmeasure} for a discussion on emission line measurements.}\label{fig:n4278_lfit}
\includegraphics[width=0.5\textwidth]{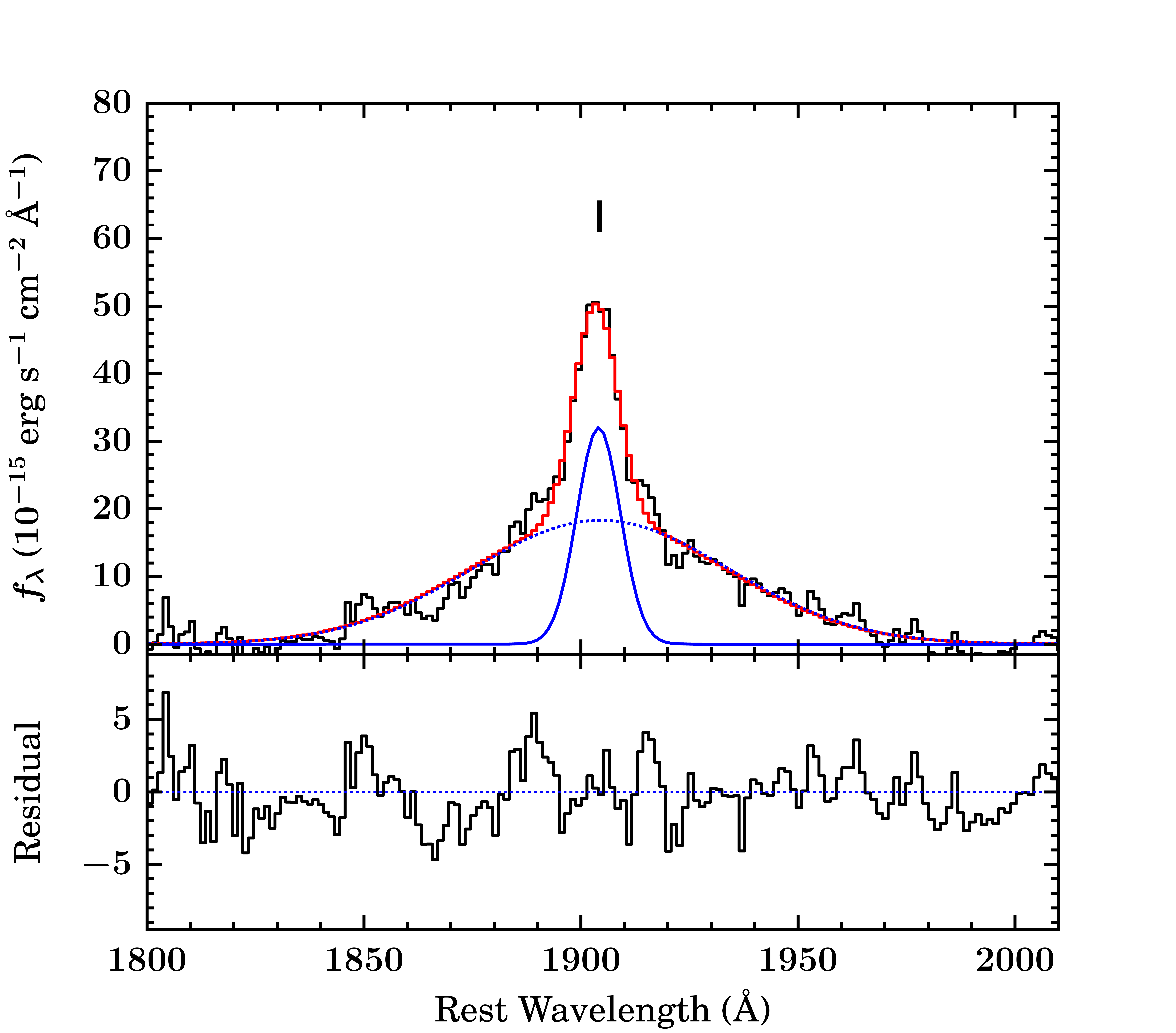}
\caption{The fit of the broad and narrow \ion{C}{3}]$~\lambda1909$ line components in the spectrum of the spatially unresolved nuclear source of NGC~4579. The top panel has the data shown in black, the total fit from \texttt{specfit} shown in red (upper smooth curve) and the emission line components shown in blue. The centroid of the emission line components overlap and are marked with a vertical hash. The dashed blue line shows the broad component, and the solid blue (lower smooth curve) line shows the narrow line component. The bottom panel shows the residual when the model is subtracted from the data, with a blue dashed line at zero to guide the eye.}\label{fig:n4579_lfit}
\end{figure} 

We also present synthetic  integrated spectra through a larger aperture in the blue and red bands, which include both the unresolved and resolved light. The results of this exercise are shown in Figures~\ref{fig:int_stack_hb} and~\ref{fig:int_stack_ha}. The synthetic spectra represent the emission from a circle centered on the nucleus with a radius of $1^{\prime\prime}$, i.e., the expected LINER-like spectrum from a ground based observation. They were produced via a weighted sum of observed spectra at different distances from the galaxy center with weights chosen according to the area of an annulus at that distance from the center. The simulated spectra also approximate the spectra we expect from a $2^{\prime\prime}\times4^{\prime\prime}$ rectangular slit used in some of the well known surveys of LINERs \citep[e.g.,][]{phill86,HoFS97III}. Both the blue and red integrated spectra are reminiscent of the characteristic LINER spectrum. All three objects have significantly stronger [\ion{O}{2}]$~\lambda$3727 emission compared to [\ion{O}{3}]$~\lambda$5007 in their integrated blue spectra. Moreover, the broad Balmer lines from the nuclei of NGC~1052 and NGC~4579 are diluted so that only a weak broad H$\alpha$ pedestal is discernible.

\subsection{Measurement of Emission Line Strengths}
\label{ssec:elmeasure}
We measured the emission line fluxes using the \texttt{iraf} task \texttt{specfit} \citep{Kriss94}, which fits the line profiles with linear combinations of Gaussians by minimizing the $\chi^2$ statistic using the simplex method. We also inspected the fit and the post fit residuals to make sure they had no systematic patterns. We regard the Gaussian components only as a means of parameterizing the line profiles and do not attach any particular physical significance to them. Near the nuclei of NGC~1052 and NGC~4579, we modeled any detected \ion{Fe}{2} and \ion{Fe}{3} emission with optical \ion{Fe}{2} templates from \citet{Boroson1992} for the blue spectra, and NUV Fe templates from \citet{Vestergaard2000} for the NUV spectra. The spectra of both the spatially unresolved nuclear sources and the central region of the extended, resolved sources in NGC~1052 and NGC~4579 have broad lines which we fitted simultaneously with the narrow lines. We show an example of a narrow line only fit in Figure~\ref{fig:n4278_lfit} and an example of a fit with both broad and narrow lines in Figure~\ref{fig:n4579_lfit}.  We report the measurements (without any extinction corrections) in Tables~\ref{table:uv_lines}--\ref{table:ha_lines}.  

\begin{figure}[t]
\vspace{-1em}
\includegraphics[width=0.5\textwidth]{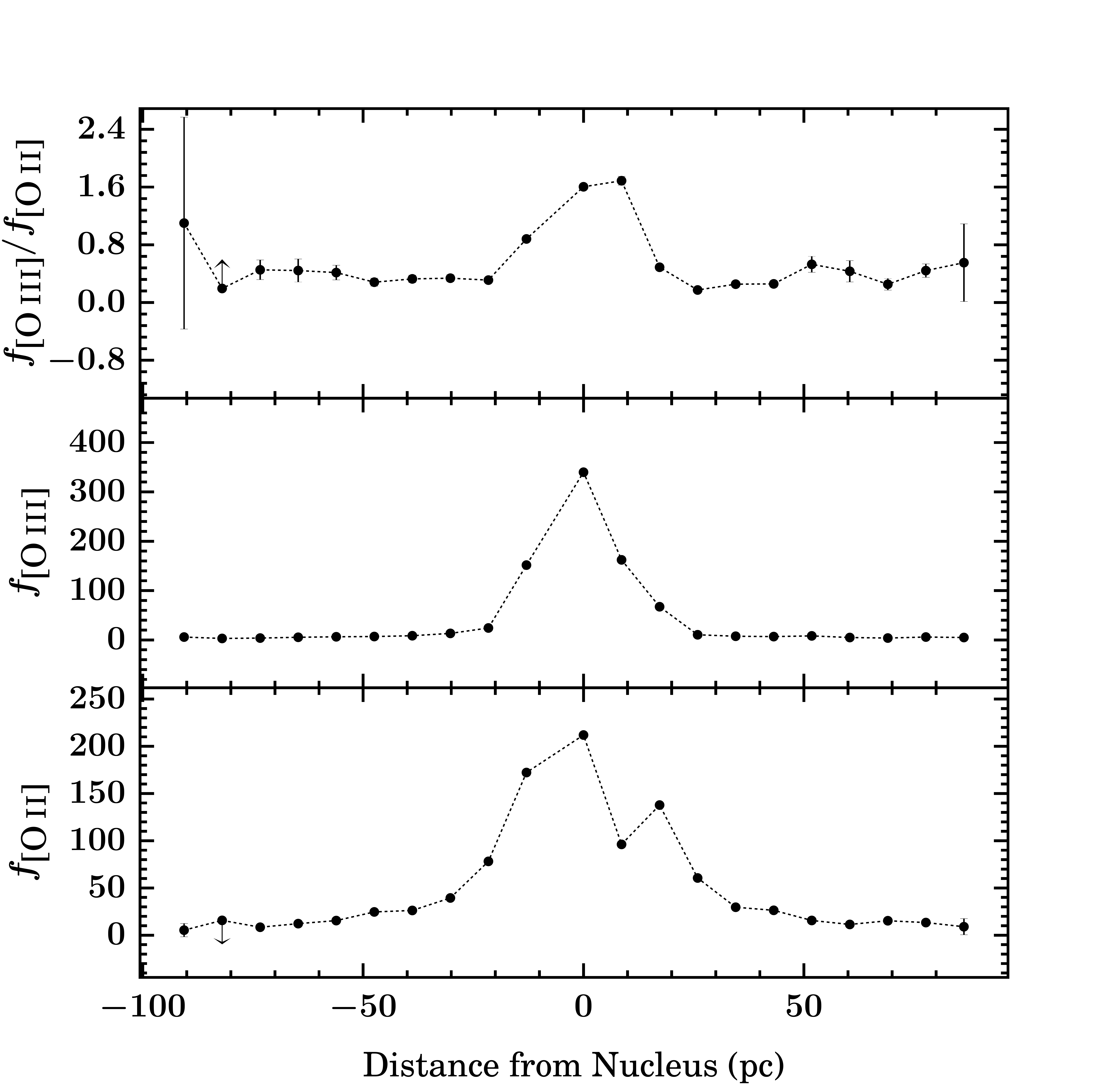}
\caption{The change of the [\ion{O}{3}]$~\lambda$5007 and [\ion{O}{2}]$~\lambda$3727 emission line fluxes in NGC~1052 with distance from the nucleus (see  Section~\ref{ssec:elmeasure}). The flux scale on the vertical axis is in units of $10^{-18}$~erg~s\textsuperscript{$-1$}~cm\textsuperscript{$-2$}~\AA\textsuperscript{$-1$}~arcsec\textsuperscript{$-2$}. \textit{Top:} The measured line ratio of $\textrm{[\ion{O}{3}]}/\textrm{[\ion{O}{2}]}$. The line ratio changes and flattens out to a non-zero value at 20~pc from the nucleus, suggesting a change in the excitation mechanism. \textit{Middle:} The change in the measured line flux of [\ion{O}{3}]$~\lambda$5007. \textit{Bottom:} The change in the measured line flux of [\ion{O}{2}]$~\lambda$3727.}\label{fig:n1052_o3o2}
\end{figure}

We constrained the widths of lines in doublets, namely \ion{Mg}{2}$~\lambda\lambda$2798,2803, [\ion{O}{3}]$~\lambda\lambda$4959,5007, [\ion{O}{1}]$~\lambda\lambda$6300,6363, [\ion{N}{2}]~$\lambda\lambda$6548,6583, and [\ion{S}{2}]~$\lambda\lambda$6716,6731 to be equal. We also imposed the corresponding physical flux ratio on the above doublets as follows: no constraint on [\ion{S}{2}]$~\lambda\lambda$6716, 6731 since the ratio depends on the density, 1/2 for the \ion{Mg}{2}$~\lambda\lambda$2798,2803 doublet, and 1/3 for all the others. Finally, if the weaker line in a doublet was not detected or blended with another line, we also constrained the wavelength ratio of the two lines. The strong blends in the H$\alpha$ complex in both the unresolved nuclear and the central resolved spectra NGC~1052 required us to constrain the [\ion{N}{2}]$~\lambda\lambda$6548,6583 widths to equal the [\ion{S}{2}]$~\lambda\lambda$6716,6731 widths. In the unresolved nuclear spectrum of NGC 1052, we further constrained the width of narrow H$\alpha$ to equal that of H$\beta$. We measured all emission lines detected with at least a 3$\sigma$ confidence, and measured upper limits for all expected lines in every spectrum where at least one line was detected. To calculate the upper limits, the shape of [\ion{O}{3}]$~\lambda$5007 from the central resolved spectrum for each object was scaled, except for [\ion{Fe}{10}]$~\lambda 6374$ and [\ion{Fe}{14}]$~\lambda 5303$ for which we used H$\beta$ for the upper limit calculation. The error bars on the line fluxes were calculated by \texttt{specfit}, \noindent and they correspond to a 1$\sigma$ confidence interval for one interesting parameter. For all doublets constrained by a fixed flux ratio, the error bar on the weaker line was obtained from the error bar on the stronger line. The measured line fluxes reflect the qualitative trend in the relative intensities discussed in Section~\ref{ssec:sdresult}. As an example, we show in Figure~\ref{fig:n1052_o3o2} the variation of the  the [\ion{O}{3}]$~\lambda$5007 and [\ion{O}{2}]$~\lambda3727$ line strengths of NGC~1052 and their ratio. The $\textrm{[\ion{O}{3}]}/\textrm{[\ion{O}{2}]}$ ratio is higher in the nucleus, and declines to a constant non-zero value at about 20~pc from the nucleus. This suggests a change in the excitation mechanism, as we discuss in Section~\ref{sec:ddresult}.

\input{uv_table.tex}

\subsection{Kinematics of the Line-Emitting Gas}
\label{ssec:kinematics}
To probe the kinematics of the line-emitting gas, we use the centroid velocities and FWHM of selected emission lines. As we noted in Section~\ref{sec:obsdata}, \citet{Walsh2008} made detailed measurements of these quantities for the three galaxies we study here using the  [\ion{N}{2}]\;$\lambda6548$ line. They use the same red spectra that we use here, which have a high enough spectral resolution to allow one to study the line profiles. We supplemented those measurements with measurements of the [\ion{S}{2}]\;$\lambda\lambda6716,6731$ and [\ion{O}{1}]\;$\lambda6363$ lines, which are not near the end of the spectra and relatively easy to deblend. Our newly acquired blue and NUV spectra have a lower spectral resolution than the red spectra, with the result that the emission lines in these spectra are not as well resolved. Therefore, we measure only the centroid of [\ion{O}{2}]\;$\lambda3727$, the strongest line at substantial distances from the nucleus.

The results of our measurements are in good agreement with those of \citet{Walsh2008}, therefore we refer the reader to their Figures~2a, 2h, 2i, and 4 for a graphical illustration. In summary, while two of the objects show some structure, none of the galaxies have regular disk rotation. NGC~1052 superficially appears to have disk rotation but has evidence of an outflow, NGC~4278 has some rotational structure but a kinematic twist near the center, and NGC~4579 does not have any evidence for regular rotation. The velocity difference between the extrema of the radial velocity curves is 400--600~km~s\textsuperscript{$-1$}. The FWHM of the lines at distances greater than 20~pc from the nucleus of each galaxy is in the range 150--350~km~s\textsuperscript{$-1$}, according to the measurements of \citet{Walsh2008}. Our measurements were made after grouping together spectra from adjacent rows, therefore the FWHM of the lines are higher as a result of the velocity gradients, falling in the range 200--500~km~s\textsuperscript{$-1$}. In the inner 20~pc the FWHM of the lines rises steeply, reaching values twice as large as those outside of 20~pc.

\clearpage

\input{hb_1052_table_1.tex}

\input{hb_4278_table_1.tex}

\input{hb_4579_table_1.tex}

\clearpage

\input{ha_table_1.tex}

\subsection{Extinction Corrections}
\label{ssec:reddening}
We calculated and applied extinction corrections to the emission line fluxes after they were measured. Since our galaxies have $z < 0.005$, we applied a single correction representing both extinction at the source and extinction in the Milky Way. We calculated $A_{\textrm{V}}$ assuming the Cardelli Law \citep{Cardelli89} and $R_{\textrm{V}}=3.1$. The observed H$\alpha$/H$\beta$ ratio in each galaxy was compared iteratively to an assumed intrinsic ratio which changed based on the dominant excitation mechanism at that physical scale.

\input{n1052_ext_corr_v1.tex}
\input{n4278_ext_corr.tex}
\input{n4579_ext_corr.tex}
We found the appropriate excitation mechanism of the spectrum by comparing reddening insensitive diagnostic line ratios to the models as we describe in Section~\ref{sec:ddresult}. We applied the correction according to the expected intrinsic ratio and compared the corrected diagnostic line ratios to the same sets of models again. If the difference between the uncorrected and corrected line ratios did not change the inferred dominant excitation mechanism at that physical distance from the nucleus, we adopted that intrinsic ratio. Additional details about the models are deferred to Section~\ref{sec:ddresult}.

The calculated values of $A_{\textrm{V}}$ and the assumed intrinsic H$\alpha$/H$\beta$ ratios for each spectrum are in Tables~\ref{table: 1052_ext_corr}--\ref{table: 4579_ext_corr}, with the exception of the central spectra of NGC~4579, where the unresolved source overwhelms the resolved emission and makes measurements of the latter impossible. These extinction correction values were also adapted for use in the NUV emission lines, which are probed on a different spatial scale than the extinction measurements. The observed extinction is patchy, and there is no overall trend with distance from the nucleus, which may cause additional uncertainty in the NUV emission line ratios discussed in Section~\ref{ssec:el_gal_results}.

We note that since the extinction corrections are derived from the ratio of two lines observed with different gratings, they are affected by the uncertainties in spatial decomposition discussed in Section~\ref{ssec:2derr}. However, we emphasize that (a) uncertainties in spatial decomposition affect only extinction estimates for spectra extracted from $\pm3$ pixel rows from the nucleus, and (b) uncertainties in extinction affect critically only one of the optical diagnostic line ratios that we use in later sections, the [\ion{O}{3}]$~\lambda5007$/[\ion{O}{2}]$~\lambda3727$ ratio. All other optical line ratios that we use involve closely separated lines and thus are insensitive to extinction corrections. The NUV line ratios are also affected but these are not crucial to our overall conclusions. We use reddening arrows to indicate the effects of extinction uncertainties in the diagnostic line ratio diagrams we present in later sections.

%Temperature and Density Estimations
%%%%%%%%%%%%%%%%%%%%%%%%%%%%%%%%%%%%%%%%%%%%%%%%%%%%%%%%%%%%%%%%%%%%%%%%%%%%%%%%%
\section{Temperature and Density Estimates}
\label{sec:tempdens}
We determined the electron temperature and density as a function of distance from the center of each galaxy using the $\textrm{[\ion{O}{3}]}\;\lambda5007/\textrm{[\ion{O}{3}]}\;\lambda4363$ and $\textrm{[\ion{S}{2}]}\;\lambda6716/\textrm{[\ion{S}{2}]}\;\lambda6731$ line ratios, respectively, at all distances \citep[e.g.,][Chapter 5]{Osterbrock06}. The temperatures predicted by AGN photoionization models \citep[$T\approx1$--$2\times10^{4}$~K,][]{Groves2004} and from shock models \citep[$T\approx4\times10^{4}$--$1\times10^{6}$~K,][]{Allen2008}, both of which are described in more detail in Section~\ref{ssec:elmodels}, are sufficiently different to make this a useful diagnostic in principle. The $\textrm{[\ion{S}{2}]}\;\lambda6716/\textrm{[\ion{S}{2}]}\;\lambda6731$ density estimate is useful for comparing our measurements to those of \citet{Walsh2008}, as well as for estimating the ionization parameter $U$ in NGC~4278 and NGC~4579 (see discussion in Section~\ref{ssec:el_gal_results}). Near the nucleus of each galaxy we also detected the [\ion{S}{2}]\;$\lambda\lambda$4069,4076 and [\ion{O}{2}]\;$\lambda$2470 lines, implying a density of $n_e\gtrsim10^{4}$~cm\textsuperscript{$-3$}, which is above the sensitivity threshold of the $\textrm{[\ion{S}{2}]}\;\lambda\lambda6716,6731$ doublet. We also used these lines to obtain an estimate on the density near the nucleus for all three objects.

The electron temperature is only estimated when the [\ion{O}{3}]$~\lambda$4363 emission line was detected or when the lower limit of [\ion{O}{3}]$~\lambda$5007/[\ion{O}{3}]$~\lambda4363>5$, as an lower limit below that threshold is not useful. The temperature derived for the nucleus of NGC~1052 is $T\approx 4\times10^{4}$~K, which can be explained by shock heating, and meaningful upper limits of $T < 3\times10^{4}$~K are obtained within the central 40~pc. NGC~4278 has very weak [\ion{O}{3}]$~\lambda$4363 emission, and only one meaningful upper limit on the temperature of $T<3.5\times10^{4}$~K was obtained. For NGC~4579 we were able to obtain limits within the central 20 pc of order $T<4\times10^4$. All of these temperature upper limits are too high to disfavor shock heating, and thus cannot be used to discriminate between the two model families in practice.

\begin{figure}[t]
\includegraphics[width=0.5\textwidth]{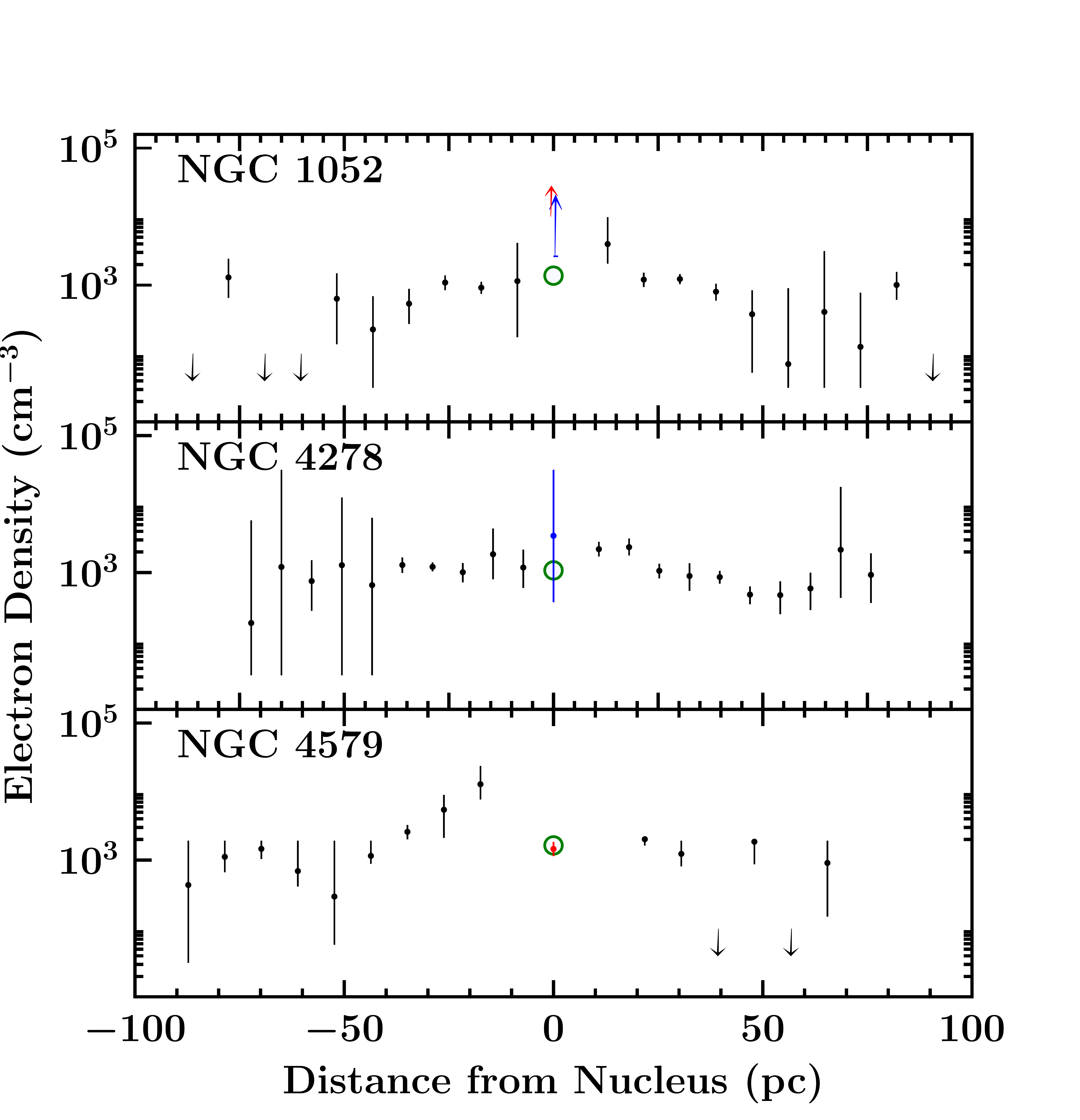}
\caption{Density estimates from the [\ion{S}{2}]$~\lambda\lambda$6716,6731 ratio for all three objects as a function of distance from the nucleus. Red points (at 0~pc) represent the estimate from the spectrum of the spatially unresolved nuclear source for NGC~1052 and NGC~4579, blue points represent the center of the extended, resolved source (at 0~pc, above the unfilled circle in NGC~1052 and NGC~4278), and the green unfilled circles represent the integrated spectrum. The details of the calculation are given in Section~\ref{sec:tempdens}. The [\ion{S}{2}]$~\lambda\lambda$~6716,6731 doublet at 0~pc and 8~pc in the extended, resolved source of NGC 4579 are not detected as described in Section~\ref{sec:lines}, so no density estimate is given. The high density found in \citet{Walsh2008} in NGC~1052 is also found here both in the center of the resolved source and the unresolved nuclear source, and is above the sensitivity of this line ratio. The estimate of the nuclear densities for all three objects are discussed in Section~\ref{sec:tempdens}.}\label{fig:s2dens}
\end{figure}

We detected the [\ion{S}{2}]$\lambda\lambda$ 6716,6731 doublet in all objects, and we are able to determine electron densities as a function of distance from the nucleus for all objects. We plot these in Figure~\ref{fig:s2dens}. The only exceptions are the central two rows in NGC~4579, where no lines were detected (see Section~\ref{sec:lines}), so we do not report density estimates at those distances. Our measured electron densities are consistent with those in \citet{Walsh2008}. The high density that was originally seen in the nucleus of NGC~1052 by \citet{Walsh2008} was found both in the resolved center of the galaxy and in the unresolved nuclear emission. We also detected the [\ion{S}{2}]\;$\lambda\lambda$4069,4076 and [\ion{O}{2}]\;$\lambda$2470 lines near the nucleus of each galaxy. By combining the temperature estimates obtained from the [\ion{O}{3}] diagnostic and the de-reddened [\ion{O}{2}]\;$\lambda$2470/[\ion{O}{2}]\;$\lambda$3727 and [\ion{S}{2}]\;$\lambda\lambda$4069,4076/[\ion{S}{2}]\;$\lambda\lambda$6716,6731 line ratios, we obtained an estimate of the density using the program PyNeb \citep{luridian2015}.

The density estimates from both the [\ion{S}{2}] and [\ion{O}{2}] diagnostics for NGC~4579 and NGC~4278 agree with those obtained via the [\ion{S}{2}]\;$\lambda\lambda$6716,6731 doublet. The high density seen in the resolved center of NGC~1052 is consistent with $n_e\sim2$--$6\times10^4$~cm\textsuperscript{$-3$} using both the [\ion{S}{2}] and [\ion{O}{2}] diagnostics, while the unresolved nuclear source requires an even higher density ($n_e\sim10^5$~cm\textsuperscript{$-3$}) to explain the [\ion{O}{2}] line ratio. This difference in densities is consistent with the need for a multi-component density model, as described by \citet{Dopita2015} and \citet{Gabel2000}.

Even though the nuclear density of NGC~1052 is extremely high, the density estimate from the integrated emission for all three galaxies is of order $10^{3}$~cm\textsuperscript{$-3$}. This difference in densities highlights the potential of obtaining misleading density estimates from spectra that sample gas with a wide range of physical conditions.

We will use the densities obtained from the spatially resolved spectra to constrain models and we will also incorporate them into a calculation of the ionization parameter, as described in Section~\ref{ssec:el_gal_results}.
%Diagnostic Diagrams
%%%%%%%%%%%%%%%%%%%%%%%%%%%%%%%%%%%%%%%%%%%%%%%%%%%%%%%%%%%%%%%%%%%%%%%%%%%%%%%%%
\section{Comparison of Emission Line Measurements with Models}
\label{sec:ddresult}
\subsection{Emission Line Diagnostic Diagrams}
\label{ssec:eldiagrams}
In order to diagnose the excitation mechanism of the emission line gas, we used the emission line intensities measured above to form extinction-corrected line ratios, which we plotted in diagnostic diagrams in the spirit of \citet*[][also known as BPT diagrams]{BPT81}.  We produced four diagnostic diagrams involving optical lines for each galaxy. We also produced three diagnostic diagrams involving NUV lines for NGC~1052 and NGC~4278. In NGC~4579 we only detected an unresolved source in the NUV band hence we did not construct the corresponding diagnostic diagrams. The diagrams, shown in Figures~\ref{fig:n1052_dd_o3o2}--\ref{fig:n4579_dd_o1ha}, are used to assess the mechanisms that excite the gas, whether excitation by fast and slow shocks, or photoionization by an AGN, pAGB stars, or young, massive stars (as in \ion{H}{2} regions). We plot the same data in each quadrant of the same figure but superpose a different model grid. 

We also compared the upper limits of $\textrm{[\ion{Fe}{10}]}\;\lambda 6374/\textrm{H}\beta$ and $\textrm{[\ion{Fe}{14}]}\;\lambda 5303/\textrm{H}\beta$ to shock models from \citet{Wilson99}. The highly ionized Fe lines were never detected, so we only have upper limits to these ratios when we detect H$\beta$. Additionally, we compared the $\textrm{[\ion{Ne}{3}]}\;\lambda3869/\textrm{[\ion{O}{2}]}\;\lambda3727$ to photoionization models for radiatively inefficient accretion flows (RIAFs) from \citet{Nagao02}\footnote{These models employ a harder spectrum of ionizing photons than the standard AGN models.} for all three galaxies, and we calculated the ionization parameter, $U$ (defined in Section~\ref{ssec:el_gal_results}), as a function of distance for NGC~4278 and NGC~4579. All comparisons to models and the calculated $U$ values are discussed in Section~\ref{ssec:el_gal_results}.  

As we noted above, four of the seven diagnostic diagrams use optical line ratios: $\textrm{[\ion{O}{3}]}/\textrm{H}\beta$ $vs.$ $\textrm{[\ion{O}{3}]}/\textrm{[\ion{O}{2}]}$, $\textrm{[\ion{N}{2}]}/\textrm{H}\alpha$, $\textrm{[\ion{S}{2}]}/\textrm{H}\alpha$, and $\textrm{[\ion{O}{1}]}/\textrm{H}\alpha$ \citep[these are commonly used diagrams; e.g.,][]{veilleux87,Kewley2006,Kauffmann03,Ho2008}, and the other three use NUV line ratios: $\textrm{\ion{C}{3}]}/\textrm{\ion{C}{2}]}$ $vs.$ $\textrm{[\ion{O}{2}]}/\textrm{\ion{C}{2}]}$,  $\textrm{[\ion{N}{2}]}/\textrm{\ion{C}{2}]}$, and $\textrm{\ion{Mg}{2}}/\textrm{\ion{C}{2}]}$. The NUV diagrams were chosen according to both the emission lines we were able to measure and the extent of the region that the models occupy in parameter space near the measured values. \ion{C}{3}]$~\lambda1909$ and \ion{C}{2}]$~\lambda2326$ are the only two lines in the NUV spectra that we could consistently measure.

In diagnostic diagrams where at least one line ratio is sensitive to extinction (i.e., the diagram involving [\ion{O}{3}]/[\ion{O}{2}] and any NUV ratios), we plot reddening arrows. These show how a point in the diagram would move if $E(B-V)$ increased by 1~mag. We note an additional source of uncertainty for the NUV lines. The measured extinction is sampled according to the angular resolution of the blue and red spectra since it comes from the H$\alpha$/H$\beta$ ratio, but the NUV line ratios are sampled at a higher angular resolution, requiring us to interpolate the extinction values.

We begin by describing below how the observed line ratios are distributed in the diagnostic diagrams, we review the models in some detail in Section~\ref{ssec:elmodels}, and we compare the models to the observations for each individual galaxy in Section~\ref{ssec:el_gal_results}.

\begin{description}
\item[NGC~1052]-- Figure~\ref{fig:n1052_dd_n2ha} most clearly shows the spatial variation of the $\textrm{[\ion{O}{3}]}/\textrm{H}\beta$ ratio. H$\beta$ becomes stronger relative to [\ion{O}{3}]$~\lambda5007$ with increasing distance from the nucleus. The spectrum of the spatially unresolved nuclear source has the largest measured $\textrm{[\ion{O}{3}]}/\textrm{H}\beta$ ratio, while the line ratio from the integrated spectrum falls in between that and the off-nuclear resolved spectra. Figure~\ref{fig:n1052_dd_o3o2} shows a similar trend for the $\textrm{[\ion{O}{3}]}/\textrm{[\ion{O}{2}]}$ ratio, which decreases with distance from the nucleus. In this case we see a clear separation in the line ratios of the unresolved nuclear source and the off-nuclear resolved emission, with the line ratio from the integrated spectrum and the nuclear spectrum of the resolved source falling between them. The $\textrm{[\ion{N}{2}]}/\textrm{H}\alpha$ and $\textrm{[\ion{S}{2}]}/\textrm{H}\alpha$ ratios are both $\approx1$, for the off-nuclear resolved spectra as shown in Figures~\ref{fig:n1052_dd_n2ha} and~\ref{fig:n1052_dd_s2ha}, while the unresolved and resolved nuclear emission have ratios $<1$. The $\textrm{[\ion{S}{2}]}/\textrm{H}\alpha$ ratio for the resolved nuclear emission is significantly smaller than 1, and separates itself from the rest of the measurements. The $\textrm{[\ion{O}{1}]}/\textrm{H}\alpha$ ratio in the spatially unresolved nuclear source is \,$\approx1$, while generally this ratio decreases with distance from the nucleus in the extended emission (see Figure~\ref{fig:n1052_dd_o1ha}). The ratio from the integrated spectrum is similar to the ratios obtained close to the nucleus. 

The NUV line ratios have significantly more scatter than the optical ones. There is no apparent trend in the $\textrm{\ion{C}{3}]}/\textrm{\ion{C}{2}]}$ ratio (Figures~\ref{fig:n1052_dd_o2c2}--\ref{fig:n1052_dd_mg2c2}). Generally, the $\textrm{[\ion{O}{2}]}/\textrm{\ion{C}{2}]}$ ratio in the unresolved source is different from that of resolved emission (Figure~\ref{fig:n1052_dd_o2c2}). All of the $\textrm{[\ion{N}{2}]}/\textrm{\ion{C}{2}]}$ ratios in the resolved emission are upper limits (Figure~\ref{fig:n1052_dd_n2c2}). Finally, the $\textrm{\ion{Mg}{2}}/\textrm{\ion{C}{2}]}$ ratio in the unresolved nuclear source is higher than what is observed in the resolved emission (Figure~\ref{fig:n1052_dd_mg2c2}), the large uncertainties notwithstanding.

\item[NGC~4278]-- The $\textrm{[\ion{O}{3}]}/\textrm{[\ion{O}{2}]}$ and $\textrm{[\ion{O}{3}]}/\textrm{H}\beta$ ratios do not change systematically with distance from the nucleus, as seen in Figure~\ref{fig:n4278_dd_o3o2}. The $\textrm{[\ion{N}{2}]}/\textrm{H}\alpha$, $\textrm{[\ion{S}{2}]}/\textrm{H}\alpha$ and $\textrm{[\ion{O}{1}]}/\textrm{H}\alpha$ ratios in the nuclear spectrum are systematically different from those of the resolved, extended emission (see Figures~\ref{fig:n4278_dd_n2ha}--\ref{fig:n4278_dd_o1ha}).  All the optical line ratios of the integrated emission are similar to the corresponding ratios observed in the extended, resolved emission (Figures~\ref{fig:n4278_dd_o3o2}--\ref{fig:n4278_dd_o1ha}).

The NUV line ratios do not show any systematic trends as a function of distance from the nucleus. The $\textrm{\ion{C}{3}]}/\textrm{\ion{C}{2}]}$ and $\textrm{\ion{Mg}{2}}/\textrm{\ion{C}{2}]}$ line ratios are clustered tightly (Figure~\ref{fig:n4278_dd_mg2c2}) while most of the $\textrm{[\ion{O}{2}]}/\textrm{\ion{C}{2}]}$ and all of the $\textrm{[\ion{N}{2}]}/\textrm{\ion{C}{2}]}$ ratios in the resolved emission are upper limits (Figures~\ref{fig:n4278_dd_o2c2}--\ref{fig:n4278_dd_n2c2}). 

\item[NGC~4579]-- Because of the overwhelming strength of the spatially unresolved nuclear source in the red spectrum, we can only measure the $\textrm{[\ion{N}{2}]}/\textrm{H}\alpha$, $\textrm{[\ion{S}{2}]}/\textrm{H}\alpha$ and $\textrm{[\ion{O}{1}]}/\textrm{H}\alpha$ ratios from the resolved emission starting at 12~pc from the nucleus. The $\textrm{[\ion{O}{3}]}/\textrm{H}\beta$ ratio of the resolved emission is similar to that measured in the unresolved nuclear source. As a result, the point representing the unresolved nuclear source is within the cluster of points representing the extended emission in most of the diagnostic diagrams. The $\textrm{[\ion{O}{3}]}/\textrm{H}\beta$ $vs$ $\textrm{[\ion{S}{2}]}/\textrm{H}\alpha$ diagram is an exception; in this diagram the point representing the unresolved nuclear source is well separated from the cluster of points representing the extended emission.

The NUV diagnostic diagrams are not plotted because the NUV spectrum of NGC~4579 is dominated by emission from the unresolved AGN. 
          
\end{description}

\subsection{Emission Line Models for Different Excitation Mechanisms}
\label{ssec:elmodels}
As shown in Figures~\ref{fig:n1052_dd_o3o2}--\ref{fig:n4579_dd_o1ha}, we used three families of models in our diagrams: shocks, photoionization from an AGN (also known as AGN narrow line region models and labeled as AGN NLR in the figures), and photoionization from hot stars. For every family of models in our diagnostic diagrams, we plot a solid outline showing the full parameter space they cover as well as a detailed grid with specific parameters that represent a narrower range of parameters that is more likely to apply to the regions of interest here. In the top two quadrants we plot grids for shock models with and without a ``precursor'' (see below). In the lower two quadrants we plot photoionization models for an AGN ionizing spectrum and for hot star ionizing spectra. The parameters for each model family and the detailed parameter grids are described below.

\begin{description}
\item[{\it Shocks}]-- This family includes two types of fast shock models (with and without a precursor, for a range metallicities) from \citet{Allen2008}, and slow shock models from \citet{Shull79}. In shock models without a precursor (hereafter simple shock models) the gas is collisionally ionized by the shock while in models with a precursor ionizing photons produced in the shock heated gas travel upstream and ionize the gas before the shock reaches it. We focused on solar metallicity models, which predict line ratios for densities in the range $n=0.1\textrm{--}10^3$~cm\textsuperscript{$-3$}, and twice solar metallicity models, which assume a density $n=1$~cm\textsuperscript{$-3$}. For each of those sets, the models are calculated for magnetic field strengths in the range $B=0.0001\textrm{--}100~\mu$G and velocities of $v=100$--1000~km~s\textsuperscript{$-1$}. The simple shock models are shown in purple in the top left quadrant of the diagnostic diagrams, while the models of shocks with a precursor are shown in blue on the top right quadrant. The detailed grid assumes a magnetic field of $B=1~\mu$G and solar metallicity, and includes the full range in density and velocity. The detailed grids were chosen to include densities similar to what was measured in Section~\ref{sec:tempdens}, which limited us to the solar metallicity tracks. We chose $B=1~\mu$G as this grid overlapped with most of the tracks with magnetic fields less than $10~\mu$G.

The slow shock models from \citet{Shull79} do not include a precursor and comprise two sets, one with constant density $n = 10$~cm\textsuperscript{$-3$}, ``cosmic'' metal abundances \citep[see Table 2 in][]{Shull79}, a magnetic field of $B=1~\mu$G, and varying velocities of $v = 40$--130~km~s\textsuperscript{$-1$}. The second set assumes the previous parameters except for a fixed velocity of $v=100$~km~s\textsuperscript{$-1$}, and then changes either the density to $n = 100$~cm\textsuperscript{$-3$}, the magnetic field is set to $B=10~\mu$G, or adopts depleted metal abundances. The slow shock models are plotted as green tracks in the top left quadrant of the diagnostic diagrams.

\item[{\it Photoionization by an AGN}]-- The AGN photoionization models plotted in green in the lower left quadrant are the dust-free models from \citet{Groves2004}. These models assume a simple power law ionizing spectrum ($F_{\nu}\propto\nu^{\alpha}$, in the range 5--1000~eV) and are governed by four parameters: metallicity spanning the range $Z=0.25\textrm{--}2\;Z_{\odot}$, density spanning the range $n=10^{2}$--$10^4$~cm\textsuperscript{$-3$}, ionization parameter\footnote{See equation~(\ref{eq:U}) in Section~\ref{ssec:el_gal_results}.} spanning the range $\log\,U=-4$ to 0, and the spectral index of the ionizing spectrum, $\alpha$, spanning the range $-1.2$ to $-2$. The detailed grid assumes a constant metallicity of  $Z=2\;Z_{\odot}$ and density $n=10^3$~cm\textsuperscript{$-3$}, consistent with the densities we calculated in Section~\ref{sec:tempdens}, and the full range of $\alpha$, but limited $\log\,U$ to the ranges of $-3$ to $-1$, making our grid similar to those in \citet{Kewley2006}. 

We compare both the RIAF and the optically thick, geometrically thin disk models of \citet{Nagao02} to the $\textrm{[\ion{Ne}{3}]}\;\lambda3869/\textrm{[\ion{O}{2}]}\;\lambda3727$ line ratio, but do not plot them in our diagrams. While the shape of the ionizing continua are different, both models are governed by the same set of parameters: density $n=10^2\textrm{--}10^5$~cm\textsuperscript{$-3$}, and ionization parameter $\log\,U=-4$ to $-1.5$. The RIAF models use a spectral energy distribution (SED) template from \citet{Kurp1999} which has no thermal blue/UV component (i.e., ``big blue bump"), employs a simple power law ionizing continuum ($f_{\nu} = \nu^{-0.89}$ with exponential cutoffs at $10^{-4}$ and $10^4\;$Ry). The optically thick, geometrically thin disk model is empirical, and has a more complex SED \citep[as described in equation 1 of][]{Nagao02}. Its distinguishing characteristics include a big blue bump with a characteristic temperature of $4.9\times10^{5}$ K, and  a power law of the form $f_{\nu}\propto\nu^{-3}$ from $100$ keV to $10^4$ Ry.
  
\item[{\it Photoionization by hot stars}]-- The hot star models include pAGB stars from \citet{Binette94}, and H \textsc{ii} models from \citet{Dopita2006} and \citet{Kewley2006}. These models are shown in the lower right quadrant of the diagnostic diagrams, with the pAGB models represented by the green shaded regions and the H \textsc{ii} models in red. The \citet{Binette94} pAGB models are based on a 13 Gyr stellar population (except for the $Z=1\;Z_{\odot}$ model, which includes both  8 and 13 Gyr populations), and are described by two parameters: metallicity spanning $Z = {1\over 3}\textrm{--}3\;Z_{\odot}$ and ionization parameter spanning $\log\,U=-3$ to $-5$. The density is assumed to be $n=10^3\;{\rm cm}^{-3}$. The age of this population allows for light from pAGB stars to be the dominant source of ionizing photons, which produces an ionizing continuum similar to a black body of $T\approx10^5$~K \citep[see][for more detail]{Binette94}. The pAGB models are only used in three out of the seven diagrams as they do not predict the relative intensities of the NUV lines or [\ion{S}{2}]$~\lambda\lambda6716,6731$. Their main difference compared to the \ion{H}{2} region models described below is the higher temperature of the stars providing the ionizing photons. 

We use H \textsc{ii} region models from two sources. The $\textrm{[\ion{O}{3}]}/\textrm{H}\beta$ $vs.$ $\textrm{[\ion{O}{3}]}/\textrm{[\ion{O}{2}]}$ and $\textrm{\ion{C}{3}]}/\textrm{\ion{C}{2}]}$ $vs.$ $\textrm{\ion{Mg}{2}}/\textrm{\ion{C}{2}]}$ diagrams use models from \citet{Dopita2006}, which cover a population age from 0.2 Myr to 4 Myr and metallicity $Z=0.2$--$2\;Z_{\odot}$. The \cite{Dopita2006} models are characterized by the central cluster mass (100--$10^6M_{\odot}$) and interstellar medium pressure ($10^4$--$10^7$~cm\textsuperscript{$-3$}~K, corresponding to $n\sim 1$--$10^3\;{\rm cm}^{-3}$), and have an ionizing continuum dominated by stars with  an effective temperature $T_{eff}\approx10^{4.5}$~K. The models from \citet{Kewley2006}, which include both the theoretical extreme starburst line from \citet{Kewley01} and the empirical composite objects line from \citet{Kauffmann03}, are shown in the $\textrm{[\ion{O}{3}]}/\textrm{H}\beta$ $vs.$ $\textrm{[\ion{N}{2}]}/\textrm{H}\alpha$, $\textrm{[\ion{S}{2}]}/\textrm{H}\alpha$, and $\textrm{[\ion{O}{1}]}/\textrm{H}\alpha$ diagrams. The \citet{Kewley01} models include both instantaneous burst as well as continuous star formation models. These models yield the maximum line ratios that can be explained by stellar photoionization, which are denoted by a solid line in the relevant diagnostic diagrams.

\end{description}

\subsection{Comparison of Models With the Data for Individual Galaxies}
\label{ssec:el_gal_results}

It is evident from the diagnostic diagrams that there is considerable degeneracy between the models. Therefore we seek the models that offer the best explanation of the data, i.e., those that explain most of the diagnostic diagrams. In the discussion below we note what range of model parameters is needed in order for the models to explain the data.

To aid us in the assessment of the AGN photoionization models, we compute the ionization parameter due to ionizing radiation from the AGN as a function of distance from the center of the galaxy via
\begin{equation}
  U = \frac{Q}{4\pi r^2 n c},
  \label{eq:U}
\end{equation}
where $Q$ is the rate of emission of hydrogen-ionizing photons, $r$ is the physical distance from the nucleus, and $n$ is the hydrogen density. We adopt the values of $Q$ obtained by \citet{Eracleous2010a} by integrating the observed spectral energy distribution and take the electron density from Section~\ref{sec:tempdens} as a proxy for the hydrogen density, accepting that there is an uncertainty of less than a factor of 2, which does not affect our conclusions.

By inspecting the diagnostic diagrams we can readily see that hot star models cannot explain the data for any of the galaxies and at any distance from the center of the galaxy. The [\ion{O}{1}]/H$\alpha$ ratio proves particularly useful in discriminating against hot star photoionization models. The other families of models have a varying degree of success as we detail below. 

\begin{figure*}
	\centering
       \includegraphics[width=0.75\textwidth]{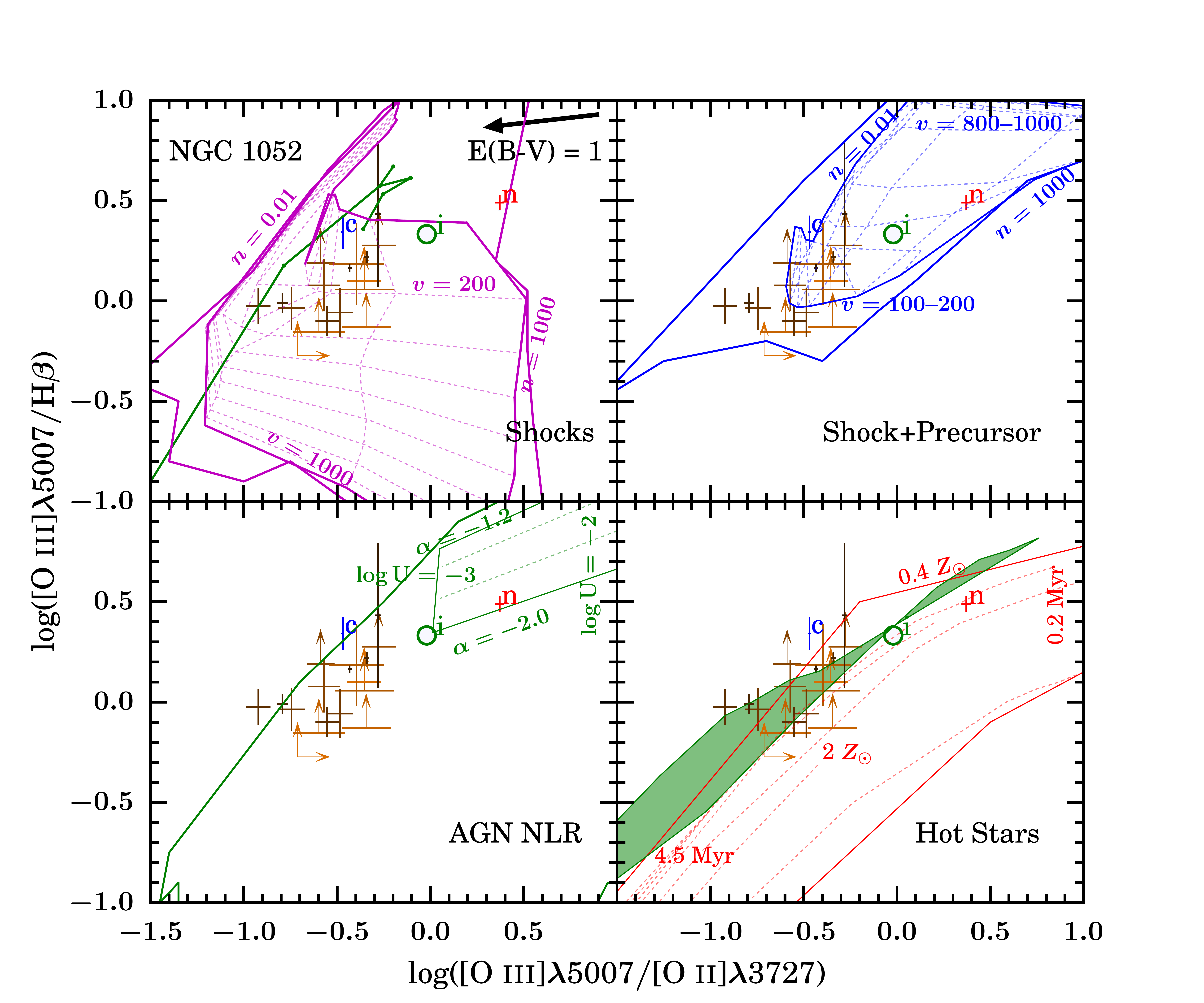}
	\caption{\footnotesize{Diagnostic diagram of $\textrm{[\ion{O}{3}]}/\textrm{H$\beta$}$ $vs.$ $\textrm{[\ion{O}{3}]}/\textrm{[\ion{O}{2}]}$ for NGC~1052. The measurements made along the slit throughout the extended, resolved source are shown in shades of orange (without labels). The darkest orange represents the measurement closest to the nucleus. The blue point labeled ``c'' represents the center of the resolved source, the red point labeled ``n'' represents the central, unresolved source, and the unfilled green circle labeled ``i'' represents the ratios in the integrated spectrum. Limits are indicated with arrows. All models, shown with solid and dotted lines, are described in Section~\ref{ssec:elmodels}. \textit{Top Left: }Comparison of data to simple shock models (without a precursor). The points connected by the solid green line represent the slow shock models of \citet{Shull79}, while the purple grid and outline are fast shock models of \citet{Allen2008}. The solid black arrow shows the change in the line ratios when an extinction correction of $E(B-V)=1\;$mag is applied. The labels show how density (in cm\textsuperscript{$-3$}) and velocity (in km~s\textsuperscript{$-1$}) change across the detailed grid. \textit{Top Right:} Comparison of data to shock plus precursor models from \citet{Allen2008}. The labels show how density (in cm\textsuperscript{$-3$}) and velocity (in km~s\textsuperscript{$-1$}) change across the detailed grid. \textit{Bottom Left: }Comparison of data to AGN NLR photoionization models from \citet{Groves2004}, with the labels showing how the ionization parameter and the power-law index of the ionizing continuum change across the detailed grid. \textit{Bottom Right: }Comparison of data to \ion{H}{2} region models from \citet{Dopita2006}. The model tracks are in red and the labels show how age and metallicity change across the grid. The green shaded region represents pAGB star models from \citet{Binette94}.}}
	\label{fig:n1052_dd_o3o2}
\end{figure*}

\begin{figure*}
	\centering
	\vspace{-3em}
       \includegraphics[width=0.75\textwidth]{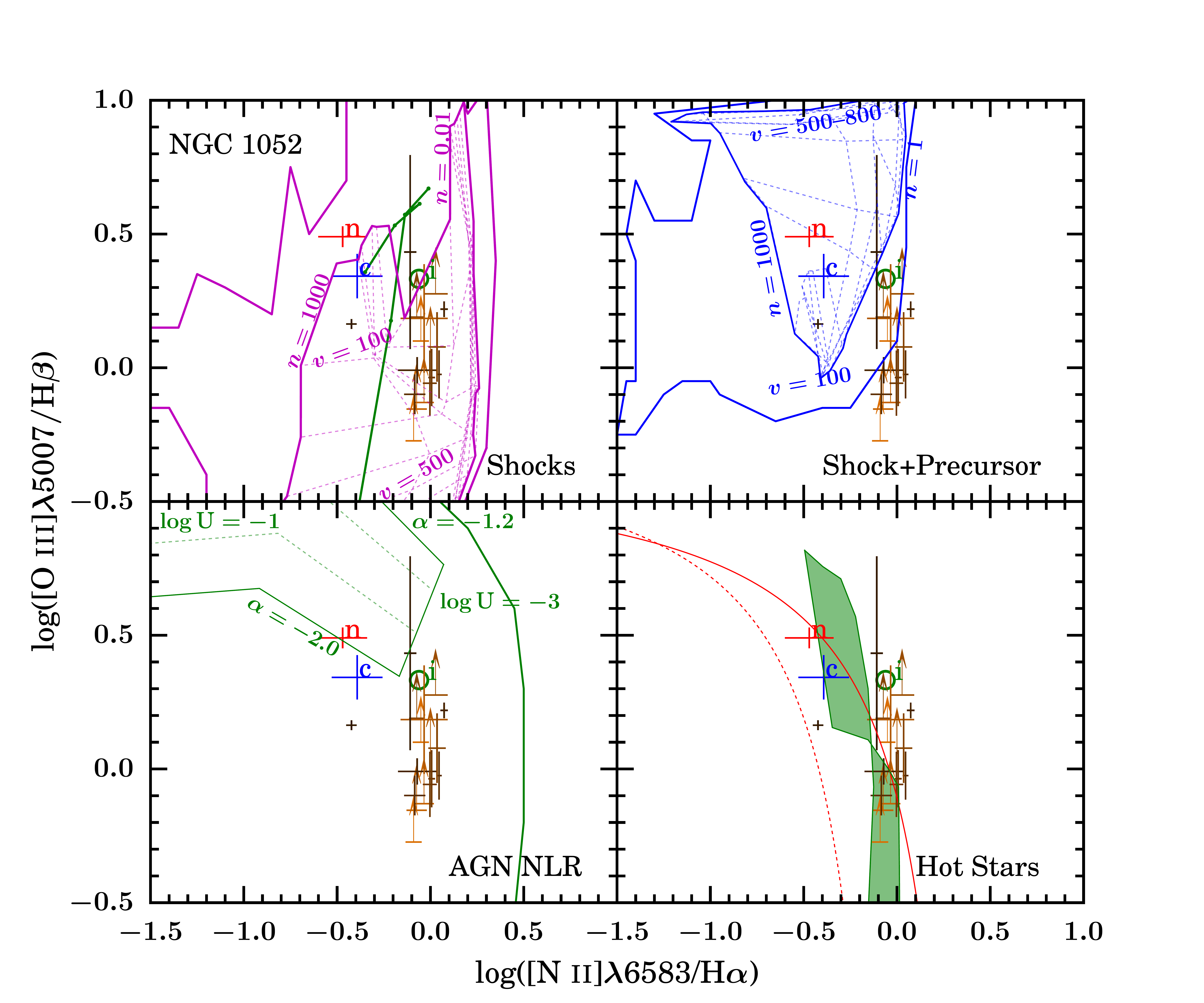}
	\caption{Diagnostic Diagram of $\textrm{[\ion{O}{3}]}/\textrm{H$\beta$}$ $vs.$ $\textrm{[\ion{N}{2}]}/\textrm{H$\alpha$}$ for NGC~1052. The conventions and model descriptions are the same as Figure~\ref{fig:n1052_dd_o3o2}. \textit{Bottom Right: } Same as Figure~\ref{fig:n1052_dd_o3o2}, except the \ion{H}{2} region models are from \citet[][solid line]{Kewley01} and \citet[][dotted line]{Kauffmann03}.}
	\label{fig:n1052_dd_n2ha}
       \includegraphics[width=0.75\textwidth]{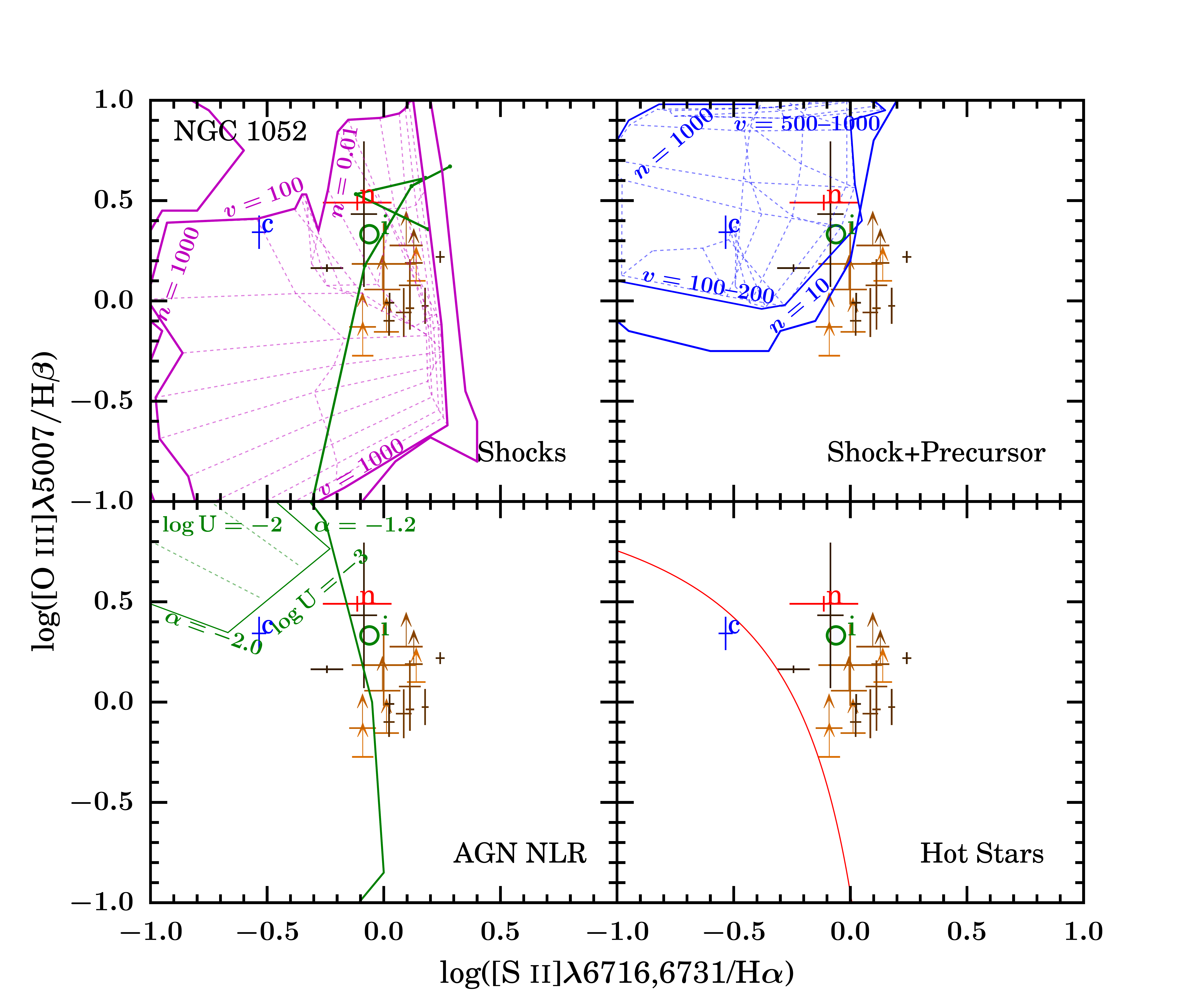}
	\caption{Diagnostic Diagram of $\textrm{[\ion{O}{3}]}/\textrm{H$\beta$}$ $vs.$ $\textrm{[\ion{S}{2}]}/\textrm{H$\alpha$}$ for NGC~1052. The conventions and model descriptions are the same as Figure~\ref{fig:n1052_dd_o3o2}. \textit{Bottom Right: }Same as Figure~\ref{fig:n1052_dd_n2ha}, except that there are no pAGB star models or \ion{H}{2} models from \citet{Kauffmann03}.}
	\label{fig:n1052_dd_s2ha}
\end{figure*}

\begin{figure*}
	\centering
	\vspace{-3em}
       \includegraphics[width=0.75\textwidth]{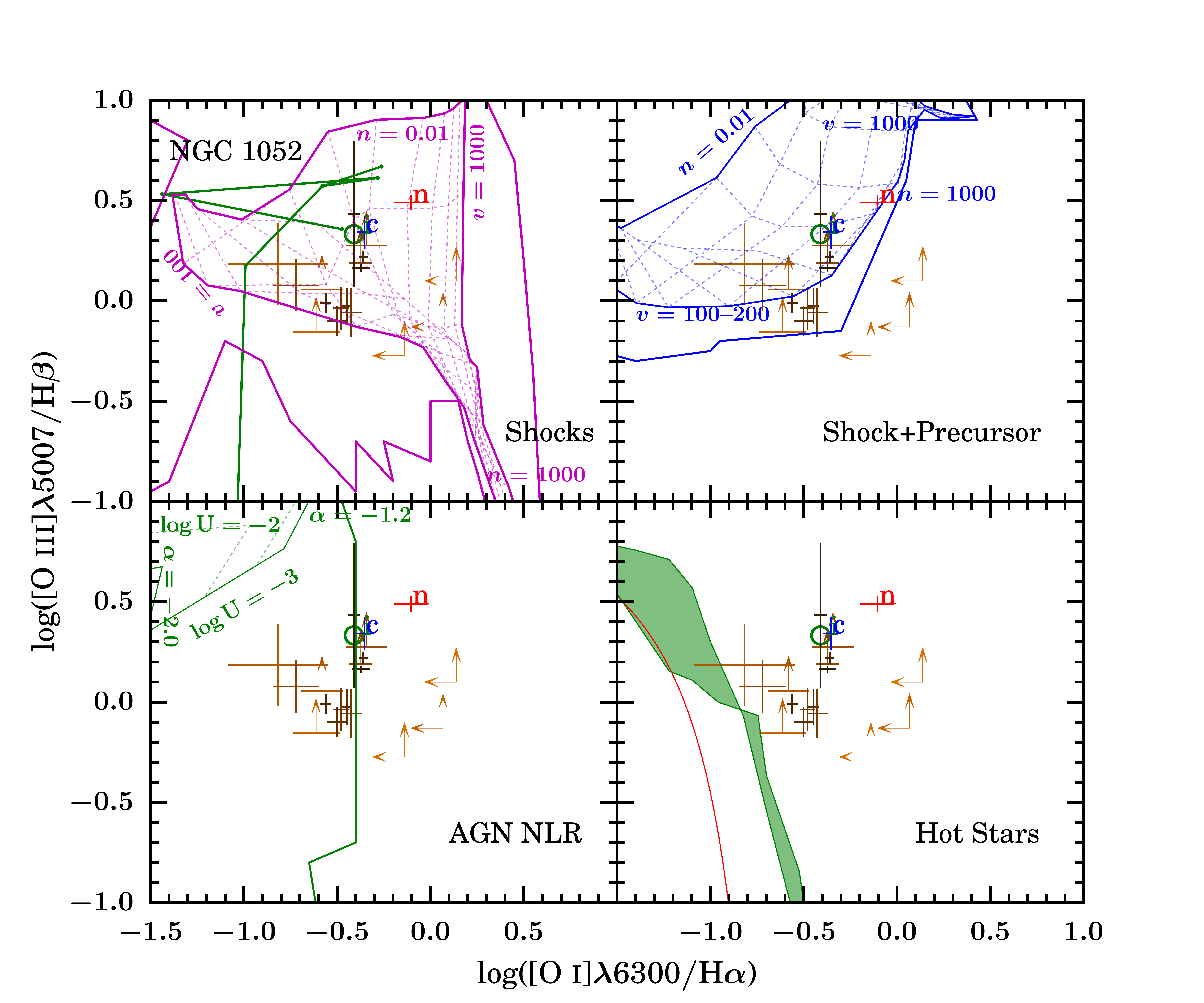}
	\caption{Diagnostic Diagram of $\textrm{[\ion{O}{3}]}/\textrm{H$\beta$}$ $vs.$ $\textrm{[\ion{O}{1}]}/\textrm{H$\alpha$}$ for NGC~1052. The conventions and model descriptions are the same as Figure~\ref{fig:n1052_dd_o3o2}. \textit{Bottom Right: } Same as Figure~\ref{fig:n1052_dd_n2ha}, but without the \citet{Kauffmann03} \ion{H}{2} region models.}
	\label{fig:n1052_dd_o1ha}
       \includegraphics[width=0.75\textwidth]{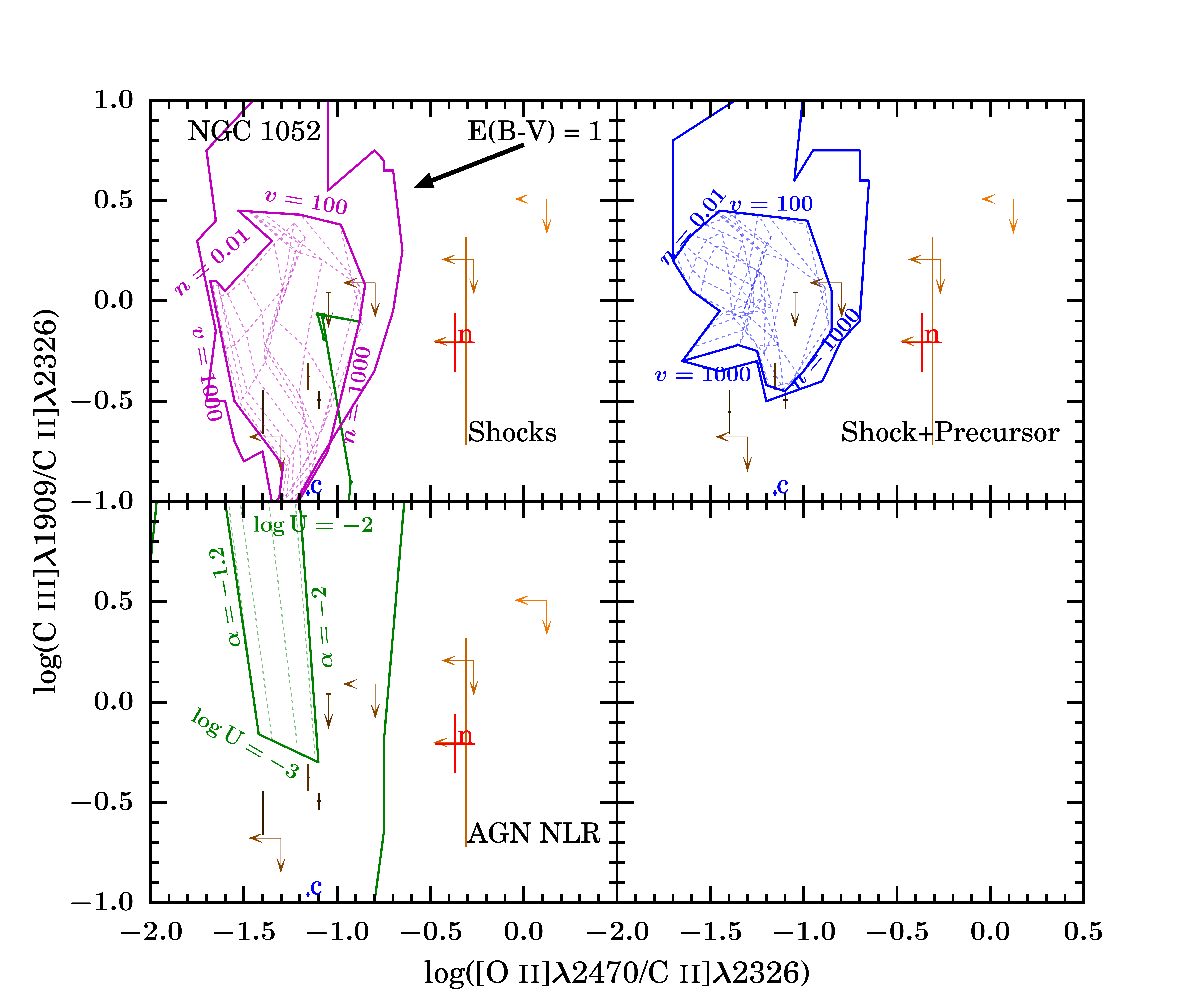}
	\caption{Diagnostic Diagram of $\textrm{\ion{C}{3}]}/\textrm{\ion{C}{2}]}$ $vs.$ $\textrm{[\ion{O}{2}]}/\textrm{\ion{C}{2}]}$ for NGC~1052. The conventions and model descriptions are the same as Figure~\ref{fig:n1052_dd_o3o2}. \textit{Top Left: }Same as Figure~\ref{fig:n1052_dd_o1ha}. The solid black arrow shows the change in the line ratios when an extinction correction of $E(B-V)=1\;$mag is applied. There are no hot star models for this diagram.}
	\label{fig:n1052_dd_o2c2}
\end{figure*}

\begin{figure*}
	\centering
	\vspace{-3em}
       \includegraphics[width=0.75\textwidth]{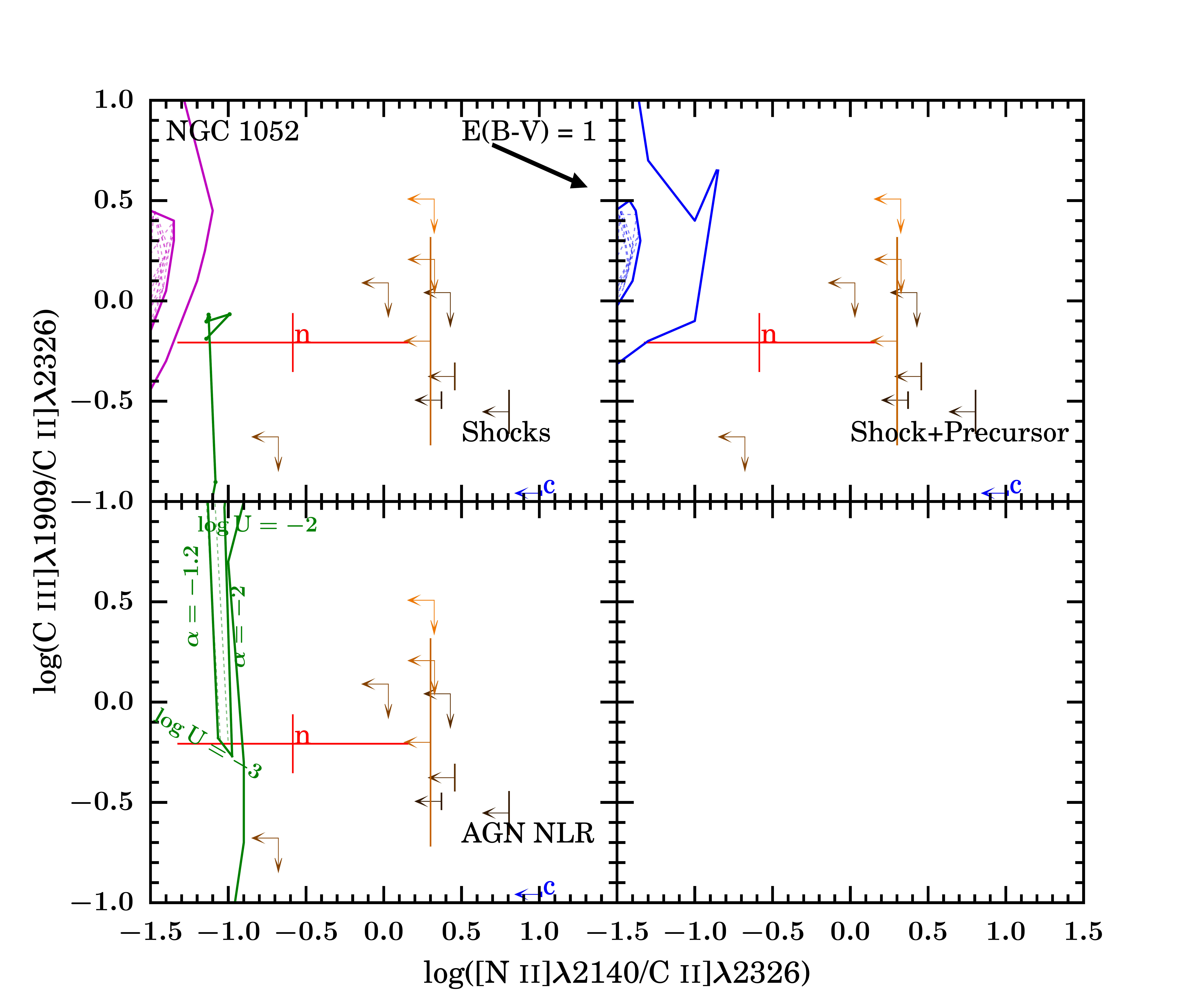}
	\caption{Diagnostic Diagram of $\textrm{\ion{C}{3}]}/\textrm{\ion{C}{2}]}$ $vs.$ $\textrm{[\ion{N}{2}]}/\textrm{\ion{C}{2}]}$ for NGC~1052. The conventions and model descriptions are the same as Figure~\ref{fig:n1052_dd_o3o2}. There are no hot star models for this diagram.}
	\label{fig:n1052_dd_n2c2}
       \includegraphics[width=0.75\textwidth]{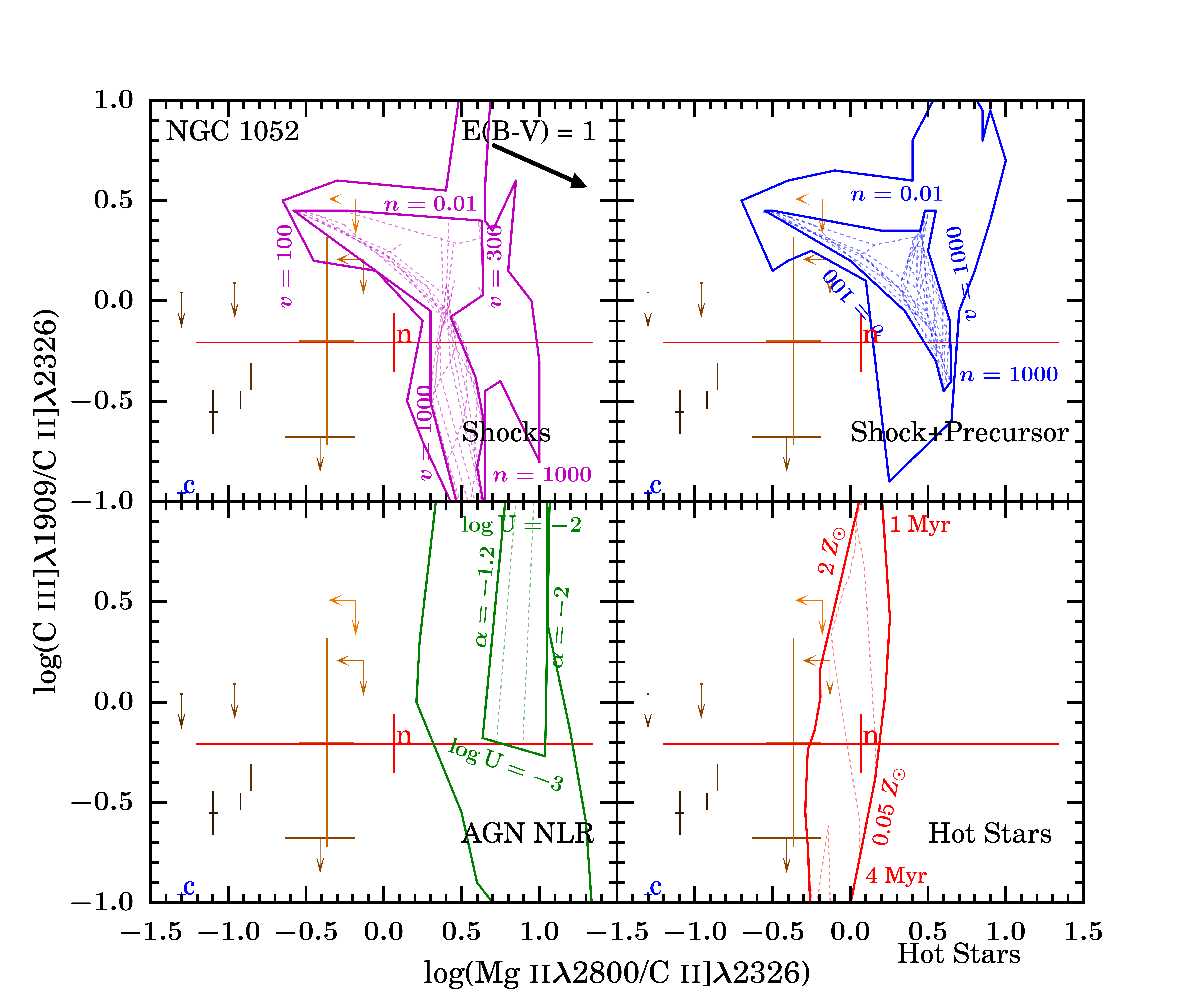}
	\caption{Diagnostic Diagram of $\textrm{\ion{C}{3}]}/\textrm{\ion{C}{2}]}$ $vs.$ $\textrm{\ion{Mg}{2}}/\textrm{\ion{C}{2}]}$ for NGC~1052. The conventions and model descriptions are the same as Figure~\ref{fig:n1052_dd_o3o2}. \textit{Bottom Right: }Same as Figure~\ref{fig:n1052_dd_o3o2}, except there are no pAGB models for this diagram.}
	\label{fig:n1052_dd_mg2c2}
\end{figure*}
\begin{figure*}
	\centering
	\vspace{-3em}
       \includegraphics[width=0.75\textwidth]{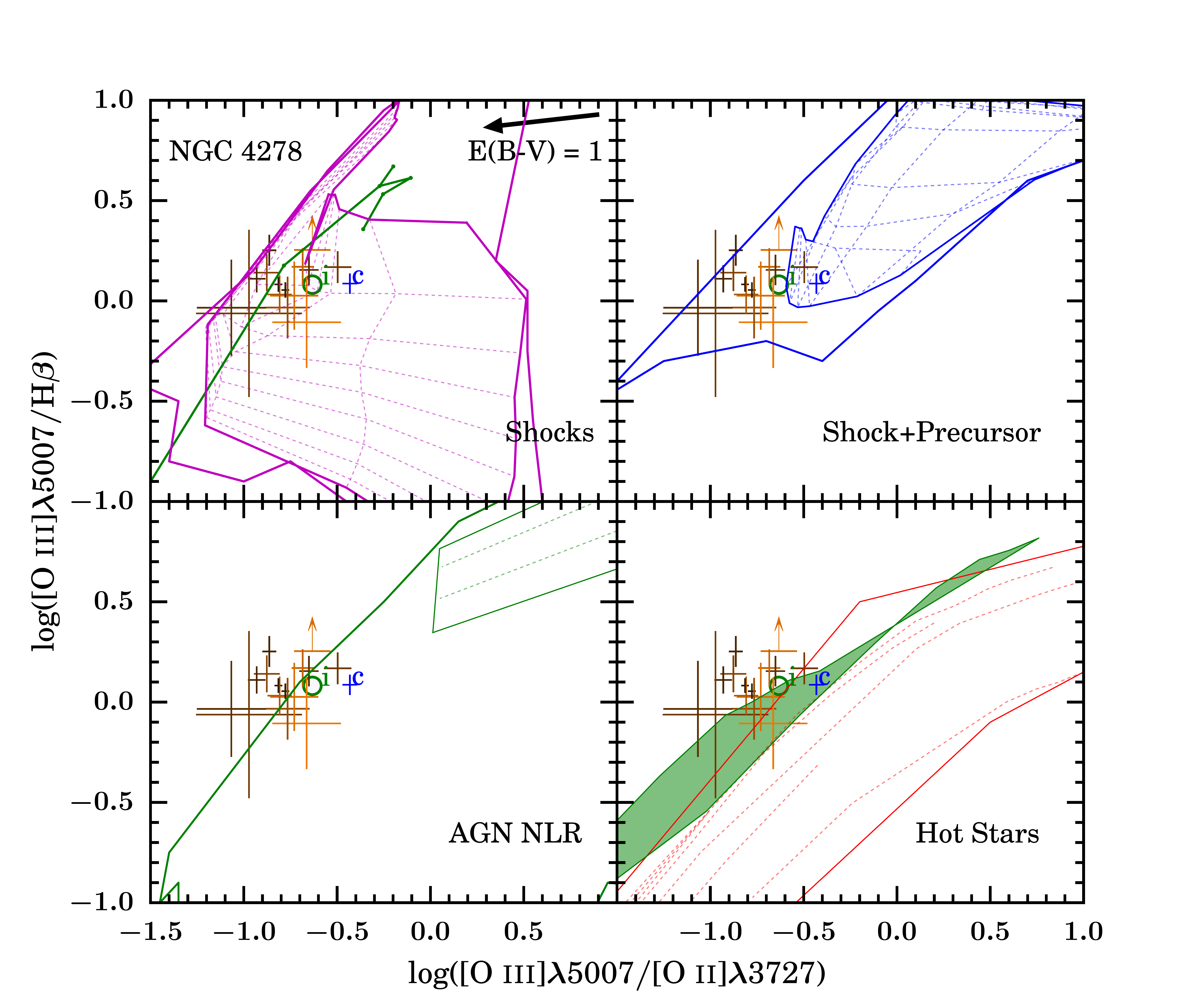}
	\caption{The same as Figure~\ref{fig:n1052_dd_o3o2}, but for NGC~4278. There was no unresolved source spectrum, hence no (red) ``n'' point is plotted.}
	\label{fig:n4278_dd_o3o2}
       \includegraphics[width=0.75\textwidth]{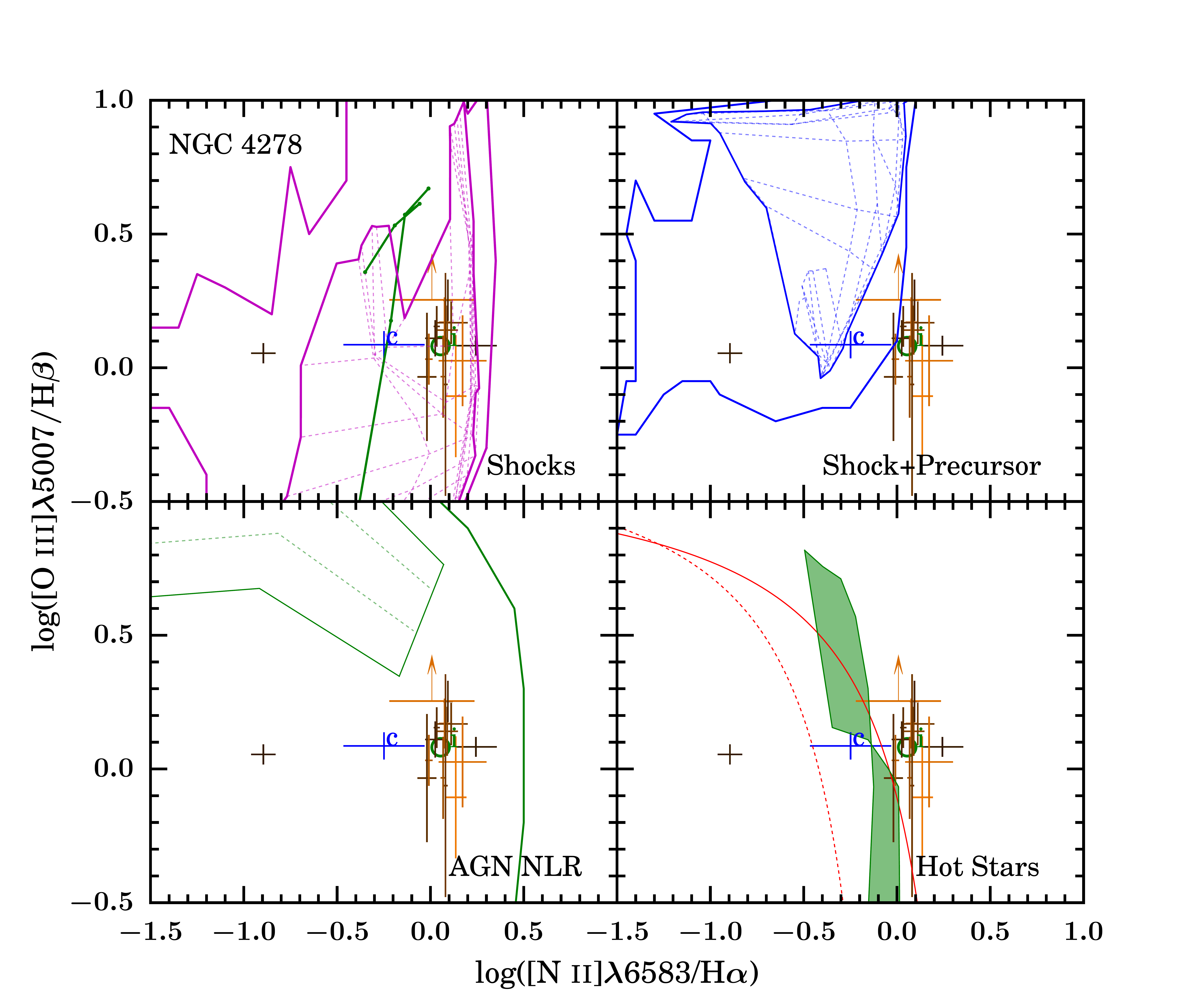}
	\caption{The same as Figure~\ref{fig:n1052_dd_n2ha}, but for NGC~4278. There was no unresolved source spectrum, hence no (red) ``n'' point is plotted.}
	\label{fig:n4278_dd_n2ha}
\end{figure*}

\begin{figure*}
	\centering
	\vspace{-3em}
       \includegraphics[width=0.75\textwidth]{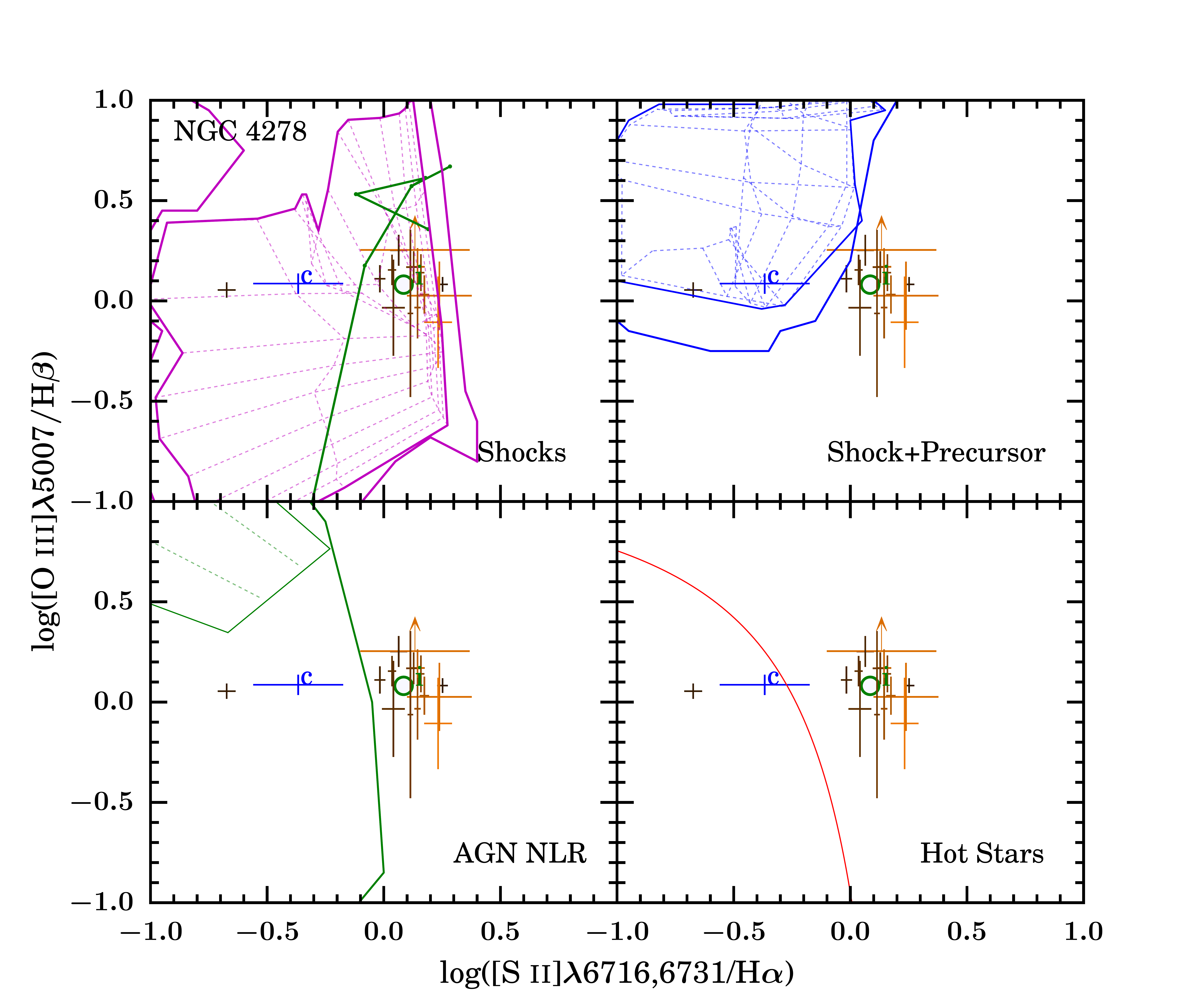}
	\caption{The same as Figure~\ref{fig:n1052_dd_s2ha}, but for NGC~4278. There was no unresolved source spectrum, hence no (red) ``n'' point is plotted.}
	\label{fig:n4278_dd_s2ha}
       \includegraphics[width=0.75\textwidth]{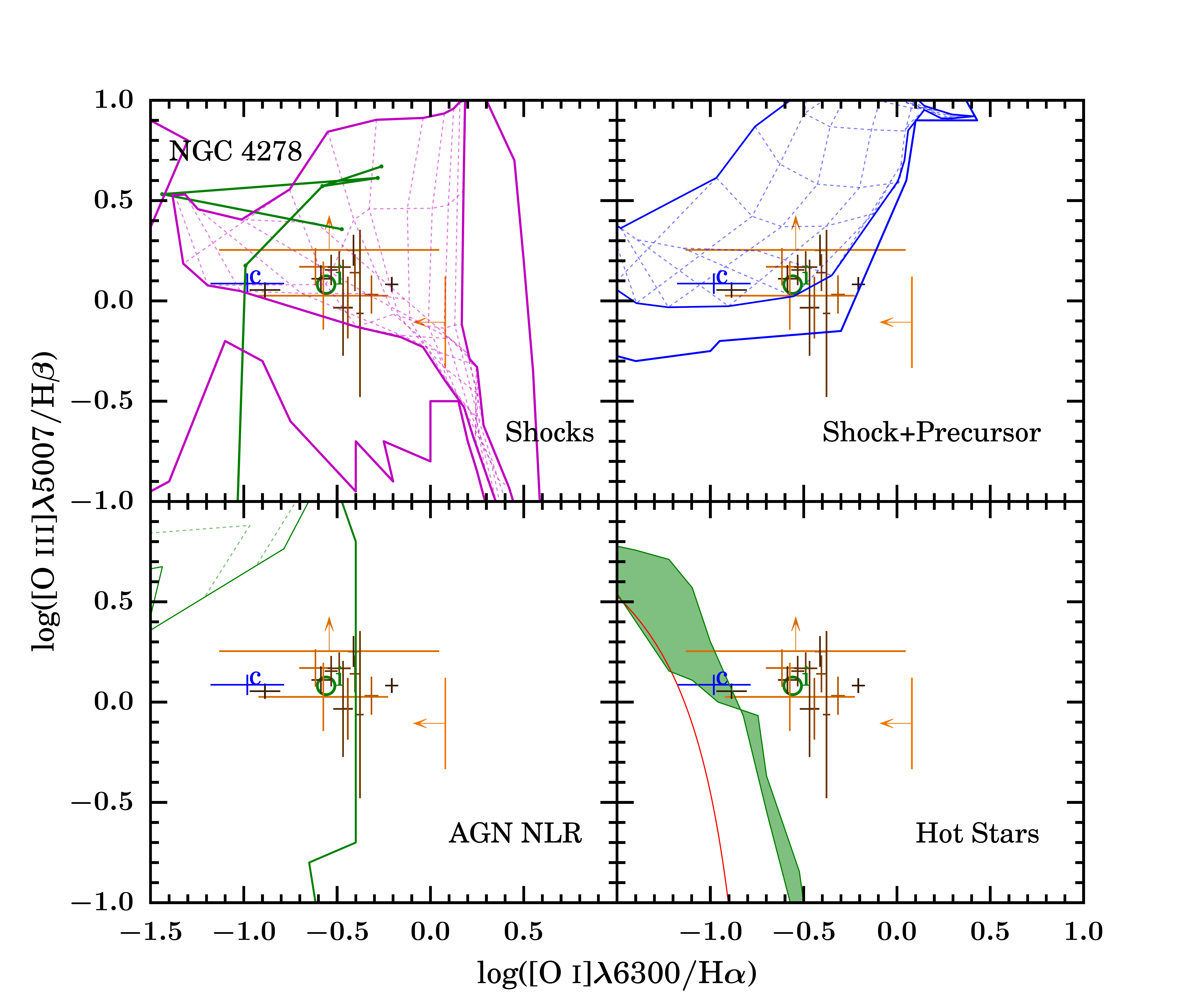}
	\caption{The same as Figure~\ref{fig:n1052_dd_o1ha}, but for NGC~4278. There was no unresolved source spectrum, hence no (red) ``n'' point is plotted.}
	\label{fig:n4278_dd_o1ha}
\end{figure*}

\begin{figure*}
	\centering
	\vspace{-3em}
       \includegraphics[width=0.75\textwidth]{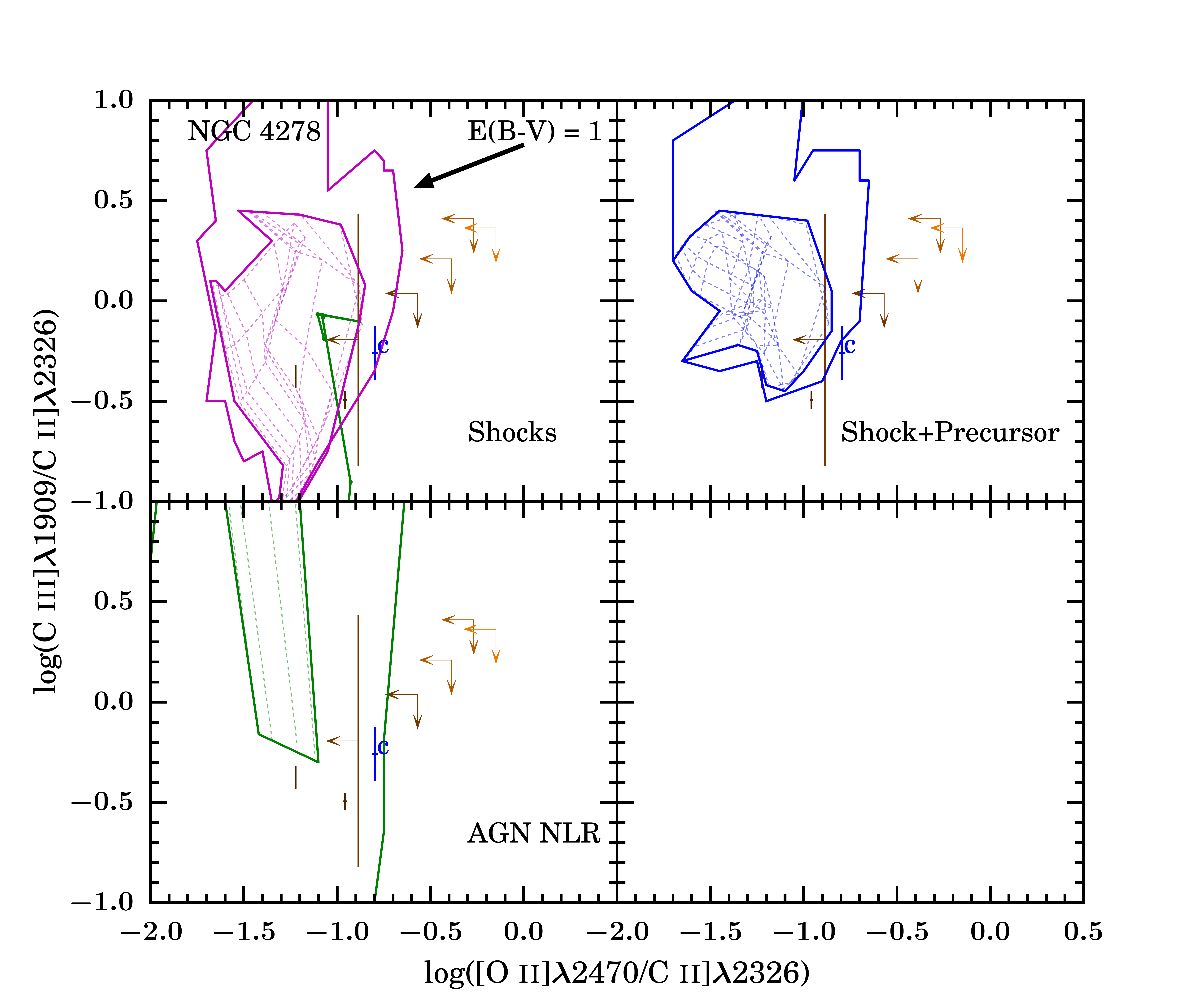}
	\caption{Same as Figure~\ref{fig:n1052_dd_o2c2}, but for NGC~4278. There was no unresolved component, hence no (red) ``n'' point is plotted. The large error bar is from the spectrum at 12~pc from the nucleus, which has a low S/N.}
	\label{fig:n4278_dd_o2c2}
       \includegraphics[width=0.75\textwidth]{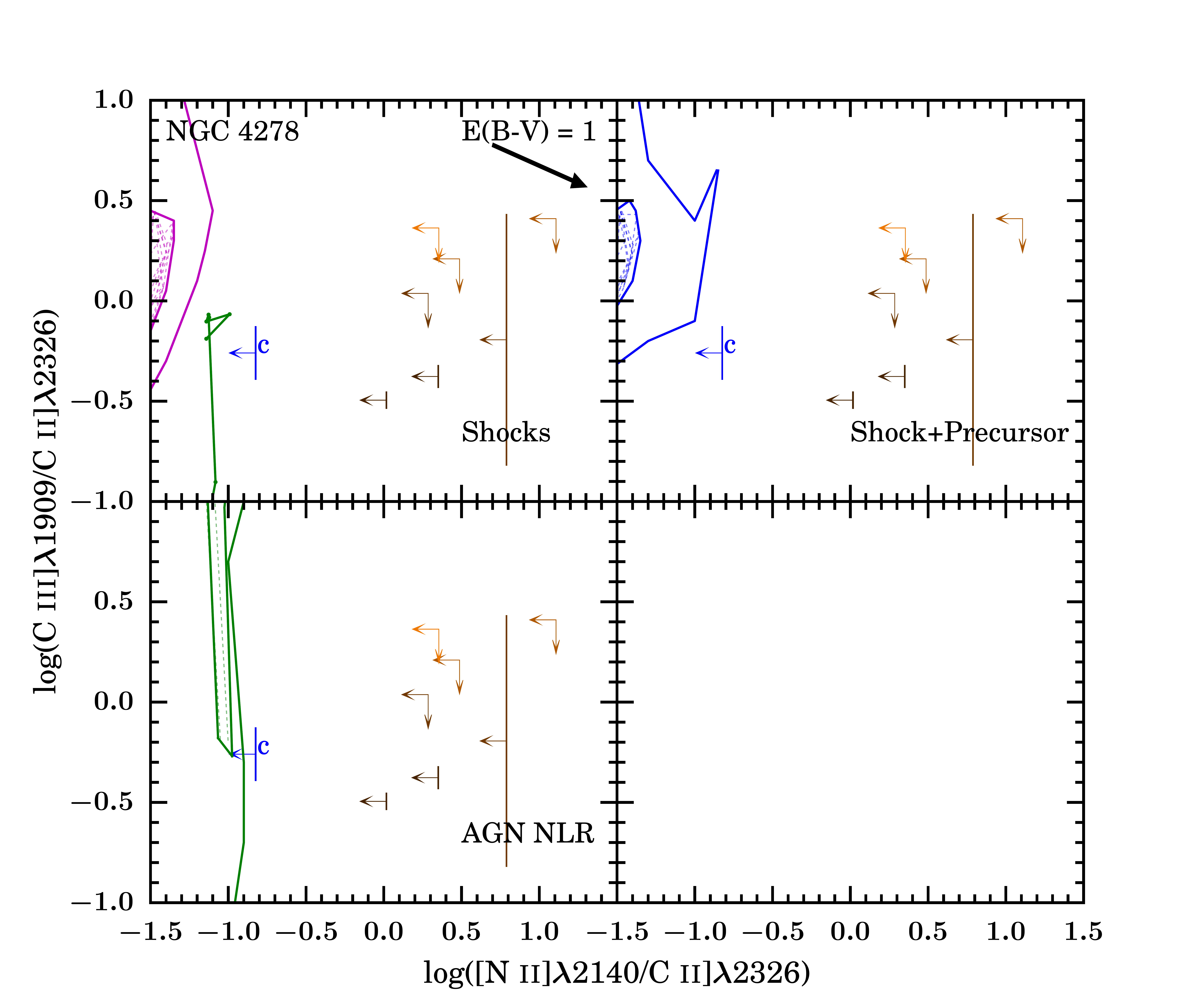}
	\caption{Same as Figure~\ref{fig:n1052_dd_n2c2}, but for NGC~4278. There was no unresolved component, hence no (red) ``n'' point is plotted. The large error bar is from the spectrum at 12~pc from the nucleus, which has a low S/N.}
	\label{fig:n4278_dd_n2c2}
\end{figure*}

\begin{figure*}
	\centering
	\vspace{-3em}
       \includegraphics[width=0.75\textwidth]{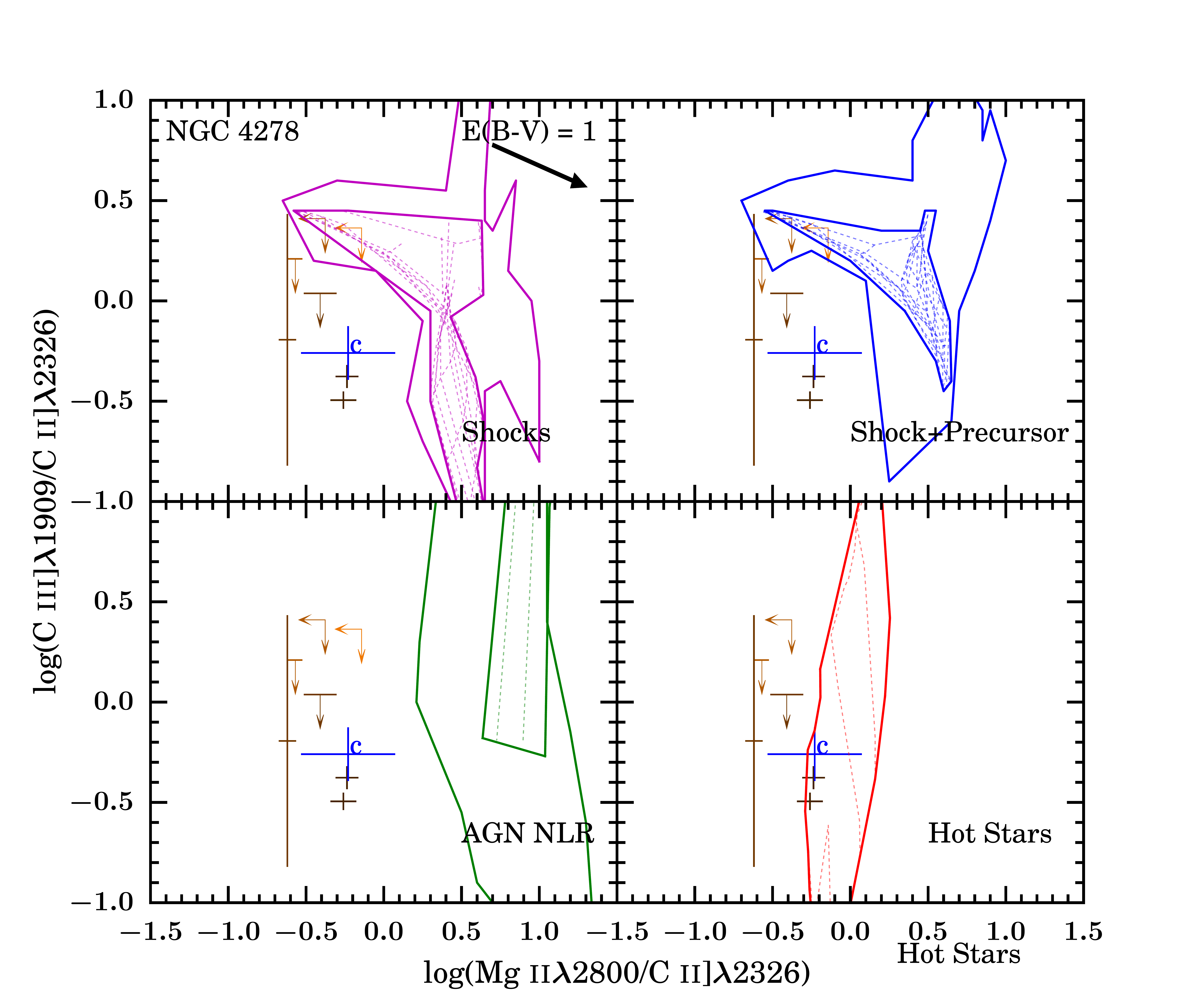}
	\caption{Same as Figure~\ref{fig:n1052_dd_mg2c2}, but for NGC~4278. There was no unresolved component, hence no (red) ``n'' point is plotted. The large error bar is from the spectrum at 12~pc from the nucleus, which has a low S/N.}
	\label{fig:n4278_dd_mg2c2}
       \includegraphics[width=0.75\textwidth]{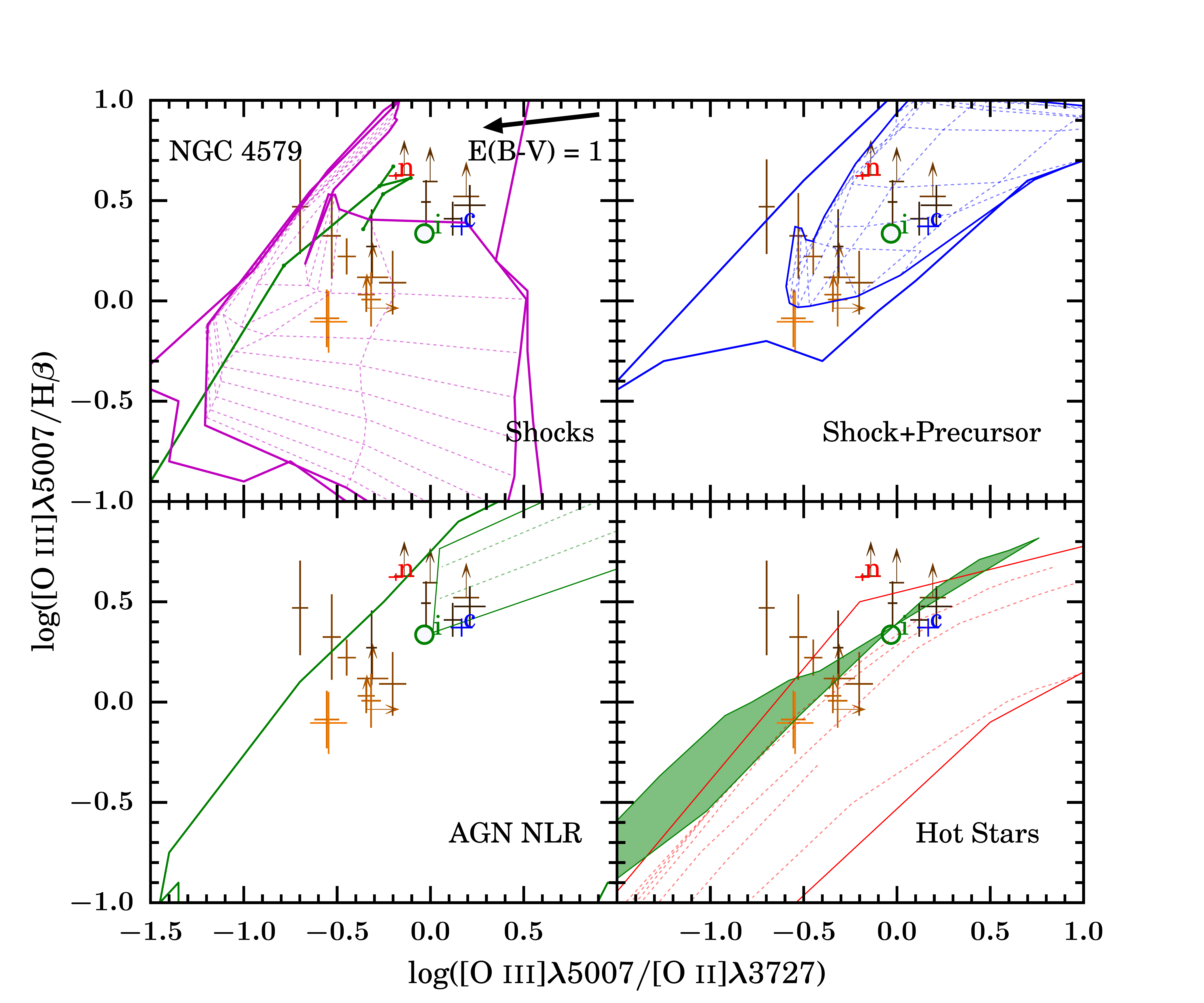}
	\caption{The same as Figure~\ref{fig:n1052_dd_o3o2}, but for NGC~4579.}
	\label{fig:n4579_dd_o3o2}
\end{figure*}

\begin{figure*}
	\centering
	\vspace{-3em}
       \includegraphics[width=0.75\textwidth]{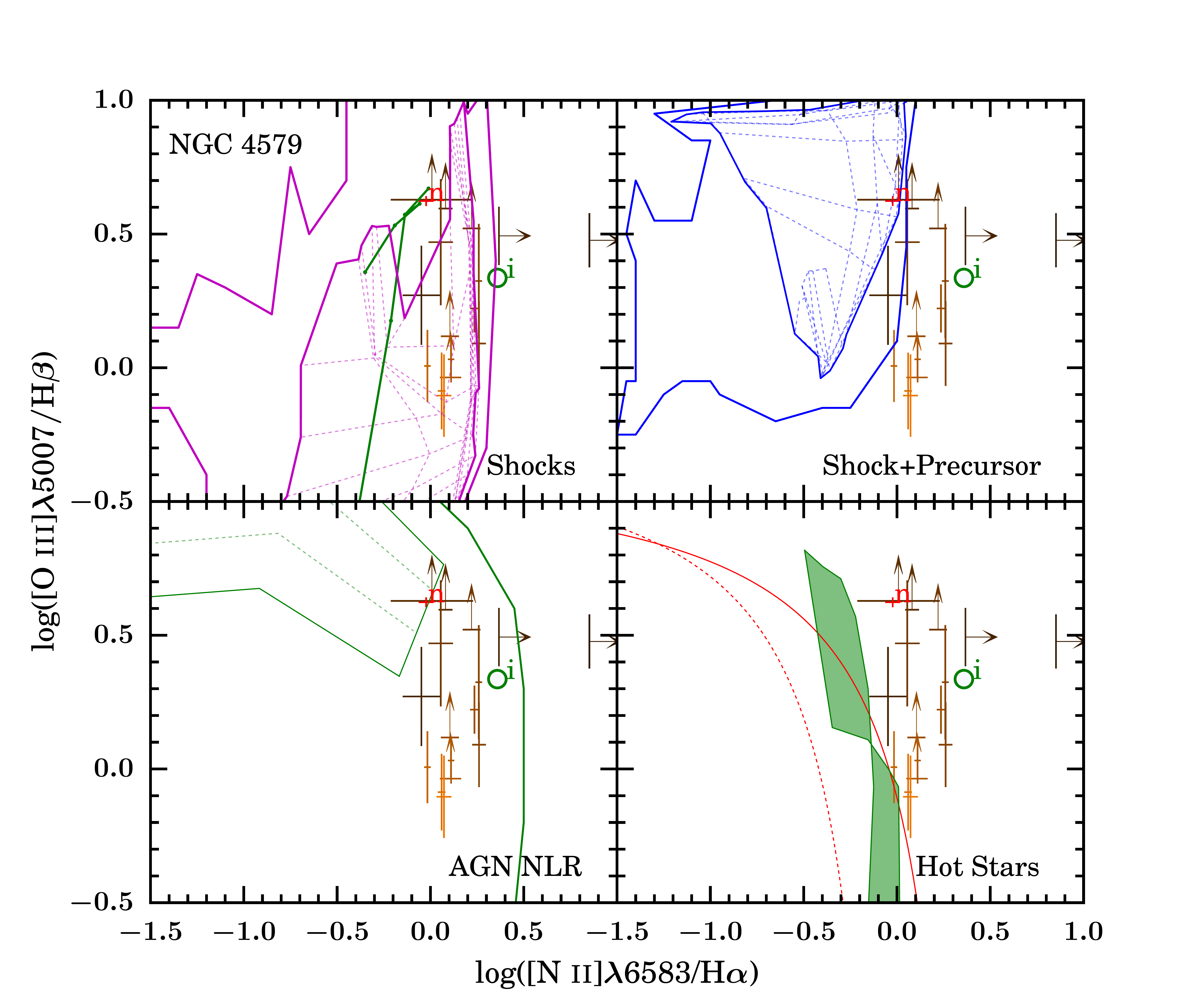}
	\caption{The same as Figure~\ref{fig:n1052_dd_n2ha}, but for NGC~4579. We did not measure emission line ratios from the resolved spectra within 12~pc from the nucleus due to the overwhelming strength of the unresolved nuclear source. See Section~\ref{ssec:2dresfit} for details.}
	\label{fig:n4579_dd_n2ha}
       \includegraphics[width=0.75\textwidth]{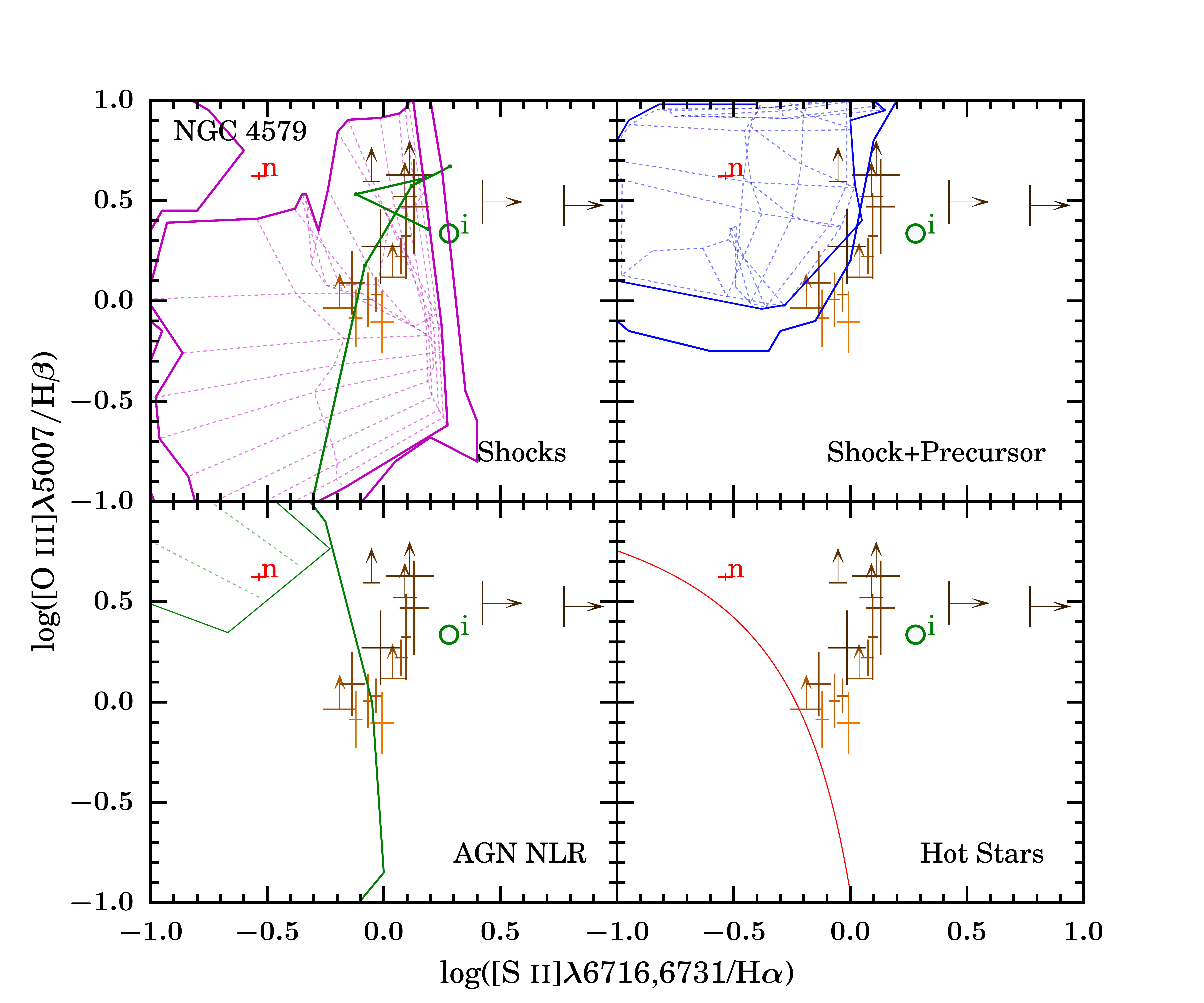}
	\caption{The same as Figure~\ref{fig:n1052_dd_s2ha}, but for NGC~4579. We did not measure emission line ratios from the resolved spectra within 12~pc from the nucleus due to the overwhelming strength of the unresolved nuclear source. See Section~\ref{ssec:2dresfit} for details.}
	\label{fig:n4579_dd_s2ha}
\end{figure*}
\begin{figure*}
\centering
       \includegraphics[width=0.75\textwidth]{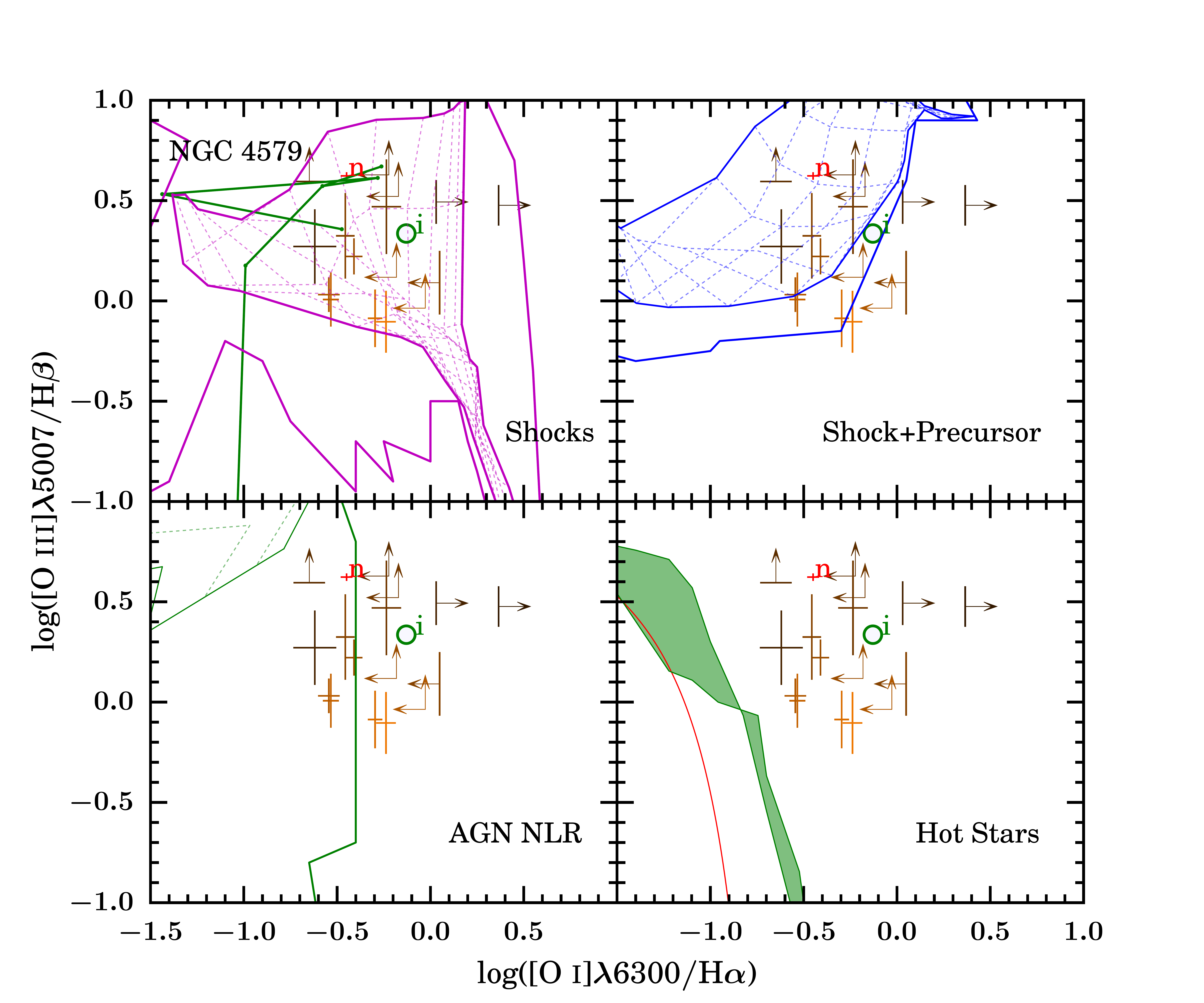}
	\caption{The same as Figure~\ref{fig:n1052_dd_o1ha}, but for NGC~4579. We did not measure emission line ratios from the resolved spectra within 12~pc from the nucleus due to the overwhelming strength of the unresolved nuclear source. See Section~\ref{ssec:2dresfit} for details.}
	\label{fig:n4579_dd_o1ha}
\end{figure*}
\begin{description}
\item[NGC~1052, Figures~\ref{fig:n1052_dd_o3o2}--\ref{fig:n1052_dd_mg2c2}]-- All but the $\textrm{\ion{C}{3}]}/\textrm{\ion{C}{2}]}$ $vs.$ $\textrm{\ion{O}{2}]}/\textrm{\ion{C}{2}]}$ diagram show that the emission from the unresolved nuclear source can be explained with the shock plus precursor model, but the required density, $n=10^{-2}\textrm{--}10^2$~cm\textsuperscript{$-3$}, is lower than the density inferred from the [\ion{S}{2}]$~\lambda\lambda6716,6731$ doublet (Section~\ref{sec:tempdens}), which disfavors this family of models. Similarly the simple shock models agree with the data in 5 of the 7 diagrams, but the required density is the same as the shock plus precursor models, disfavoring all shock models for the unresolved nuclear source emission for NGC 1052. Photoionization from an AGN agrees with the data in 5 of the 7 diagrams as long as $-1.9\leq\alpha\leq-1.2$, $-3\leq \log\,U \leq-2.8$, $Z=2\textrm{--}4\;Z_{\odot}$, and $n=10^3\textrm{--}10^4$~cm\textsuperscript{$-3$}.

In 6 of the 7 diagrams, the line ratios of the spatially resolved spectra can be explained by simple shock models with  $Z=1$--$2\;Z_{\odot}$, $n=10^{-2}$--$10^2$~cm\textsuperscript{$-3$}, $B=0.5$--$18~\mu$G, and velocities $v = 100\textrm{--}500$~km~s\textsuperscript{$-1$}. Photoionization from an AGN explains the data in 4 out of the 7 diagrams, as long as $-1.9\leq\alpha\leq-1.2$, $-3.65\leq\log\,U\leq-3.4$, $Z=1$--$2\;Z_{\odot}$, and $n=10^2$--$10^4$~cm\textsuperscript{$-3$}. The shock plus precursor models agree with the data in 4 out of 7 diagrams, as long as $Z=1$--$2\;Z_{\odot}$, $B=0.2$--$32~\mu$G, $v = 100$--500~km~s\textsuperscript{$-1$}, and $n=10^{-2}$--$10^2$~cm\textsuperscript{$-3$}. The densities required by the simple shock and the shock plus precursor models are consistent, within uncertainties, with the values inferred from the [\ion{S}{2}] doublet ratio.

The spatially resolved nuclear emission often follows the spatially resolved extended emission as described above, except for Figures~\ref{fig:n1052_dd_n2ha}--\ref{fig:n1052_dd_s2ha}. When these diagrams are taken into account, the spatially resolved nuclear emission is consistent with simple shock models in 5 out of the 7 diagrams, as long as $B=0.5~\mu$G, $v=100$--300~km~s\textsuperscript{$-1$}, $n=10^{-2}$--$10^3$~cm\textsuperscript{$-3$}, and $Z=1\;Z_{\odot}$. Photoionization from an AGN explains the data in 5 out of the 7 diagrams, as long as $-1.9\leq\alpha\leq-1.2$, $-3.6\leq\log\,U\leq-3.4$, $Z=1$--$2\;Z_{\odot}$ and $n=10^2$--$10^3$~cm\textsuperscript{$-3$}. The shock plus precursor models also agree with the data in 5 out of the 7 diagrams, as long as $B=0.2-16~\mu$G, $v=300$-400~km~s\textsuperscript{$-1$}, $n=10^{-2}$--$10^3$~cm\textsuperscript{$-3$}, and $Z=1\;Z_{\odot}$.

The upper limits for [\ion{Fe}{10}]$~\lambda 6374$ and [\ion{Fe}{14}]$~\lambda 5303$ agree with shock models from \citet{Wilson99} for $v = 200\textrm{--}350$~km~s\textsuperscript{$-1$}. The $\textrm{[\ion{Ne}{3}]}/\textrm{[\ion{O}{2}]}$ line ratio near the nucleus are consistent with both families of AGN photoionization models of \citet{Nagao02}.

\item[NGC~4278, Figures~\ref{fig:n4278_dd_o3o2}--\ref{fig:n4278_dd_mg2c2}]-- The spatially resolved spectra can be explained by simple shock models with $Z=1$--$2\;Z_{\odot}$, $n=10^{-2}$--$10^3$~cm\textsuperscript{$-3$}, $B=0.15$--$20~\mu$G, and $v = 100$--500~km~s\textsuperscript{$-1$}. The line ratios from the spatially resolved spectra can also be explained by shock plus precursor models in five of the seven diagrams, with $Z=1$--$2\;Z_{\odot}$, $n=10^{-2}$--$10^3$~cm\textsuperscript{$-3$}, $B=1$--$1000~\mu$G, $v = 100$--500~km~s\textsuperscript{$-1$}. Models of photoionization by an AGN agree with the data in five of the seven diagrams, as long as $-2.0\leq\alpha\leq-1.75$, $-3.65\leq\log\,U\leq-3.45$, $Z=0.5$--$2\;Z_{\odot}$ and $n=10^2$--$10^3$~cm\textsuperscript{$-3$}. The point representing the nuclear spectrum of NGC~4278 is often separated from the other data points in the diagnostic diagrams, and agrees with almost all models. We will discuss this point further in Section~\ref{ssec:phys_interp}.

The upper limits for [\ion{Fe}{10}]$~\lambda 6374$ and [\ion{Fe}{14}]$~\lambda 5303$ agree with shock models from \citet{Wilson99} for $v = 200\textrm{--}350$~km~s\textsuperscript{$-1$}. We could not detect $\textrm{[\ion{Ne}{3}]}/\textrm{[\ion{O}{2}]}$. The values of $\log\,U$ change from $\approx -1$ close to the nucleus to $\approx -3$ at 30~pc from the nucleus. 

\item[NGC~4579, Figures~\ref{fig:n4579_dd_o3o2}--\ref{fig:n4579_dd_o1ha}]-- The NUV spectrum is dominated by the light from the AGN, so we consider only the four optical diagnostic diagrams for this galaxy. In all diagrams, the line ratios from the spatially unresolved nuclear source can be explained by the simple shock and shock plus precursor models with $Z=1$--$2\;Z_{\odot}$, $n=0.1$--320~cm\textsuperscript{$-3$}, $B=0.0001$--$40~\mu$G, and $v = 300$--800~km~s\textsuperscript{$-1$}, but the low density required disfavors these models given the density estimate described in Section~\ref{sec:tempdens}. Photoionization by an AGN can explain the data in three of the four diagrams, if $-1.8\leq\alpha\leq-1.2$, $-3.2\leq\log\,U\leq-2.7$, $Z=0.5$--$4\;Z_{\odot}$, and $n=10^2$--$10^4$~cm\textsuperscript{$-3$}.
\end{description}
In all diagnostic diagrams, the line ratios from the spatially resolved spectra can be explained by simple shock models with $Z=1$--$2\;Z_{\odot}$, $n=10^{-2}$--$10^3$~cm\textsuperscript{$-3$}, $B=0.0001$--$40~\mu$G, and $v=100$--500~km~s\textsuperscript{$-1$}. The upper limits for [\ion{Fe}{10}]$~\lambda 6374$  and [\ion{Fe}{14}]$~\lambda 5303$ agree with shock models from \citet{Wilson99} for $v = 200$--350~km~s\textsuperscript{$-1$}. 

The [\ion{Ne}{3}]\;$\lambda3869$ line is detected only very close to the nucleus; when this line is detected the $\textrm{[\ion{Ne}{3}]}/\textrm{[\ion{O}{2}]}$ ratio agrees with the predictions of photoionization models for both families of AGN models described in \citet{Nagao02}. Our estimated value of $\log\,U$ drops from $\approx -2$ near the nucleus to $\approx -3.8$ at $\approx 80$~pc from the nucleus.

Our conclusion is that simple shock models can explain the relative strengths of most of the emission lines from regions we can resolve, i.e., on scales of a few parsec or more in NGC~1052 and NGC~4579. NGC~4278 is strongly affected by the small scale jets associated with its AGN, and will be discussed in more detail in Section~\ref{ssec:phys_interp}. The only exception are the $\textrm{\ion{C}{3}}]/\textrm{\ion{C}{2}]}$ $vs.$ $\textrm{\ion{Mg}{2}}/\textrm{\ion{C}{2}]}$ diagrams, in Figures~\ref{fig:n1052_dd_mg2c2} and~\ref{fig:n4278_dd_mg2c2}, where the data are in poor agreement with all the models. However, the model results for the \ion{Mg}{2}\;$\lambda$2800 doublet, the only resonance lines that we use here, should be regarded with caution. They were obtained with the MAPPINGS~III code, which does not evaluate multiple scatterings for resonance lines and ensuing attenuation. Thus, the strength of the \ion{Mg}{2}\;$\lambda$2800 doublet could be overestimated, as described in \citet{Groves2004}. The above conclusion is corroborated by the kinematics of the line-emitting gas, presented in Section~\ref{ssec:kinematics}. The velocity of the gas is similar to the shock velocities needed to explain the relative strengths of the lines. Thus, the picture is reminiscent of the case of M87 where the gas within the central $\sim20$~pc is photoionized \citep{Sabra03}, while the off-nuclear gas is arranged in a disk and the lines appear to be the result of shock excitation \citep{Dopita1996,Dopita1997}.
      
The line ratios measured from the integrated spectra of all three objects fall in regions of the diagnostic diagrams that can be described by more than one model. In view of the conclusions above we can attribute this ambiguity to the fact that the large ($\sim100\;$pc-size) region probed by the integrated spectrum includes a mixture of photoionized gas and gas excited by shocks.

The $\textrm{[\ion{S}{2}]}/\textrm{H}\alpha$ ratio and, to some degree, also the  $\textrm{[\ion{N}{2}]}/\textrm{H}\alpha$ and $\textrm{[\ion{O}{1}]}/\textrm{H}\alpha$ ratios, provide crucial constraints on the models for the excitation of the line emitting gas. Our preference for shock models is based largely on their ability to reproduced the observed values of these ratios when other models cannot. Because of the importance of the  $\textrm{[\ion{S}{2}]}/\textrm{H}\alpha$ ratio, we examine more carefully the predictions for its value made by a variety of photoionization models. We observe this ratio to be clustered around 1.0 in NGC~1052 and around 1.2 in NGC~4278, and to span the range 1.0--1.4 in NGC~4579. These values are too large for AGN NLR models to explain, regardless of what the other line ratios (or physical parameters) are. Because of these large values, {\it composite} photoionization involving a range of densities and ionization parameters from the sets that we tested \citep[cf.,][]{Barth1996} cannot reproduce $\textrm{[\ion{S}{2}]}/\textrm{H}\alpha$ ratios as high as those observed. The dusty, isobaric models of \citet{Dopita2002b} and \cite{Groves2004} can produce such high values of $\textrm{[\ion{S}{2}]}/\textrm{H}\alpha$ for $\log U \approx -3$ but they cannot simultaneously reproduce the $\textrm{[\ion{N}{2}]}/\textrm{H}\alpha$ and $\textrm{[\ion{O}{3}]}/\textrm{H}\beta$ ratios. The \citet{Ho1993} models can reproduce the observed values $\textrm{[\ion{S}{2}]}/\textrm{H}\alpha$ for $\log U\approx -4$ but they require $n < 10^3$~cm\textsuperscript{$-3$}, in contradiction with what we infer from the relative strengths of the lines in the [\ion{S}{2}]$~\lambda\lambda6716,6731$ doublet. The models of \citet{Binette1996} and their extension by \citet{Sabra03} employ a mix of ionization-bounded and matter-bounded clouds and can only reproduce the observed $\textrm{[\ion{S}{2}]}/\textrm{H}\alpha$ for very low densities of $n \sim 50$~cm\textsuperscript{$-3$}, which does not agree with the density constraints from the lines in the [\ion{S}{2}]$~\lambda\lambda6716,6731$ doublet.

We further compared the measured $\textrm{[\ion{O}{1}]}/\textrm{[\ion{O}{3}]}$, $\textrm{[\ion{O}{2}]}/\textrm{[\ion{O}{3}]}$, and $\textrm{[\ion{S}{2}]}/\textrm{[\ion{O}{3}]}$ line ratios of the spatially unresolved nuclear source in NGC~1052 and NGC~4579 to the two families of photoionization models in \citet{Nagao02}. We found that these ratios fall in the regions of the diagnostic diagrams where the predictions of the RIAF and optically thick, geometrically thin disk models overlap given the densities inferred from the [\ion{S}{2}] doublet ratio. Thus we cannot distinguish between the models for the SED of the ionizing continuum based on the data presented here \citep[see][and references therein for a discussion of SED shapes and models]{Maoz2007,Ho2008,Eracleous2010b,Nemmen2014}.

We conclude our comparison of the data with models by noting an important caveat. All the models that we have compared with the data are ``single-zone'' models, i.e., they employ fixed values of their physical parameters. However, it is quite possible that the spatial bins from which we extracted the spectra include gas with a range of densities and other parameters. This is most likely in the central parts of each galaxy where the density gradient is high. It is also possible that photoionized and shocked gas co-exist. Our understanding of the physical processes can, therefore, be advanced by constructing ``multi-zone'' and ``multi-process'' models and comparing them to the data.

%We also note that the $\textrm{[\ion{O}{1}]}/\textrm{H}\alpha$ ratio proves particularly useful in discriminating against models invoking photoionization by hot stars.

%Discussion of individual objects
%%%%%%%%%%%%%%%%%%%%%%%%%%%%%%%%%%%%%%%%%%%%%%%%%%%%%%%%%%%%%%%%%%%%%%%%%%%%%%%%%
\section{Discussion}
\label{sec:discussion}
The LINER-like spectra of the galaxies in our sample appear to result from a combination of photoionization from the AGN and shocks. In NGC~1052 and NGC~4579, the two cases where we isolated the spatially unresolved nuclear light from the extended, resolved light, we found a transition from photoionization by an AGN to excitation by shocks at $\sim20$~pc from the nucleus. Below, we consider physical interpretations for each galaxy and then discuss implications for the LINER population as a whole.
\subsection{Physical Interpretation of Diagnostic Diagrams}
\label{ssec:phys_interp}
\begin{description}
\item[NGC~1052]--  The emission from the spatially unresolved nuclear source is best explained by photoionization from the AGN. In 5 out of the 7 diagrams there is good agreement between the data and AGN photoionization models, within uncertainties, while all other models are disfavored by the low densities required in order for other model families to explain the data.

The extended, resolved emission from NGC~1052 is consistent with excitation by shocks with velocities around 300--500~km~s\textsuperscript{$-1$}. Previous studies \citep{Pogge00,Walsh2008} found evidence for strong outflows and ionized regions associated with jet-like features. Studies that incorporate radio observations found strong evidence for shocks in a turbulent, rotating disk on both parsec and sub-parsec scales, suggesting that the jet is interacting with the circumnuclear gas \citep{Sugai2005,Dopita2015}.

In view of the jet-disk interaction, the spectrum of the spatially-resolved nuclear source may contain some contributions from shocked gas, not just gas that is photoionized by the AGN. Indeed, we see that 5 out of the 7 diagnostic diagrams for this galaxy have spatially resolved nuclear emission line ratios that agree with the AGN photoionization models. We also find that 5 out of the 7 diagnostic diagrams of the spatially resolved nuclear source also agree with the shock plus precursor models, which is consistent with the turbulent rotating disk idea presented in \citet{Sugai2005} and \citet{Dopita2015}. The ambiguity in the explanations for the spatially resolved nuclear emission could imply that the source of the shocked gas could be very close to the central engine in this galaxy.

\item[NGC~4278]-- All of the observed emission line ratios are well described by shocks with velocities of 100--500~km~s\textsuperscript{$-1$}, except for the central spectrum of the resolved light, which agrees with simple shock, shock plus precursor, and AGN photoionization models.  While the presence of jets indicates that there is an AGN in NGC~4278, the inferred ionization parameter we obtain in Section~\ref{ssec:el_gal_results} ($\log\,U\approx-1$) is considerably different than than that required for AGN photoionization models to reproduce the diagnostic diagrams ($\log\,U\approx-3$). This mismatch calls into question photoionization from the AGN as an explanation for the observed line ratios and leads us to favor shock models as a better explanation of the observed spectra. We speculate that our direct view of the gas that is photoionized by the AGN is obscured. 

NGC~4278 is well known to have sub-parsec scale jets \citep{Giroletti2005,Helboldt07} which lie completely within the slit used in our observations, as well as a warped velocity field within $0.\!\!^{\prime\prime} 5$ \citep{Walsh2008}. \citet{Giroletti2005} found that the jets in NGC~4278 have a similar structure to those seen in radio-loud AGN, but are disrupted before reaching kiloparsec scales. We conclude the shocks created by the disrupted jets are responsible for powering the emission lines from the gas that we can observe directly.

\item[NGC~4579] -- The unresolved nuclear emission is best explained by photoionization from the AGN. In three out of the four diagnostic diagrams there is good agreement between the data and AGN photoionization models. The only diagram in which agreement is marginal is the $\textrm{[\ion{O}{3}]}/\textrm{H}\beta$ $vs$ $\textrm{[\ion{O}{3}]}/\textrm{[\ion{O}{2}]}$ diagram (Figure~\ref{fig:n4579_dd_o3o2}), where the data and models can be brought into agreement by a small adjustment of the extinction correction. The value of $\log\,U$ we determine in Section~\ref{ssec:el_gal_results}  agrees with the value required by the photoionization models. We note that the models of \citet{Groves2004} do not encompass the value of $\alpha$ that is measured for NGC~4579 \citep[$\alpha\approx-0.9$,][]{Eracleous2002}, which is another possible cause of the small discrepancy between the models and the data in this diagram. Our conclusion that AGN photoionization is the dominant mechanism in the unresolved nuclear source is corroborated by the NUV spectrum in Figure~\ref{fig:unres_stack_uv}, which includes broad lines and a featureless continuum, just as in an a luminous active galaxy galaxy.

The resolved emission is only consistent with simple shock models with velocities around 100--500~km~s\textsuperscript{$-1$}. While jets are not clearly detected in this object, previous studies have suggested that jets are responsible for the sub-parsec radio emission seen in NGC~4579 \citep[][and references therein]{Falcke2000,Ulvestad2001}. Additionally irregular gas rotation was found both in the central $1^{\prime\prime}$ \citep{Walsh2008}, as well as larger scales \citep{Gonzalez1996}. Thus, it appears possible that there is an outflow from the AGN that interacts with circumnuclear gas.
\end{description}
\subsection{Implications for the LINER Population}
\label{ssec:imp_pop}
We find that, in order to completely describe the emission line spectra of the three LINERs in our small sample we must invoke a combination of photoionization from the AGN and excitation by shocks at larger distances from the nucleus, in agreement with previous work on M87 by \citet{Dopita1997} and \citet{Sabra03}. Both of these processes are ultimately associated with the AGN and derive their power from accretion onto the central supermassive black hole. The idea that AGNs can interact with their surroundings in other ways besides photoionization is not new \citep[e.g.,]{Cecil1995,Capetti1997,Falcke1998,Ferruit1999,Dopita2002a}, but it does not always receive its due attention because photoionization is the dominant mechanism producing the emission lines in more luminous AGNs \citep{Laor1998}.

Conversely, discussions of the power sources of LINERs typically seek an explanation in terms of a single mechanism \citep[see][and references therein]{Ho2008}, but often a single mechanism does not provide a complete explanation of  the observed line strengths. For example, photoionization from an AGN often does explain the {\it relative} line strengths but not the energy budget \citep[e.g.,][and references therein]{Flohic2006,Eracleous2010a}. The central engines in our objects have properties similar to a less luminous, accretion-powered engine, whose lower luminosity allows the contribution from shocks to become non-negligible as predicted in \citet{Laor1998}. 

Thus, we conclude that for LINERs that have an AGN, the effect of shocks caused by jets or other types of outflows cannot be ignored. Another possibility is that energetic particles emanating from a RIAF \citep{Narayan95,Blanford99} may deposit their energy in circumnuclear gas and produce emission lines with relative intensities similar to shocks. The shocks and photoionization (and perhaps other processes) will combine in delivering energy to the line emitting gas, which means the models relying on just one of these mechanisms will not adequately describe the observed emission lines. The relative contributions of shocks and photoionization to the power budget will depend on the individual properties of the AGN and vary from object to object. As we noted in \S1, there is no way to assess quantitatively the contribution of jets to the power budget, and therefore we cannot determine whether the jets are powerful enough to explain the emission line luminosities. Hence we must infer how much they contribute from the emission line ratios and by spatially resolving and separating the regions where different excitation mechanisms dominate.

Clearly, the combination of AGN photoionization and AGN-driven shocks will not apply to all objects. The LIERs in non-active galaxies studied by \citet{Yan2012}, \citet{Belfiore2016}, and others, reside in galaxies without an AGN and represent emission from most of the volume of the galaxy. In such cases photoionization by hot stars, especially pAGB stars, is an essential process to consider since an AGN is not a plausible power source \citep[e.g.,][]{Binette94,Papaderos2013,Singh2013}. By extension, pAGB stars may also be the source of ionizing photons that power the line emission in nearby LINERs, as noted by \cite{Ho2008} and \citet{Eracleous2010a}. The photoionization may be accompanied by shocks from the fast winds of pAGB stars and proto-planetary nebulae \citep[e.g.,][]{Bujarabal94,Sevenster01,Vandesteene03} with theoretically-predicted speeds reaching several hundred ~km~s\textsuperscript{$-1$} \citep[e.g.,][]{Frank94}. Thus the combination of line emission mechanisms observed in LINERs may also apply to LIERs, although the origin of the ionizing photons and shocks is different.

%CONCLUSION
%%%%%%%%%%%%%%%%%%%%%%%%%%%%%%%%%%%%%%%%%%%%%%%%%%%%%%%%%%%%%%%%%%%%%%%%%%%%%%%%%
\section{Summary}
\label{sec:summary}

In this work, we have analyzed spatially resolved spectra of three nearby luminous LINERs: NGC~1052, NGC~4278 and NGC~4579. The spectra sample the compact nuclear emission line regions with a spatial resolution of better than 10~pc. All three objects have multiple indicators of an AGN, as well as observed H$\alpha$ line strengths on the scale of $\sim100$ pc that cannot be explained by photoionization from the AGN alone \citep{Eracleous2010a}. To amass a wide range of diagnostic lines on small spatial scales, we obtained new spectra in the blue (2900--5700~\AA) and NUV (1570--3180~\AA) bands and combined them with archival red spectra (6300--6860~\AA) from \citet{Walsh2008} obtained with a similar \textit{HST} setup . 

In order to separate the resolved light from the bright nucleus, we decomposed the 2--D spectra into a spatially unresolved nuclear source and an extended, resolved source for NGC~1052 and NGC~4579. We attempted a similar decomposition with NGC~4278, but we found that the strength of the spatially unresolved nuclear source was statistically consistent with zero. We also created an ``integrated'' spectrum, which emulates what might be observed from a ground based telescope.

We measured and compared diagnostic emission line ratios from the spectra of the extended resolved light, the spectrum of the spatially unresolved nuclear source, and integrated spectrum to models for different excitation mechanisms: shocks, photoionization from an AGN, and photoionization by hot stars (young and old). The physical model that best describes the data comprises an AGN that photoionizes the gas near the nucleus and shocks that ionize the gas at larger distances from the nucleus. The relative importance of the photoionization and shocks depends on the properties of the AGN and the source of the shocks, which varies from object to object. In the LINERs of our small sample the luminosity of the AGN is low enough that it does not overpower excitation by shocks. This physical model will not apply in cases of extended LIERs or passive red galaxies. In such galaxies, photoionization from pAGB stars and, perhaps, shocks associated with outflows from these stars may be responsible for the observed emission-line spectra.

In closing, we note that a single mechanism may not fully explain observed line strengths in LINER emission on scales of order of $~100$~pc or larger. Similarly, to fully understand the power source of extended LIERs, spatially resolved spectra are required to track any changes in the excitation mechanisms on small spatial scales.

%ACKNOWLEDGEMENTS
%%%%%%%%%%%%%%%%%%%%%%%%%%%%%%%%%%%%%%%%%%%%%%%%%%%%%%%%%%%%%%%%%%%%%%%%%%%%%%%%% 
\acknowledgements
We thank the anonymous referee for insightful comments that helped us improve this paper. We thank Ed Moran as well as John Biretta and Tala Monroe of the STIS Instrument team for their help with some of the technical aspects of this work. This work was supported by funding from the Alfred P. Sloan Foundation's Minority Ph.D. (MPHD) Program, awarded to MM in 2014--15. This material is based upon work supported by the National Science Foundation Graduate Research Fellowship Program under Grant No. DGE1255832. Any opinions, findings, and conclusions or recommendations expressed in this material are those of the authors and do not necessarily reflect the views of the National Science Foundation. Support for program number HST-GO-12595.001-A was provided by NASA through a grant from the Space Telescope Science Institute, which is operated by the Association of Universities for Research in Astronomy, Inc., under NASA contract NAS5-26555. This work was performed in part at Aspen Center for Physics, which is supported by National Science Foundation grant PHY-1607611. This research has made use of the NASA/IPAC Extragalactic Database (NED) which is operated by the Jet Propulsion Laboratory, California Institute of Technology, under contract with the National Aeronautics and Space Administration. LCH was supported by the National Key R\&D Program of China (2016YFA0400702) and the National Science Foundation of China (11473002, 11721303). This research has made use of the AstroBetter blog and Wiki, NASA's Astrophysics Data System, and the NASA/IPAC Extragalactic Database (NED), which is operated by the Jet Propulsion Laboratory, California Institute of Technology, under contract with the National Aeronautics and Space Administration. Research by A.J.B. is supported in part by NSF grant AST-1412693.

%BIBLIOGRAPHY
%%%%%%%%%%%%%%%%%%%%%%%%%%%%%%%%%%%%%%%%%%%%%%%%%%%%%%%%%%%%%%%%%%%%%%%%%%%%%%%%%

\bibliographystyle{apj}
\bibliography{all.111916.bib}

%FIGURES 
%%%%%%%%%%%%%%%%%%%%%%%%%%%%%%%%%%%%%%%%%%%%%%%%%%%%%%%%%%%%%%%%%%%%%%%%%%%%%%%%%
\label{lastpage}
\end{document}

%% file: tgt_prop.tex
\begin{deluxetable}{cccccc}
\tabletypesize{\scriptsize}
   \setlength{\tabcolsep}{3pt}
\tablecaption{Science Targets and Basic Properties\label{table: redshift}}
%\tablewidth{0pc}
\tablecolumns{4}
\tablehead{
\colhead{Object} & \colhead{} & \colhead{Distance} & \colhead{CCD Spatial} & \colhead{MAMA Spatial} & \colhead{Morphological}\vspace{-4pt} \\
{(NGC)} & {$z$} & {(Mpc)} & {Scale (pc/pix)} & {Scale (pc/pix)} & {Type}}
\startdata
{1052} & {0.0050\tablenotemark{a}} & {17.8\tablenotemark{b}} & {4.31} & {2.16} & {E4} \\
{4278} & {0.0023\tablenotemark{a}} & {14.9\tablenotemark{c}} & {3.61} & {1.81} & {E1--2}\\
{4579} & {0.0050\tablenotemark{d}} & {16.8\tablenotemark{e}} & {4.07} & {2.04} & {SBb}
\enddata
\tablenotetext{a}{Redshift from \citet{van2012}.}
\tablenotetext{b}{Distance from \citet{Tully1988}; in agreement with the surface brightness fluctuation method \citep[see][]{Jensen03}.}
\tablenotetext{c}{Distance from the surface brightness fluctuation method \citep[see][]{Jensen03}.}
\tablenotetext{d}{Redshift from \citet{UZC1999}.}
\tablenotetext{e}{Distance from \citet{Tully1988}; in agreement with the tip of the red giant branch method \citep[see][]{Tully2013}.}
\end{deluxetable}

%% file: obs_sum.tex
\begin{deluxetable*}{lllccccc}
\tablecaption{Summary of New Observations of Science Targets\label{table: dates}}
\tabletypesize{\scriptsize}
\tablewidth{0pt}
\tablecolumns{6}
\tablehead{
\colhead{Object} & \colhead{Spectrum} & \colhead{Instrument} & \colhead{Wavelength} & \colhead{Position Angle} & \colhead{UT Date} & \colhead{Number of} & \colhead{Exposure}\vspace{-8pt} \\
\colhead{(NGC)} & \colhead{Observed} & \colhead{Used} & \colhead{Range (\AA)}  & \colhead{(\textsuperscript{$\circ$}E of N)} & {} & \colhead{Exposures} & \colhead{Time (s)}}
\startdata
{1052} & {red (H$\alpha$)} & {STIS/CCD} & {6300$-$6860} & {13.8} & {1999$-$01$-$02} & {1} & {1974} \\
       & {blue (H$\beta$)} & {STIS/CCD} & {2900$-$5700}  & {11.1}  & {2012$-$11$-$29} & {6} & {4407}\\
       & {NUV}  & {STIS/NUV-MAMA} & {1570$-$3180}     & {11.0} & {2012$-$12$-$06} & {4} & {10736}\\
{4278} & {red (H$\alpha$)} & {STIS/CCD} & {6300$-$6860} & {88.0} & {2000$-$05$-$11} & {1} & {3128} \\
       & {blue (H$\beta$)} & {STIS/CCD} & {2900$-$5700}  & {85.1}  & {2012$-$05$-$20} & {6} & {4437}\\
       & {NUV} & {STIS/NUV-MAMA} & {1570$-$3180}     & {85.0} & {2012$-$05$-$20} & {2} & {6014}\\
{4579} & {red (H$\alpha$)} & {STIS/CCD} & {6300$-$6860} & {95.4} & {1999$-$04$-$21} & {1} & {2692} \\
       & {blue (H$\beta$)} & {STIS/CCD} & {2900$-$5700}  & {92.5}  & {2012$-$05$-$19} & {6} & {4425}\\
       & {NUV} & {STIS/NUV-MAMA}  & {1570$-$3180}    & {92.4} & {2012$-$05$-$19} & {2} & {5990}\\
\enddata
\end{deluxetable*}

%% file: stellar_obs.tex
\begin{deluxetable*}{lllccc}
\tablecaption{Summary of Observations of PSF Stars\label{table: stars}}
\tabletypesize{\scriptsize}
\tablewidth{0pc}
\tablecolumns{5}
\tablehead{
\colhead{Object} & \colhead{Spectrum} & \colhead{Instrument} & \colhead{Wavelength}  & \colhead{UT Date}  & \colhead{Exposure}\vspace{-8pt}\\
\colhead{} & \colhead{Observed} & \colhead{Used} & \colhead{Range (\AA)}   & {}  & \colhead{Time (s)}}
\startdata
{Sirius B} & {red (H$\alpha$)} & {STIS/CCD} & {6300$-$6860} & {2004$-$02$-$06} & {90} \\
{BD+75D325} & {blue (H$\beta$)} & {STIS/CCD} & {2900$-$5700} & {2000$-$08$-$10} & {29} \\
{GRW+70D5824} & {NUV} & {STIS/NUV-MAMA} & {1570$-$3180} & {1997$-$10$-$31} & {150}
\enddata
\end{deluxetable*}

%% file: uv_table.tex
\begin{deluxetable*}{lrccccccc}
\tablecaption{Emission-Line Measurements from the NUV Spectra\label{table:uv_lines}}
\tabletypesize{\scriptsize}
\tablewidth{0pc}
\tablecolumns{9}
\tablehead{
\colhead{} & \colhead{} & \multicolumn{7}{c}{Emission-Line Flux\tablenotemark{b}}\vspace{-12pt}\\
\colhead{} & \colhead{Dist.\tablenotemark{a}} & \multicolumn{7}{c}{\hrulefill}\vspace{-8pt}\\
          \colhead{Galaxy} & \colhead{(pc)} & \colhead{\ion{C}{3}]$\lambda$1909}& \colhead{\ion{N}{2}$\lambda$2142}& \colhead{\ion{C}{2}]$\lambda$2326}& \colhead{[\ion{O}{2}]$\lambda$2470}& \colhead{\ion{Mg}{2}$\lambda$2798}& \colhead{\ion{Mg}{2}$\lambda$2803}& \colhead{\ion{Mg}{1}$\lambda$2852}}
\startdata 
{NGC~1052} & {$-19$} &{$< 10$} & {$< 6$} & \phantom{1}{$5.1\pm0.6$} & {$< 5$} & {$< 7$} & {$< 8$} & {$< 9$}\\
{  } & {$-15$} & {$< 20$} & {$< 5$} & {$12.2\pm0.8$} & {$< 5$} & \phantom{1}{$4.0\pm0.6$} & \phantom{1}{$2.0\pm0.3$} & {$< 9$}\\
{  } & {$-11$} & {$12\pm2$} & \phantom{1}{$< 10$} & {$17.2\pm0.9$} & \phantom{1}{$5.0\pm0.7$} & \phantom{1}{$8.7\pm0.8$} & \phantom{1}{$4.3\pm0.4$} & {$< 5$}\\
{  } & {$-6$} & {$16\pm2$} & \phantom{1}{$< 10$} & {$27.0\pm0.9$} & {$11\pm1$} & {$20\pm1$} &{$9.78\pm0.5$} & {$< 8$}\\
{  } & {$0$} & \phantom{1}{$9\pm3$} & {$< 7$} & {$24\pm1$} & {$12\pm1$} & {$17\pm1$} & \phantom{1}{$8.6\pm0.5$} & {$< 8$}\\
{  } & {$4$} & \phantom{1}{$9\pm1$} & {$< 7$} & \phantom{1}{$20\pm10$} & \phantom{1}{$7.1\pm0.9$} & \phantom{1}{$10\pm10$} & \phantom{1}{$10\pm10$} & {$< 7$}\\
{  } & {$9$} & {$< 20$} & {$< 1$} & {$13\pm2$} & {$< 1$} & \phantom{1}{$3.8\pm0.7$} & \phantom{1}{$1.9\pm0.4$} & {$< 9$}\\
{  } & {$13$} & {$< 20$} & {$< 9$} & \phantom{1}{$7\pm1$} & {$< 7$} & \phantom{1}{$6\pm1$} & \phantom{1}{$3.2\pm0.6$} & {$< 6$}\\
{  } & {$17$} & \phantom{1}{$5\pm2$} & \phantom{1}{$< 10$} & \phantom{1}{$7.1\pm0.8$} & {$< 5$} & \phantom{1}{$2.2\pm0.7$} & \phantom{1}{$1.1\pm0.4$} & {$< 8$}\\
{  } & {$22$} & {$< 20$} & {$< 8$} & \phantom{1}{$4.7\pm0.7$} & {$< 7$} & {$< 4$} & {$< 8$} & {$< 8$}\\
{  } & {Unres} & {$91\pm3$} & \phantom{1}{$< 30$} & {$130\pm20$} & \phantom{1}{$70\pm10$} & {$160\pm60$} & \phantom{1}{$80\pm30$} & \phantom{1}{$< 40$}\\
\\
{NGC~4278} & {$-13$} & {$< 20$} & \phantom{1}{$< 10$} & \phantom{1}{$8.8\pm0.8$} & {$< 9$} & \phantom{1}{$3.8\pm0.8$} & \phantom{1}{$1.9\pm0.4$} & \phantom{1}{$< 10$}\\
{  } & {$-9$} & {$< 20$} & \phantom{1}{$< 10$} & {$15\pm1$} & \phantom{1}{$10\pm10$} & \phantom{1}{$9\pm2$} & \phantom{1}{$5\pm1$} & \phantom{1}{$< 20$}\\
{  } & {$-5$} & {$16\pm3$} & \phantom{1}{$< 20$} & {$38\pm2$} & \phantom{1}{$8\pm1$} & {$28\pm2$} & {$14\pm1$} & \phantom{1}{$< 10$}\\
{  } & {$0$} & {$21\pm3$} & \phantom{1}{$< 30$} & {$83\pm8$} & {$23\pm2$} & {$60\pm3$} & {$30\pm1$} & \phantom{1}{$< 20$}\\
{  } & {$4$} & {$< 40$} & \phantom{1}{$< 20$} & {$46\pm2$} & {$16\pm2$} & {$55\pm3$} & {$27\pm1$} & \phantom{1}{$< 10$}\\
{  } & {$7$} & {$13\pm5$} & \phantom{1}{$< 20$} & {$11\pm2$} & \phantom{1}{$< 20$} & {$18\pm5$} & \phantom{1}{$9\pm3$} & \phantom{1}{$< 20$}\\
{  } & {$11$} & {$< 20$} & \phantom{1}{$< 10$} & \phantom{1}{$4.5\pm0.9$} & {$< 9$} & \phantom{1}{$< 20$} & {$< 9$} & \phantom{1}{$< 20$}\\
{  } & {$18$} & {$< 20$} & \phantom{1}{$< 10$} & \phantom{1}{$8\pm1$} & \phantom{1}{$< 10$} & \phantom{1}{$< 20$} & {$< 9$} & \phantom{1}{$< 20$}\\
\\
{NGC~4579} & {Unres} & {$40\pm1$} & {$2.2\pm0.6$} & {$34\pm3$} & \phantom{1}{$8\pm1$} & {$34\pm4$} & {$17\pm2$} & {$16\pm2$}
\enddata
\tablenotetext{a}{Distance with respect to nuclear row in 2--D spectrum, while ``Unres'' refers to the unresolved nuclear source spectrum.}
\tablenotetext{b}{Flux is presented in units of $10^{-17}$~erg~s\textsuperscript{$-1$}~cm\textsuperscript{$-2$}.}
\end{deluxetable*}

%% file: hb_1052_table_1.tex
\movetabledown=6.3in
\begin{rotatetable}
    \begin{deluxetable}{rccccccccccccccccc}
\tabletypesize{\scriptsize}
   \setlength{\tabcolsep}{2pt}
 \tablecaption{Emission-Line Measurements from the NGC~1052 Blue Spectra  \label{table:hb_1052}}
\tablewidth{0pc}
\tablecolumns{18}
\tablehead{
\colhead{} & \multicolumn{17}{c}{Emission-Line Flux\tablenotemark{b}}\vspace{-12pt}\\
\colhead{Dist.\tablenotemark{a}} & \multicolumn{17}{c}{\hrulefill}\vspace{-8pt}\\
\colhead{} & \colhead{[\ion{O}{2}]}& \colhead{[\ion{Ne}{3}]} & \colhead{\ion{H8+He}{1}} & \colhead{[\ion{Ne}{3}]} & \colhead{\ion{H7+He}{2}} & \colhead{[\ion{S}{2}]} & \colhead{H$\delta$} & \colhead{H$\gamma$} & \colhead{[\ion{O}{3}]} & \colhead{\ion{He}{2}} &  \colhead{H$\beta$} & \colhead{[\ion{O}{3}]} & \colhead{[\ion{O}{3}]} & \colhead{[\ion{Fe}{7}]} & \colhead{[\ion{Fe}{6}]} & \colhead{[\ion{N}{1}]} & \colhead{[\ion{Fe}{14}]}\vspace{-8pt}\\
\colhead{(pc)} & \colhead{3727}& \colhead{3869} & \colhead{3890} & \colhead{3967} & \colhead{3970} & \colhead{4069,4076} & \colhead{4101} & \colhead{4341} & \colhead{4363} & \colhead{4686} &  \colhead{4861} & \colhead{4959} & \colhead{5007} & \colhead{5160} & \colhead{5177} & \colhead{5200} & \colhead{5303}}

\startdata 
{$-91$} & \phantom{2}{$5\pm7$} & {$< 8$} & {$< 8$} & {$< 3$} & {$< 8$} & {$< 6$} & {$< 6$} & {$< 3$} & {$< 3$} & {$< 4$} & \phantom{1}{$< 10$} & {$< 2$} & {$< 6$} & {$< 3$} & {$< 3$} & {$< 3$} & {$< 5$}\\
{$-82$} & {$< 20$} & {$< 6$} & {$< 9$} & {$< 2$} & {$< 4$} & {$< 5$} & {$< 4$} & {$< 4$} & {$< 4$} & {$< 5$} & {$< 6$} & {$1.0\pm0.3$} & {$3\pm1$} & {$< 3$} & {$< 3$} & {$< 5$} & {$< 6$}\\
{$-73$} & \phantom{1}{$8\pm2$} & {$< 6$} & {$< 6$} & {$< 2$} & {$< 4$} & {$< 5$} & {$< 5$} & {$< 4$} & {$< 4$}  & {$< 5$} & {$< 5$} & {$1.3\pm0.3$} & {$3.8\pm0.9$} & {$< 5$} & {$< 5$} & {$< 5$} & {$< 5$}\\
{$-65$} & {$12\pm2$} & {$< 8$} & {$< 8$} & {$< 3$} & {$< 8$} & {$< 7$} & {$< 7$} & {$< 4$} & {$< 4$} & {$< 4$} & {$< 5$} & {$1.8\pm0.5$} & {$5\pm2$} & {$< 4$} & {$< 4$} & {$< 3$} & {$< 5$}\\
{$-56$} &{$15\pm3$} & {$< 5$} & {$< 5$} & {$< 2$} & {$< 9$} & {$< 7$} & {$< 6$} & {$< 5$} & {$< 5$}  & {$< 4$} & {$< 7$} & {$< 2$} & {$< 6$} & {$< 4$} & {$< 4$} & {$2\pm1$} & {$< 5$}\\
{$-47$} & {$25\pm2$} & \phantom{1}{$< 10$} & \phantom{1}{$< 10$} & {$< 4$} & {$< 7$} & {$< 8$} & {$< 4$} & {$< 4$} & {$< 4$} & {$< 7$} & {$6\pm1$} & {$2.3\pm0.4$} & {$7\pm1$} & {$< 5$} & {$< 5$} & {$< 5$} & {$< 5$}\\
{$-39$} & {$26\pm2$} & {$< 5$} & {$< 5$} & {$< 2$} & {$< 7$} & {$< 7$} & {$< 7$} & {$2\pm1$} & {$< 6$} & {$< 4$} & {$9\pm2$} & {$2.9\pm0.4$} & {$9\pm1$} & {$< 8$} & {$< 8$} & {$< 7$} & {$< 7$}\\
{$-30$} & \phantom{2}{$39\pm3$}\phantom{1} & {$< 9$} & {$< 9$} & {$< 3$} & {$< 8$} & {$< 8$} & {$< 8$} & {$8\pm1$} & {$< 6$} & {$< 5$} & {$16\pm2$}\phantom{1} & {$4.4\pm0.6$} & {$13\pm2$}\phantom{1} & {$5\pm2$} & {$< 4$} & {$7\pm6$} & {$< 6$}\\
{$-22$} & \phantom{2}{$78\pm2$}\phantom{1} & \phantom{2}{$4\pm1$} & {$5\pm5$} & {$1.3\pm0.4$} & {$3\pm1$} & \phantom{1}{$5\pm1$} & {$< 6$} & {$10\pm1$}\phantom{1} & {$< 3$} & {$< 6$} & {$23\pm2$}\phantom{1} & {$8.1\pm0.5$} & {$24\pm1$}\phantom{1} & {$6\pm3$} & {$< 7$} & {$10\pm2$}\phantom{1} & {$< 8$}\\
{$-13$} & {$172\pm3$}\phantom{2} & {$21\pm4$} & {$4\pm3$} & {$7\pm1$} & {$< 9$} & {$25\pm3$} & {$< 8$} & {$32\pm3$}\phantom{1} & {$< 5$} & {$< 7$} & {$95\pm4$}\phantom{1} & {$50.6\pm0.9$}\phantom{1} & {$152\pm3$}\phantom{10} & {$8\pm2$} & {$< 9$} & {$17\pm2$}\phantom{1} & \phantom{1}{$< 10$}\\
{$0$} & {$212\pm3$}\phantom{2} & {$43\pm2$} & {$< 8$} & {$14.3\pm0.8$}\phantom{1} & {$7\pm7$} & {$51\pm2$} & {$14\pm2$} & {$28\pm4$}\phantom{1} & {$14\pm3$}\phantom{1} & {$8\pm3$} & {$130\pm30$}\phantom{1} & {$113\pm1$}\phantom{10} & {$340\pm4$}\phantom{10} & {$15\pm5$}\phantom{1} & \phantom{1}{$< 10$} & {$38\pm9$}\phantom{1} & {$< 6$}\\
{$9$} & {$96\pm3$} & {$23\pm2$} & {$6\pm4$} & {$7.7\pm0.6$} & {$8\pm2$} & {$19\pm1$} & \phantom{1}{$6\pm1$} & {$14\pm5$}\phantom{1} & {$< 6$} & {$< 5$} & {$50\pm40$} & {$54\pm1$}\phantom{1} & {$162\pm4$}\phantom{10} & {$10\pm8$}\phantom{1} & {$< 4$} & {$30\pm10$} & {$< 6$}\\
{$17$} & {$138\pm2$}\phantom{1} & {$22\pm2$} & {$4\pm2$} & {$7.4\pm0.7$} & {$6\pm2$} & {$19\pm3$} & {$14\pm3$} & {$24\pm2$}\phantom{1} & {$< 3$} & {$< 6$} & {$40\pm3$}\phantom{1} & {$22.4\pm0.6$}\phantom{1} & {$67\pm2$}\phantom{1} & {$5\pm2$} & {$< 6$} & {$19\pm2$}\phantom{1} & {$< 6$}\\
{$26$} & {$61\pm2$} & \phantom{1}{$6\pm2$} & {$2\pm2$} & {$2.1\pm0.5$} & {$3\pm2$} & {$< 8$} & {$< 8$} & {$5.0\pm0.3$} & {$< 5$} & {$< 4$} & {$11\pm2$}\phantom{1} & {$3.5\pm0.5$} & {$10\pm2$}\phantom{1} & {$< 7$} & {$< 7$} & {$7\pm2$} & {$< 6$}\\
{$34$} & {$30\pm2$} & \phantom{2}{$4\pm1$} & {$< 7$} & {$1.4\pm0.4$} & {$< 8$} & {$< 6$} & \phantom{1}{$4\pm1$} & {$< 4$} & {$< 4$}  & {$< 5$} & {$8\pm2$} & {$2.5\pm0.4$} & {$8\pm1$} & {$< 7$} & {$< 7$} & {$5\pm1$} & {$< 6$}\\
{$43$} & {$26.3\pm0.9$} & \phantom{1}{$< 10$} & \phantom{1}{$< 10$} & {$< 4$} & {$< 6$} & {$< 7$} & {$< 7$} & {$< 3$} & {$< 3$}  & {$< 7$} & {$< 4$} & {$2.3\pm0.4$} & {$7\pm1$} & {$< 5$} & {$< 5$} & {$< 5$} & {$< 5$}\\
{$52$} & {$16\pm2$} & \phantom{1}{$< 10$} & \phantom{1}{$< 10$} & {$< 3$} & {$< 7$} & {$< 4$} & {$< 5$} & {$< 5$} & {$< 5$} & {$< 3$} & {$< 4$} & {$2.8\pm0.5$} & {$8\pm1$} & {$< 6$} & {$< 6$} & {$< 4$} & {$< 6$}\\
{$60$} & {$11\pm3$} & \phantom{1}{$< 10$} & \phantom{1}{$< 10$} & {$< 3$} & {$< 9$} & {$< 6$} & {$< 5$} & {$< 5$} & {$< 5$} & {$< 3$} & {$3\pm1$} & {$1.6\pm0.4$} & {$5\pm1$} & {$< 5$} & {$< 5$} & {$< 4$} & {$< 3$}\\
{$69$} & {$15\pm2$} & {$< 7$} & {$< 7$} & {$< 2$} & {$< 9$} & {$< 5$} & {$< 7$} & {$< 5$} & {$< 5$} & {$< 5$} & {$< 6$} & {$1.3\pm0.4$} & {$4\pm1$} & {$< 5$} & {$< 5$} & {$< 5$} & {$< 5$}\\
{$78$} & {$13\pm2$} & {$< 8$} & {$< 8$} & {$< 3$} & {$< 7$} & {$< 5$} & {$< 4$} & {$< 4$} & {$< 4$} & {$< 5$} & {$< 5$} & {$2.0\pm0.3$} & {$6\pm1$} & {$< 4$} & {$< 4$} & {$< 4$} & {$< 4$}\\
{$86$} & \phantom{2}{$9\pm9$} & {$< 8$} & {$< 8$} & {$< 3$} & {$< 7$} & {$< 5$} & {$< 4$} & {$< 4$} & {$< 4$} & {$< 5$} & {$< 5$} & {$< 2$} & {$< 5$} & {$< 4$} &{$< 4$} & {$< 4$} & {$< 4$}\\
{Unres} & {$1030\pm20$}\phantom{1} & {$100\pm20$} & {$290\pm20$}\phantom{1} & {$33\pm6$}\phantom{1} & {$90\pm40$} & {$810\pm30$}\phantom{1} & \phantom{1}{$20\pm70$} & {$500\pm100$} & {$400\pm300$} & {$40\pm20$} & {$951\pm70$} & {$1000\pm60$}\phantom{1} & {$3000\pm200$} & {$130\pm10$} & \phantom{1}{$< 40$} & {$190\pm20$} & \phantom{1}{$< 60$}\\
{Integ} & {$8800\pm200$} & {$1700\pm300$} & {$800\pm100$} & {$560\pm90$}\phantom{1} & {$400\pm300$} & {$2000\pm200$}\phantom{1} & \phantom{1}{$500\pm200$} & {$2000\pm1000$} & \phantom{10}{$< 600$} & \phantom{10}{$< 600$} & {$3900\pm300$} & {$2810\pm70$}\phantom{1} & {$8400\pm200$} & \phantom{1}{$800\pm400$} & \phantom{10}{$< 500$} & {$1500\pm300$} & \phantom{10}{$< 700$}\\
\enddata
\tablenotetext{a}{Distance with respect to nuclear row in 2--D spectrum, while ``Unres'' and ``Integ'' refer to the unresolved nuclear source and integrated spectra, respectively.}
\tablenotetext{b}{Flux is presented in units of $10^{-17}$~erg~s\textsuperscript{$-1$}~cm\textsuperscript{$-2$}.}
\end{deluxetable}
\end{rotatetable}

%% file: hb_4278_table_1.tex
\movetabledown=6.3in
\begin{rotatetable}
    \begin{deluxetable}{rccccccccccccccccc}
\tabletypesize{\scriptsize}
   \setlength{\tabcolsep}{2pt}
 \tablecaption{Narrow Emission-Line Measurements from the NGC~4278 Blue Spectra\label{table:hb_4278}}
\tablewidth{0pc}
\tablecolumns{18}
\tablehead{
\colhead{} & \multicolumn{17}{c}{Emission-Line Flux\tablenotemark{b}}\vspace{-12pt}\\
\colhead{Dist.\tablenotemark{a}} & \multicolumn{17}{c}{\hrulefill}\vspace{-8pt}\\
\colhead{} & \colhead{[\ion{O}{2}]}& \colhead{[\ion{Ne}{3}]} & \colhead{\ion{H8+He}{1}} & \colhead{[\ion{Ne}{3}]} & \colhead{\ion{H7+He}{2}} & \colhead{[\ion{S}{2}]} & \colhead{H$\delta$} & \colhead{H$\gamma$} & \colhead{[\ion{O}{3}]} & \colhead{\ion{He}{2}} &  \colhead{H$\beta$} & \colhead{[\ion{O}{3}]} & \colhead{[\ion{O}{3}]} & \colhead{[\ion{Fe}{7}]} & \colhead{[\ion{Fe}{6}]} & \colhead{[\ion{N}{1}]} & \colhead{[\ion{Fe}{14}]}\vspace{-8pt}\\
\colhead{(pc)} & \colhead{3727}& \colhead{3869} & \colhead{3890} & \colhead{3967} & \colhead{3970} & \colhead{4069,4076} & \colhead{4101} & \colhead{4341} & \colhead{4363} & \colhead{4686} &  \colhead{4861} & \colhead{4959} & \colhead{5007} & \colhead{5160} & \colhead{5177} & \colhead{5200} & \colhead{5303}}
\startdata 
{$-72$} & {$20\pm3$} & {$< 10$} & {$< 10$} & {$< 4$} & {$< 10$} & {$< 10$} & {$< 10$} & {$< 6$}\phantom{1} & {$< 6$}\phantom{1}  & {$< 7$}\phantom{1} & {$6\pm2$} & {$1.5\pm0.6$} & {$4\pm2$} & {$< 8$}\phantom{1} & {$< 8$}\phantom{1} & {$< 6$}\phantom{1} & {$< 4$}\phantom{1}\\
{$-65$} & {$26\pm3$} & {$< 10$} & {$< 8$}\phantom{1} & {$< 3$} & {$< 10$} & {$< 10$} & {$< 10$} & {$< 9$}\phantom{1} & {$< 9$}\phantom{1} & {$< 9$}\phantom{1} & {$5\pm1$} & {$1.8\pm0.5$} & {$6\pm2$} & {$< 10$} & {$< 10$} & {$< 9$}\phantom{1} & {$< 6$}\phantom{1}\\
{$-58$} & {$42\pm2$} & {$< 10$} & {$< 10$} & {$< 4$} & {$< 10$} & {$< 10$} & {$< 10$} & {$2.9\pm0.9$} & {$< 7$}\phantom{1} & {$< 6$}\phantom{1} & {$6\pm1$} & {$3.1\pm0.4$} & {$9\pm1$} & {$< 6$}\phantom{1} & {$< 6$}\phantom{1} & {$< 6$}\phantom{1} & {$< 4$}\phantom{1}\\
{$-51$} & {$54\pm2$} & {$3\pm2$} & {$< 9$}\phantom{1} & {$1.1\pm0.6$} & {$< 10$} & {$< 10$} & {$< 8$}\phantom{1} & {$4.7\pm0.9$} & {$< 6$}\phantom{1} & {$< 7$}\phantom{1} & {$8\pm1$} & {$3.0\pm0.4$} & {$9\pm1$} & {$< 5$}\phantom{1} & {$< 5$}\phantom{1} & {$6\pm2$} & {$< 5$}\phantom{1}\\
{$-43$} & {$53\pm5$} & {$< 20$} & {$< 20$} & {$< 6$} & {$< 20$} & {$< 10$} & {$< 10$} & {$5\pm2$} & {$< 5$}\phantom{1}  & {$< 8$}\phantom{1} & {$11\pm3$}\phantom{1} & {$3.5\pm0.9$} & {$10\pm3$}\phantom{1} & {$< 7$}\phantom{1} & {$< 7$}\phantom{1} & {$< 10$} & {$< 5$}\phantom{1}\\
{$-36$} & {$57\pm3$} & {$< 20$} & {$< 20$} & {$< 5$} & {$4\pm2$} & {$< 10$} & {$< 10$} & {$4.3\pm0.8$} & {$< 10$} & {$< 7$}\phantom{1} & {$9\pm1$} & {$4.5\pm0.7$} & {$13\pm2$}\phantom{1} & {$< 8$}\phantom{1} & {$< 8$}\phantom{1} & {$7\pm2$} & {$< 6$}\phantom{1}\\
{$-29$} & {$51\pm5$} & {$< 10$} & {$< 10$} & {$< 4$} & {$< 8$}\phantom{1} & {$< 10$} & {$< 9$}\phantom{1} & {$5\pm1$} & {$< 10$} & {$< 10$} & {$11\pm1$}\phantom{1} & {$5.4\pm0.7$} & {$16\pm2$}\phantom{1} & {$< 10$} & {$< 10$} & {$< 10$} & {$< 5$}\phantom{1}\\
{$-22$} & {$57\pm2$} & {$< 20$} & {$< 20$} & {$< 5$} & {$< 10$} & {$< 10$} & {$< 10$} & {$< 10$} & {$< 10$} & {$< 9$}\phantom{1} & {$10\pm1$}\phantom{1} & {$4.9\pm0.6$} & {$15\pm2$}\phantom{1} & {$< 9$}\phantom{1} & {$< 9$}\phantom{1} & {$< 7$}\phantom{1} & {$< 7$}\phantom{1}\\
{$-15$} & {$78\pm2$} & {$3.2\pm0.2$} & {$< 10$} & {$1.05\pm0.06$} & {$3\pm1$} & {$6\pm1$} & {$< 10$} & {$< 8$}\phantom{1} & {$< 7$}\phantom{1} & {$< 10$} & {$11\pm1$}\phantom{1} & {$4.8\pm0.5$} & {$14\pm2$}\phantom{1} & {$< 8$}\phantom{1} & {$< 8$}\phantom{1} & {$< 7$}\phantom{1} & {$< 5$}\phantom{1}\\
{$-7$} & {$166\pm2$}\phantom{1} & {$9\pm3$} & {$< 10$} & {$3.2\pm0.9$} & {$4.4\pm0.5$} & {$13\pm2$}\phantom{1} & {$< 10$} & {$8\pm1$} & {$< 20$} & {$< 10$} & {$31\pm2$}\phantom{1} & {$13.0\pm0.6$}\phantom{1} & {$39\pm2$}\phantom{1} & {$< 10$} & {$< 10$} & {$< 10$} & {$< 10$}\\
{$0$} & {$240\pm20$} & {$< 10$} & {$63\pm4$} & {$< 3$} & {$30\pm2$}\phantom{1} & {$75\pm3$}\phantom{1} & {$8\pm2$} & {$40\pm10$} & {$< 10$} & {$< 6$}\phantom{1} & {$98\pm10$} & {$41\pm2$}\phantom{1} & {$124\pm6$}\phantom{10} & {$< 10$} & {$< 10$} & {$< 10$} & {$< 10$}\\
{$11$} & {$85\pm2$} & {$< 20$} & {$< 20$} & {$< 6$} & {$< 20$} & {$45\pm3$}\phantom{1} & {$< 10$} & {$39\pm2$}\phantom{1} & {$< 10$} & {$< 9$\phantom{1}} & {$35\pm3$}\phantom{1} & {$15.1\pm0.6$}\phantom{1} & {$45\pm2$}\phantom{1} & {$< 10$} & {$< 10$} & {$< 7$}\phantom{1} & {$< 10$}\\
{$18$} & {$80\pm2$} & {$< 10$} & {$< 10$} & {$< 4$} & {$< 10$} & {$< 10$} & {$< 10$} & {$9\pm2$} & {$< 10$} & {$< 8$}\phantom{1} & {$10\pm2$}\phantom{1} & {$6.5\pm0.5$} & {$20\pm2$}\phantom{1} & {$< 8$}\phantom{1} & {$< 8$}\phantom{1} & {$< 10$} & {$< 5$}\phantom{1}\\
{$25$} & {$102\pm8$}\phantom{1} & {$< 20$} & {$< 20$} & {$< 7$} & {$< 10$} & {$< 20$} & {$< 10$} & {$9\pm5$} & {$< 10$} & {$< 8$}\phantom{1} & {$16\pm6$}\phantom{1} & {$5\pm2$} & {$16\pm7$}\phantom{1} & {$< 7$}\phantom{1} & {$< 7$}\phantom{1} & {$9\pm7$} & {$< 7$}\phantom{1}\\
{$33$} & \phantom{1}{$60\pm10$} & {$< 10$} & {$< 10$} & {$< 5$} & {$< 10$} & {$< 20$} & {$< 20$} & {$< 8$}\phantom{1} & {$< 8$}\phantom{1} & {$< 8$}\phantom{1} & {$13\pm9$}\phantom{1} & {$4\pm2$} & {$12\pm7$}\phantom{1} & {$< 10$} & {$< 10$} & {$< 10$} & {$< 4$}\phantom{1}\\
{$40$} & \phantom{1}{$30\pm20$} & {$< 10$} & {$< 10$} & {$< 4$} & {$< 8$}\phantom{1} & {$< 7$}\phantom{1} & {$< 7$}\phantom{1} & {$< 20$} & {$< 20$}  & {$< 10$} & {$< 7$}\phantom{1} & {$< 2$} & {$< 6$}\phantom{1} & {$< 10$} & {$< 10$} & {$< 10$} & {$< 6$}\phantom{1}\\
{$47$} & \phantom{1}{$30\pm20$} & {$< 20$} & {$< 20$} & {$< 5$} & {$< 9$}\phantom{1} & {$< 10$} & {$< 10$} & {$< 10$} & {$< 10$} & {$< 8$}\phantom{1} & {$< 5$}\phantom{1} & {$< 2$} & {$< 5$}\phantom{1} & {$< 6$}\phantom{1} & {$< 6$}\phantom{1} & {$< 6$}\phantom{1} & {$< 4$}\phantom{1}\\
{$54$} & \phantom{1}{$20\pm10$} & {$< 20$} & {$< 20$} & {$< 6$} & {$< 10$} & {$< 10$} & {$< 10$} & {$< 10$} & {$< 10$} & {$< 9$}\phantom{1} & {$< 10$} & {$< 3$} & {$< 8$}\phantom{1} & {$< 8$}\phantom{1} & {$< 8$}\phantom{1} & {$< 8$}\phantom{1} & {$< 4$}\phantom{1}\\
{$61$} & \phantom{1}{$16\pm2$}\phantom{1} & {$< 20$} & {$< 20$} & {$< 5$} & {$< 20$} & {$< 10$} & {$< 10$} & {$< 10$} & {$< 10$} & {$< 8$\phantom{1}} & {$< 10$} & {$< 4$} & {$< 10$} & {$< 9$}\phantom{1} & {$< 9$}\phantom{1} & {$< 9$}\phantom{1} & {$< 5$}\phantom{1}\\
{$69$} & \phantom{1}{$15\pm2$}\phantom{1} & {$< 10$} & {$< 10$} & {$< 4$} & {$< 10$} & {$< 9$}\phantom{1} & {$< 9$}\phantom{1} & {$< 10$} & {$< 10$} & {$< 9$}\phantom{1} & {$< 2$}\phantom{1} & {$1.2\pm0.2$} & {$3.5\pm0.7$} & {$< 7$}\phantom{1} & {$< 7$}\phantom{1} & {$< 7$}\phantom{1} & {$< 5$}\phantom{1}\\
{$76$} & {$11\pm2$} & {$< 8$}\phantom{1} & {$< 8$}\phantom{1} & {$< 3$} & {$< 9$}\phantom{1} & {$< 9$}\phantom{1} & {$< 9$}\phantom{1} & {$< 10$} & {$< 10$} & {$< 10$} & {$< 8$}\phantom{1} & {$< 3$} & {$< 10$} & {$< 9$}\phantom{1} & {$< 9$}\phantom{1} & {$< 9$}\phantom{1} & {$< 4$}\phantom{1}\\
{Integ} & {$13000\pm1000$} & \phantom{10}{$< 2000$} & \phantom{10}{$< 2000$} & \phantom{10}{$< 600$} & \phantom{10}{$< 2000$} & \phantom{10}{$< 2000$} & \phantom{10}{$< 2000$} & \phantom{10}{$< 1000$} & \phantom{10}{$< 1000$} & \phantom{10}{$< 1000$} & {$2000\pm1000$} & {$1000\pm400$} & {$3000\pm1000$} & \phantom{10}{$< 2000$} & \phantom{10}{$< 2000$} & \phantom{10}{$< 2000$} & \phantom{10}{$< 1000$}
\enddata
\tablenotetext{a}{Distance with respect to nuclear row in 2--D spectrum, while ``Integ'' refers to the integrated spectrum.}
\tablenotetext{b}{Flux is presented in units of $10^{-17}$~erg~s\textsuperscript{$-1$}~cm\textsuperscript{$-2$}.}
\end{deluxetable}
\end{rotatetable}

%% file: hb_4579_table_1.tex
\movetabledown=6.3in
\begin{rotatetable}
    \begin{deluxetable}{rccccccccccccccccc}
\tabletypesize{\scriptsize}
   \setlength{\tabcolsep}{2pt}
 \tablecaption{Narrow Emission-Line Measurements from the NGC~4579 Blue Spectra\label{table:hb_4579}}
\tablewidth{0pc}
\tablecolumns{18}
\tablehead{
\colhead{} & \multicolumn{17}{c}{Emission-Line Flux\tablenotemark{b}}\vspace{-12pt}\\
\colhead{Dist.\tablenotemark{a}} & \multicolumn{17}{c}{\hrulefill}\vspace{-8pt}\\
\colhead{} & \colhead{[\ion{O}{2}]}& \colhead{[\ion{Ne}{3}]} & \colhead{\ion{H8+He}{1}} & \colhead{[\ion{Ne}{3}]} & \colhead{\ion{H7+He}{2}} & \colhead{[\ion{S}{2}]} & \colhead{H$\delta$} & \colhead{H$\gamma$} & \colhead{[\ion{O}{3}]} & \colhead{\ion{He}{2}} &  \colhead{H$\beta$} & \colhead{[\ion{O}{3}]} & \colhead{[\ion{O}{3}]} & \colhead{[\ion{Fe}{7}]} & \colhead{[\ion{Fe}{6}]} & \colhead{[\ion{N}{1}]} & \colhead{[\ion{Fe}{14}]}\vspace{-8pt}\\
\colhead{(pc)} & \colhead{3727}& \colhead{3869} & \colhead{3890} & \colhead{3967} & \colhead{3970} & \colhead{4069,4076} & \colhead{4101} & \colhead{4341} & \colhead{4363} & \colhead{4686} &  \colhead{4861} & \colhead{4959} & \colhead{5007} & \colhead{5160} & \colhead{5177} & \colhead{5200} & \colhead{5303}}
\startdata 
{$-85$} & {$5\pm1$} & {$< 6$} & {$< 6$}\phantom{10}  & {$< 2$} & {$< 4$} & {$< 4$} & {$< 4$}\phantom{10}  & {$< 3$} & {$< 3$}\phantom{10} & {$< 3$}\phantom{10} & {$< 4$} & \phantom{1}{$< 1$} & {$< 4$} & {$< 4$}\phantom{1} & {$< 4$}\phantom{10} & {$< 2$}\phantom{1} & {$< 4$}\phantom{1}\\
{$-77$} & {$4\pm1$} & {$< 4$} & {$< 4$}\phantom{10}  & {$< 1$} & {$< 5$} & {$< 4$} & {$< 4$}\phantom{10}  & {$< 4$} & {$< 4$}\phantom{10} & {$< 2$}\phantom{10} & {$< 4$} & \phantom{1}{$< 1$} & {$< 4$} & {$< 3$}\phantom{1} & {$< 3$}\phantom{10} & {$< 2$}\phantom{1} & {$< 4$}\phantom{1}\\
{$-69$} & {$9\pm1$} & {$< 6$} & {$< 6$}\phantom{10}  & {$< 2$} & {$< 6$} & {$< 3$} & {$< 5$}\phantom{10}  & {$< 3$} & {$< 3$}\phantom{10} & {$< 4$}\phantom{10} & {$< 3$} & \phantom{1}{$< 1$} & {$< 3$} & {$< 4$}\phantom{1} & {$< 4$}\phantom{10} & {$< 3$}\phantom{1} & {$< 4$}\phantom{1}\\
{$-61$} & {$< 6$} & {$5\pm2$} & {$< 4$}\phantom{10}  & \phantom{1}{$1.5\pm0.5$} & {$< 5$} & {$< 4$} & {$< 5$}\phantom{10}  & {$< 1$} & {$< 1$}\phantom{10} & {$< 3$}\phantom{10} & {$< 3$} & \phantom{1}{$0.9\pm0.3$} & \phantom{1}{$3\pm1$} & {$< 5$}\phantom{1} & {$< 5$}\phantom{10} & {$< 4$}\phantom{1} & {$< 6$}\phantom{1}\\
{$-53$} & {$13\pm1$}  & {$< 5$} & {$< 5$}\phantom{10}  & {$< 2$} & {$< 5$} & {$< 6$} & {$< 4$}\phantom{10}  & {$< 4$} & {$< 4$}\phantom{10} & {$< 4$}\phantom{10} & {$< 5$} & \phantom{1}{$2.1\pm0.3$} & \phantom{1}{$6\pm1$} & {$< 2$}\phantom{1} & {$< 2$}\phantom{10} & {$< 2$}\phantom{1} & {$< 5$}\phantom{1}\\
{$-45$} & {$12\pm1$}  & {$< 5$} & {$< 5$}\phantom{10}  & {$< 2$} & {$< 7$} & {$< 3$} & {$< 4$}\phantom{10}  & {$< 2$} & {$< 2$}\phantom{10} & {$< 4$}\phantom{10} & \phantom{1}{$6\pm2$} & \phantom{1}{$2.5\pm0.4$} & \phantom{1}{$8\pm1$} & {$< 3$}\phantom{1} & {$< 3$}\phantom{10} & {$< 3$}\phantom{1} & {$< 4$}\phantom{1}\\
{$-37$} & {$10\pm1$}  & {$< 5$} & {$< 5$}\phantom{10}  & {$< 2$} & {$< 7$} & {$< 6$} & {$< 5$}\phantom{10}  & {$< 5$} & {$< 5$}\phantom{10} & {$< 4$}\phantom{10} & {$< 5$} & \phantom{1}{$5.2\pm0.6$} & {$16\pm2$} & {$< 4$}\phantom{1} & {$< 4$}\phantom{10} & {$< 3$}\phantom{1} & {$< 5$}\phantom{1}\\
{$-28$} & {$23\pm1$}  & {$< 6$} & {$< 6$}\phantom{10}  & {$< 2$} & {$< 6$} & {$< 3$} & {$< 4$}\phantom{10}  & {$< 3$} & {$< 3$}\phantom{10} & {$< 4$}\phantom{10} & {$< 6$} & \phantom{1}{$7.6\pm0.5$} & {$23\pm2$} & {$< 5$}\phantom{1} & {$< 5$}\phantom{10} & {$< 6$}\phantom{1} & {$< 6$}\phantom{1}\\
{$-20$} & {$42\pm2$}  & {$< 5$} & {$< 5$}\phantom{10}  & {$< 2$} & {$< 5$} & {$< 7$} & {$< 4$}\phantom{10}  & \phantom{1}{$9\pm2$} & {$< 4$}\phantom{10} & {$< 3$}\phantom{10} & {$15\pm7$} & {$10.0\pm0.5$} & {$30\pm2$} & {$< 6$}\phantom{1} & {$< 6$}\phantom{10} & {$< 6$}\phantom{1} & {$< 7$}\phantom{1}\\
{$-12$} & {$120\pm20$}  & {$55\pm4$} & {$< 10$}\phantom{1} & {$18\pm1$} & {$11\pm5$}\phantom{1} & {$50\pm10$} & {$< 6$}\phantom{10}  & {$113\pm5$}\phantom{1}  & {$< 5$}\phantom{10} & {$< 8$}\phantom{10} & \phantom{1}{$60\pm10$} & {$63\pm2$} & {$188\pm5$}\phantom{1} & {$10\pm3$}\phantom{1} & {$< 8$}\phantom{10} & {$25\pm3$}\phantom{1} & {$< 10$}\\
{$0$} & {$510\pm70$}  & {$390\pm10$} & {$< 60$}\phantom{1} & {$129\pm3$}\phantom{1}  &  {$70\pm20$} & {$218\pm9$}\phantom{10}  & {$96\pm6$}\phantom{1}  & {$260\pm20$} &  {$< 30$}\phantom{1} & {$< 20$}\phantom{1} & {$320\pm30$} & {$248\pm3$}\phantom{1} & {$744\pm8$}\phantom{1} & {$90\pm10$} & {$< 10$}\phantom{1} & {$60\pm10$} & {$< 20$}\\
{$8$} & {$260\pm30$}  & {$169\pm8$}\phantom{1}  & {$< 20$}\phantom{1} & {$56\pm3$} & {$19\pm8$}\phantom{1} & {$90\pm20$} & {$< 20$}\phantom{1} &  \phantom{10}{$< 100$} &  {$< 100$} & {$< 10$}\phantom{1} & {$130\pm30$} & {$113\pm4$}\phantom{1} & {$340\pm10$} & {$< 20$} & {$< 20$}\phantom{1} & {$< 20$} & {$< 20$}\\
{$16$} & {$62\pm3$}  & {$< 7$} & {$< 7$}\phantom{10}  & {$< 2$} & {$< 7$} & {$< 6$} & {$< 6$}\phantom{10}  & {$< 4$} & {$< 4$}\phantom{10} & {$< 4$}\phantom{10} & {$19\pm5$} & {$19.5\pm0.9$} & {$58\pm3$} & {$< 7$}\phantom{1} & {$< 7$}\phantom{10} & {$< 7$}\phantom{1} & {$< 9$}\phantom{1}\\
{$24$} & {$40\pm2$}  & {$< 4$} & {$< 4$}\phantom{10}  & {$< 1$} & {$< 7$} & {$< 6$} & {$< 5$}\phantom{10}  & {$< 4$} & {$< 4$}\phantom{10} & {$< 4$}\phantom{10} & \phantom{1}{$< 7$} & \phantom{1}{$9\pm1$} & {$29\pm3$} & {$< 6$}\phantom{1} & {$< 6$}\phantom{10} & {$< 6$}\phantom{1} & {$< 9$}\phantom{1}\\
{$33$} & {$41\pm1$}  & {$< 6$} & {$< 6$}\phantom{10}  & {$< 2$} & {$< 3$} & {$< 5$} & {$< 4$}\phantom{10}  & {$< 4$} & {$< 4$}\phantom{10}  & {$< 6$}\phantom{10} & \phantom{1}{$5\pm3$} & \phantom{1}{$5.4\pm0.5$} & {$16\pm2$} & {$< 5$}\phantom{1} & {$< 5$}\phantom{10} & {$< 5$}\phantom{1} & {$< 5$}\phantom{1}\\
{$41$} & {$35\pm1$}  & {$< 5$} & {$< 5$}\phantom{10}  & {$< 2$} & {$< 7$} & {$< 5$} & {$< 5$}\phantom{10}  & {$< 4$} & {$< 4$}\phantom{10} & {$< 5$}\phantom{10} & \phantom{1}{$7\pm3$} & \phantom{1}{$5.1\pm0.5$} & {$15\pm2$} & {$< 5$}\phantom{1} & {$< 5$}\phantom{10} & {$< 5$}\phantom{1} & {$< 6$}\phantom{1}\\
{$49$} & {$31\pm1$}  & {$< 5$} & {$< 5$}\phantom{10} & {$< 2$} & {$< 5$} & {$< 5$} & {$< 5$}\phantom{10}  & {$< 5$} & {$< 5$}\phantom{10} & {$< 3$}\phantom{10} & \phantom{1}{$8\pm1$} & \phantom{1}{$4.6\pm0.5$} & {$14\pm1$} & {$< 4$}\phantom{1} & {$< 4$}\phantom{10} & {$< 4$}\phantom{1} & {$< 5$}\phantom{1}\\
{$57$} & {$29\pm1$}  & {$< 5$} & {$< 5$}\phantom{10}  & {$< 2$} & {$< 5$} & {$< 5$} & {$< 6$}\phantom{10}  & \phantom{1}{$7\pm1$} & {$< 5$}\phantom{10} & {$< 4$}\phantom{10} & {$12\pm2$} & \phantom{1}{$4.4\pm0.4$} & {$13\pm1$} & {$< 5$}\phantom{1} & {$< 5$}\phantom{10} & {$< 5$}\phantom{1} & {$< 6$}\phantom{1}\\
{$65$} & {$24\pm2$}  & {$< 6$} & {$< 6$}\phantom{10}  & {$< 2$} & {$< 6$} & {$< 5$} & {$< 5$}\phantom{10}  & {$< 6$} & {$< 6$}\phantom{10} & {$< 2$}\phantom{10} & {$12\pm4$} & \phantom{1}{$4.1\pm0.4$} & {$12\pm1$} & {$< 5$}\phantom{1} & {$< 5$}\phantom{10} & {$< 5$}\phantom{1} & {$< 5$}\phantom{1}\\
{$73$} & {$21\pm1$}  & {$< 4$} & {$< 4$}\phantom{10}  & {$< 1$} & {$< 4$} & {$< 4$} & {$< 4$}\phantom{10}  & {$< 6$} & {$< 6$}\phantom{10} & {$< 4$}\phantom{10} & \phantom{1}{$8\pm2$} & \phantom{1}{$2.1\pm0.3$} & \phantom{1}{$6.4\pm0.9$} & {$< 3$}\phantom{1} & {$< 3$}\phantom{10} & {$< 3$}\phantom{1} & {$< 4$}\phantom{1}\\
{$81$} & {$16\pm1$}  & {$< 4$} & {$< 4$}\phantom{10}  & {$< 1$} & {$< 3$} & {$< 4$} & {$< 4$}\phantom{10}  & {$< 4$} & {$< 4$}\phantom{10} & {$< 4$}\phantom{10} & \phantom{1}{$6\pm2$} & \phantom{1}{$1.6\pm0.3$} & \phantom{1}{$5\pm1$} & {$< 3$}\phantom{1} & {$< 3$}\phantom{10} & {$< 3$}\phantom{1} & {$< 6$}\phantom{1}\\
{Unres} & {$981\pm90$}  & \phantom{1}{$800\pm200$} & {$600\pm100$}\phantom{1} & {$270\pm60$} & {$230\pm90$}\phantom{1}  & {$690\pm90$} & {$< 40$}\phantom{1} & {$971\pm50$} &  {$< 40$}\phantom{1} & {$< 60$}\phantom{1} & {$430\pm20$} & {$680\pm10$} & {$2050\pm40$}\phantom{1} & {$120\pm50$}\phantom{1} & {$< 70$}\phantom{1} & {$200\pm200$} & {$< 90$}\\
{Integ} & {$7100\pm200$}  & {$2700\pm300$}  &  {$< 700$} & {$910\pm90$} & {$400\pm200$} & {$1200\pm200$} &  {$< 500$} & {$2100\pm300$} & {$< 500$} & {$< 500$} & {$3100\pm300$} & {$2200\pm100$} & {$6600\pm300$} & {$500\pm300$} & {$< 300$} & {$900\pm200$} & \phantom{1}{$< 300$}\\
\enddata
\tablenotetext{a}{Distance with respect to nuclear row in 2--D spectrum, while ``Unres'' and ``Integ'' refer to the unresolved nuclear source and integrated spectra, respectively.}
\tablenotetext{b}{Flux is presented in units of $10^{-17}$~erg~s\textsuperscript{$-1$}~cm\textsuperscript{$-2$}.}
\end{deluxetable}
\end{rotatetable}

%% file: ha_table_1.tex
\startlongtable
\begin{deluxetable*}{lccccccccc}
\tabletypesize{\scriptsize}
\setlength{\tabcolsep}{2pt}
\tablecaption{Emission-Line Measurements from the Red Spectra\label{table:ha_lines}}
\tablewidth{0pc}
\tablecolumns{10}
\tablehead{
\colhead{} & \colhead{} & \multicolumn{8}{c}{Emission-Line Flux\tablenotemark{b}}\vspace{-12pt}\\
\colhead{} & \colhead{Dist.\tablenotemark{a}} & \multicolumn{8}{c}{\hrulefill}\vspace{-8pt}\\
          \colhead{Galaxy} & \colhead{(pc)} & \colhead{[\ion{O}{1}]$\lambda$6300}& \colhead{[\ion{O}{1}]$\lambda$6363}& \colhead{[\ion{Fe}{10}]$\lambda$6374}& \colhead{[\ion{N}{2}]$\lambda$6548}& \colhead{H$\alpha$$\lambda$6563}& \colhead{[\ion{N}{2}]$\lambda$6583}& \colhead{[\ion{S}{2}]$\lambda$6716}& \colhead{[\ion{S}{2}]$\lambda$6731}}
\startdata
{NGC 1052} & {$-91$} & {$6.4\pm0.9$} & {$2.1\pm0.3$} & {$< 10$ }\phantom{10} & {$1.8\pm0.2$} & {$8\pm1$} & {$5.4\pm0.7$} & {$5.3\pm0.6$} & {$3.5\pm0.6$}\\
{  } & {$-82$} & {$< 9$} & {$< 3$} & {$< 9$\,\,\,\,}\phantom{10} & {$3.5\pm0.2$} & {$12.9\pm0.9$}\phantom{1} & {$10.5\pm0.7$}\phantom{1} & {$5.4\pm0.6$} & {$5.1\pm0.6$}\\
{  } & {$-73$} & \phantom{1}{$< 10$} & {$< 5$} & {$< 10$ }\phantom{10} & {$3.6\pm0.3$} & {$12\pm1$}\phantom{1} & {$11\pm1$}\phantom{1} & {$6.5\pm0.8$} & {$3.0\pm0.6$}\\
{  } & {$-65$} & {$2.9\pm0.6$} & {$1.0\pm0.2$} & {$< 7$\,\,\,\,}\phantom{10} & {$3.7\pm0.4$} & {$15\pm2$}\phantom{1} & {$11\pm1$}\phantom{1} & {$7\pm1$} & {$3.8\pm0.7$}\\
{  } & {$-56$} & {$5\pm1$} & {$1.6\pm0.3$} & {$< 10$ }\phantom{10} & {$6.1\pm0.6$} & {$20\pm2$}\phantom{1} & {$18\pm2$}\phantom{1} & {$7.7\pm0.8$} & {$6.3\pm0.8$}\\
{  } & {$-47$} & \phantom{1}{$< 10$} & {$< 4$} & {$< 20$ }\phantom{10} & {$6.5\pm0.4$} & {$18\pm2$}\phantom{1} & {$20\pm1$}\phantom{1} & {$14\pm1$}\phantom{1} & {$9\pm1$}\\
{  } & {$-39$} & {$11\pm1$}\phantom{1} & {$3.7\pm0.4$} & {$< 10$ }\phantom{10} & {$9.7\pm0.6$} & {$29\pm2$}\phantom{1} & {$29\pm2$}\phantom{1} & {$20\pm1$}\phantom{1} & {$16\pm1$}\phantom{1}\\
{  } & {$-30$} & {$18\pm2$}\phantom{1} & {$5.9\pm0.6$} & {$< 10$ }\phantom{10} & {$16\pm2$}\phantom{1} & {$57\pm3$}\phantom{1} & {$47\pm6$}\phantom{1} & {$32\pm1$}\phantom{1} & {$29\pm1$}\phantom{1}\\
{  } & {$-22$} & {$31\pm2$}\phantom{1} & {$10.2\pm0.5$}\phantom{1} & {$< 20$ }\phantom{10} & {$34\pm8$}\phantom{1} & {$119\pm5$ }\phantom{1} & {$100\pm20$}\phantom{1} & {$70\pm2$}\phantom{1} & {$61\pm2$}\phantom{1}\\
{  } & {$-13$} & {$230\pm20$}\phantom{1} & {$78\pm8$}\phantom{1} & {$< 20$ }\phantom{10} & {$76\pm3$}\phantom{1} & {$600\pm30$}\phantom{1} & {$230\pm10$}\phantom{1} & {$190\pm50$}\phantom{1} & {$170\pm30$}\phantom{1}\\
{  } & {$0$} & {$560\pm40$}\phantom{1} & {$190\pm10$}\phantom{1} & {$< 40$ }\phantom{10} & {$200\pm60$}\phantom{1} & {$1460\pm80$ }\phantom{1} & {$600\pm200$} & {$180\pm10$}\phantom{1} & {$290\pm20$}\phantom{1}\\
{  } & {$9$} & {$149\pm5$ }\phantom{1} & {$50\pm2$}\phantom{1} & {$< 20$ }\phantom{10} & {$113\pm4$ }\phantom{1} & {$430\pm30$}\phantom{1} & {$340\pm10$}\phantom{1} & {$160\pm20$}\phantom{1} & {$210\pm40$}\phantom{1}\\
{  } & {$17$} & {$56\pm2$}\phantom{1} & {$18.6\pm0.8$}\phantom{1} & {$< 10$ }\phantom{10} & {$51\pm2$}\phantom{1} & {$128\pm4$ }\phantom{1} & {$152\pm5$ }\phantom{1} & {$116\pm5$ }\phantom{2} & {$109\pm5$ }\phantom{2}\\
{  } & {$26$} & {$24\pm1$}\phantom{1} & {$7.9\pm0.4$} & {$< 20$ }\phantom{10} & {$25.7\pm0.6$}\phantom{1} & {$69\pm2$}\phantom{1} & {$77\pm2$}\phantom{1} & {$55\pm2$}\phantom{1} & {$52\pm1$}\phantom{1}\\
{  } & {$34$} & {$10.1\pm0.9$}\phantom{1} & {$3.4\pm0.3$} & {$< 10$ }\phantom{10} & {$10.6\pm0.3$}\phantom{1} & {$31\pm1$}\phantom{1} & {$31.9\pm0.9$}\phantom{1} & {$22.3\pm0.8$}\phantom{1} & {$19.1\pm0.8$}\phantom{1}\\
{  } & {$43$} & {$8\pm1$} & {$2.6\pm0.3$} & {$< 10$ }\phantom{10} & {$5.1\pm0.4$} & {$18\pm1$}\phantom{1} & {$15\pm1$}\phantom{1} & {$14\pm1$}\phantom{1} & {$11\pm1$}\phantom{1}\\
{  } & {$52$} & {$6\pm1$} & {$2.1\pm0.5$} & {$< 20$ }\phantom{10} & {$4.9\pm0.4$} & {$14\pm2$}\phantom{1} & {$15\pm1$}\phantom{1} & {$10\pm1$}\phantom{1} & {$7\pm1$}\\
{  } & {$60$} & {$2\pm1$} & {$0.7\pm0.4$} & {$< 10$ }\phantom{10} & {$4.0\pm0.6$} & {$13\pm3$}\phantom{1} & {$12\pm2$}\phantom{1} & {$7\pm2$} & {$6\pm2$}\\
{  } & {$69$} & {$3.0\pm0.8$} & {$1.0\pm0.3$} & {$< 10$ }\phantom{10} & {$3.5\pm0.3$} & {$12\pm1$}\phantom{1} & {$10.4\pm0.9$}\phantom{1} & {$7.5\pm0.8$} & {$5.2\pm0.7$}\\
{  } & {$78$} & \phantom{1}{$< 20$} & {$< 5$} & {$< 20$ }\phantom{10} & {$3.3\pm0.2$} & {$11.1\pm0.8$}\phantom{1} & {$9.8\pm0.6$} & {$8.0\pm0.5$} & {$7.2\pm0.6$}\\
{  } & {$86$} & \phantom{1}{$< 10$} & {$< 3$} & {$< 10$ }\phantom{10} & {$3.2\pm0.3$} & {$7.6\pm0.6$} & {$9.7\pm0.8$} & {$7.8\pm0.8$} & {$4.7\pm0.7$}\\
{  } & {Unres} & {$2700\pm400$}\phantom{1} & {$900\pm100$} & {$< 100$}\phantom{1} & {$400\pm100$} & {$3500\pm500$}\phantom{1} & {$1200\pm300$}\phantom{1} & {$500\pm400$} & {$2200\pm800$}\phantom{1}\\
{  } & {Integ} & {$6000\pm1000$} & {$2100\pm400$}\phantom{1} & {$< 1000$} & {$4600\pm400$}\phantom{1} & {$16000\pm2000$}\phantom{1} & {$14000\pm1000$}\phantom{1} & {$7000\pm1000$} & {$7000\pm1000$}\\
\\
{NGC 4278} & {$-72$} & {$< 9$} & {$< 3$} & {$< 9$\,\,\,\,}\phantom{10} & {$3.5\pm0.3$} & {$7.6\pm0.8$} & {$10.4\pm0.8$}\phantom{1} & {$6.9\pm0.8$} & {$6.1\pm0.7$}\\
{  } & {$-65$} & {$3\pm2$} & {$0.8\pm0.6$} & {$< 10$ }\phantom{10}  & {$4.8\pm0.8$} & {$9\pm2$} & {$14\pm2$}\phantom{1} & {$8\pm2$} & {$9\pm2$}\\
{  } & {$-58$} & {$5\pm1$} & {$1.7\pm0.3$} & {$< 10$ }\phantom{10}  & {$8.3\pm0.4$} & {$21\pm1$}\phantom{1} & {$25\pm1$}\phantom{1} & {$16\pm1$}\phantom{1} & {$13.0\pm0.9$}\phantom{1}\\
{  } & {$-51$} & {$13\pm1$}\phantom{1} & {$4.3\pm0.3$} & {$< 10$ }\phantom{10}  & {$8.8\pm0.3$} & {$26.9\pm0.8$}\phantom{1} & {$26\pm1$}\phantom{1} & {$23\pm1$}\phantom{1} & {$18\pm1$}\phantom{1}\\
{  } & {$-43$} & {$13.4\pm0.8$\phantom{1}} & {$4.5\pm0.3$} & {$< 10$ }\phantom{10}  & {$14.7\pm0.3$}\phantom{1} & {$37.6\pm0.9$}\phantom{1} & {$44.1\pm0.9$}\phantom{1} & {$29.7\pm0.8$}\phantom{1} & {$23.3\pm0.7$}\phantom{1}\\
{  } & {$-36$} & {$15.3\pm0.9$}\phantom{1} & {$5.1\pm0.3$} & {$< 9$\,\,\,\,}\phantom{10} & {$17\pm2$}\phantom{1} & {$41\pm1$}\phantom{1} & {$51\pm7$}\phantom{1} & {$33\pm1$}\phantom{1} & {$28.4\pm0.9$}\phantom{1}\\
{  } & {$-29$} & {$12\pm2$}\phantom{1} & {$4.0\pm0.5$} & {$< 20$ }\phantom{10} & {$12\pm2$}\phantom{1} & {$37\pm2$}\phantom{1} & {$35\pm7$}\phantom{1} & {$26\pm2$}\phantom{1} & {$23\pm2$}\phantom{1}\\
{  } & {$-22$} & {$10.0\pm0.7$}\phantom{1} & {$3.3\pm0.2$} & {$< 10$ }\phantom{10} & {$12.5\pm0.3$}\phantom{1} & {$35\pm1$}\phantom{1} & {$37\pm1$}\phantom{1} & {$19.8\pm0.7$}\phantom{1} & {$18.0\pm0.7$}\phantom{1}\\
{  } & {$-15$} & {$12\pm1$}\phantom{1} & {$3.9\pm0.4$} & {$< 10$ }\phantom{10} & {$17\pm2$}\phantom{1} & {$47\pm2$}\phantom{1} & {$50\pm6$}\phantom{1} & {$22\pm1$}\phantom{1} & {$24\pm1$}\phantom{1}\\
{  } & {$-7$} & {$25\pm2$}\phantom{1} & {$8.2\pm0.6$} & {$< 10$ }\phantom{10} & {$24\pm6$}\phantom{1} & {$41\pm2$}\phantom{1} & {$70\pm20$} & {$36\pm2$}\phantom{1} & {$40\pm2$}\phantom{1}\\
{  } & {$0$} & {$40\pm10$} & {$13\pm4$}\phantom{1} & {$< 20$ }\phantom{10} & {$70\pm30$} & {$400\pm100$} & {$220\pm90$}\phantom{1} & {$70\pm40$} & {$90\pm40$}\\
{  } & {$11$} & {$32\pm6$}\phantom{1} & {$11\pm2$}\phantom{1} & {$< 20$ }\phantom{10} & {$12\pm2$}\phantom{1} & {$270\pm10$}\phantom{1} & {$35\pm5$}\phantom{1} & {$32\pm3$}\phantom{1} & {$30\pm3$}\phantom{1}\\
{  } & {$18$} & {$18\pm1$}\phantom{1} & {$6.1\pm0.4$} & {$< 10$ }\phantom{10} & {$21\pm1$}\phantom{1} & {$50\pm2$}\phantom{1} & {$62\pm3$}\phantom{1} & {$29\pm5$}\phantom{1} & {$31\pm4$}\phantom{1}\\
{  } & {$25$} & {$26\pm1$}\phantom{1} & {$8.7\pm0.5$} & {$< 10$ }\phantom{10} & {$26\pm1$}\phantom{1} & {$81\pm9$}\phantom{1} & {$78\pm4$}\phantom{1} & {$49\pm3$}\phantom{1} & {$44\pm2$}\phantom{1}\\
{  } & {$33$} & {$19.7\pm0.8$}\phantom{1} & {$6.6\pm0.3$} & {$< 10$ }\phantom{10} & {$20.1\pm0.4$}\phantom{1} & {$50\pm1$}\phantom{1} & {$60\pm1$}\phantom{1} & {$34.6\pm0.9$}\phantom{1} & {$32.4\pm0.8$}\phantom{1}\\
{  } & {$40$} & {$11.1\pm0.8$}\phantom{1} & {$3.7\pm0.3$} & {$< 30$}\phantom{10} & {$8.9\pm0.3$} & {$25\pm1$}\phantom{1} & {$26.6\pm0.9$}\phantom{1} & {$17.3\pm0.8$}\phantom{1} & {$16.4\pm0.8$}\phantom{1}\\
{  } & {$47$} & {$4\pm2$} & {$1.3\pm0.8$} & {$< 10$ }\phantom{10} & {$6\pm4$} & {$15\pm4$}\phantom{1} & {$20\pm10$} & {$14\pm4$\phantom{1}} & {$11\pm4$}\phantom{1}\\
{  } & {$54$} & {$2\pm1$} & {$0.5\pm0.4$} & {$< 10$ }\phantom{10} & {$5\pm1$} & {$14\pm3$}\phantom{1} & {$14\pm4$}\phantom{1} & {$9\pm3$} & {$9\pm3$}\\
{  } & {$61$} & {$3.7\pm0.7$} & {$1.2\pm0.2$} & {$< 30$ }\phantom{10} & {$4.5\pm0.4$} & {$12\pm1$}\phantom{1} & {$13\pm1$}\phantom{1} & {$8.0\pm0.8$} & {$6.8\pm0.8$}\\
{  } & {$69$} & {$3\pm4$} & {$1\pm1$} & {$< 10$ }\phantom{10} & {$3\pm1$} & {$9\pm4$} & {$10\pm4$}\phantom{1} & {$7\pm4$} & {$6\pm4$}\\
{  } & {$76$} & \phantom{1}{$< 20$} & {$< 8$} & {$< 20$ }\phantom{10} & {$1.7\pm0.5$} & {$9\pm2$} & {$5\pm2$} & {$5\pm2$} & {$4\pm2$}\\
{  } & {Integ} & {$2500\pm200$}\phantom{2} & {$820\pm70$}\phantom{1} & {$< 3000$} & {$3400\pm100$}\phantom{1} & {$8900\pm300$}\phantom{1} & {$10100\pm300$ }\phantom{1} & {$5600\pm200$}\phantom{1} & {$5100\pm200$}\phantom{1}\\
\\
{NGC 4579} & {$-85$} & \phantom{1}{$< 10$} & {$< 4$} & {$< 10$ }\phantom{10} & {$2.4\pm0.5$} & {$4\pm2$} & {$7\pm1$} & {$< 7$} & {$< 7$}\\
{  } & {$-77$} & {$< 9$} & {$< 3$} & {$< 9$\,\,\,\,}\phantom{10} & \phantom{1}{$3\pm20$}  & \phantom{1}{$4\pm14$} & {$10\pm50$} & \phantom{1}{$< 14$} & \phantom{1}{$< 14$}\\
{  } & {$-69$} & {$< 9$} & {$< 3$} & {$< 9$\,\,\,\,}\phantom{10} & {$4.0\pm0.7$} & {$4\pm2$} & {$12\pm2$}\phantom{1} & {$< 6$} & {$< 6$}\\
{  } & {$-61$} & \phantom{1}{$< 10$} & {$< 3$} & {$< 10$ }\phantom{10} & {$4.7\pm0.3$} & {$11\pm1$}\phantom{1} & {$14\pm1$}\phantom{1} & {$3.8\pm0.6$} & {$3.4\pm0.6$}\\
{  } & {$-53$} & {$< 9$} & {$< 3$} & {$< 9$\,\,\,\,}\phantom{10} & {$6.0\pm0.4$} & {$14\pm1$}\phantom{1} & {$18\pm1$}\phantom{1} & {$9.9\pm0.9$} & {$5.5\pm0.8$}\\
{  } & {$-45$} & \phantom{1}{$< 20$} & {$< 5$} & {$< 9$\,\,\,\,}\phantom{10} & {$8.3\pm0.4$} & {$13.6\pm0.9$}\phantom{1} & {$25\pm1$}\phantom{1} & {$4.9\pm0.7$} & {$5.1\pm0.7$}\\
{  } & {$-37$} & \phantom{1}{$< 10$} & {$< 4$} & {$< 10$ }\phantom{10} & {$8.8\pm0.5$} & {$6\pm2$} & {$26\pm1$}\phantom{1} & {$12\pm1$}\phantom{1} & {$7.6\pm0.9$}\\
{  } & {$-28$} & {$6\pm1$} & {$2.2\pm0.4$} & {$< 10$ }\phantom{10} & {$11.6\pm0.5$}\phantom{1} & {$29\pm2$}\phantom{1} & {$35\pm2$}\phantom{1} & {$13.1\pm0.8$}\phantom{1} & {$12.4\pm0.9$}\phantom{1}\\
{  } & {$-20$} & {$14\pm3$}\phantom{1} & {$4.7\pm0.9$} & {$< 20$ }\phantom{10} & {$18\pm2$}\phantom{1} & {$60\pm10$} & {$54\pm7$}\phantom{1} & {$29\pm1$}\phantom{1} & {$31\pm1$}\phantom{1}\\
{  } & {$-12$} & {$39\pm2$}\phantom{1} & {$12.9\pm0.6$}\phantom{1} & {$< 10$ }\phantom{10} & {$39\pm1$}\phantom{1} & {$< 20$} & {$118\pm4$ }\phantom{1} & {$52\pm2$}\phantom{1} & {$47\pm2$}\phantom{1}\\
{  } & {$0$\tablenotemark{c}} & {...} & {...} & {...} & {...} & {...} & {...} & {...} & {...}\\
{  } & {$8$\tablenotemark{c}} & {...} & {...} &  {...} &  {...} & {...} & {...} & {...} & {...}\\
{  } & {$16$} & {$16\pm7$}\phantom{1} & {$5\pm2$} & {$< 20$ }\phantom{10} & {$11\pm2$}\phantom{1} & {$< 10$} & {$34\pm6$}\phantom{1} & {$14\pm1$}\phantom{1} & {$25\pm2$}\phantom{1}\\
{  } & {$24$} & \phantom{1}{$< 10$} & {$< 4$} & {$< 10$ }\phantom{10} & {$8\pm4$} & {$22\pm4$}\phantom{1} & {$20\pm10$} & {$12\pm3$}\phantom{1} & {$17\pm4$}\phantom{1}\\
{  } & {$33$} & {$14\pm2$}\phantom{1} & {$4.6\pm0.5$} & {$< 10$ }\phantom{10} & {$9.6\pm0.6$} & {$25\pm4$}\phantom{1} & {$29\pm2$}\phantom{1} & {$16.5\pm0.9$}\phantom{1} & {$19\pm1$}\phantom{1}\\
{  } & {$41$} & {$< 9$} & {$< 3$} & {$< 9$\,\,\,\,}\phantom{10} & {$16.6\pm0.3$}\phantom{1} & {$27\pm1$}\phantom{1} & {$49.7\pm0.9$}\phantom{1} & {$18.1\pm0.7$}\phantom{1} & {$16.7\pm0.7$}\phantom{1}\\
{  } & {$49$} & \phantom{1}{$< 10$} & {$< 4$} & {$< 10$ }\phantom{10} & {$15.9\pm0.4$}\phantom{1} & {$28\pm1$}\phantom{1} & {$48\pm1$}\phantom{1} & {$19\pm1$}\phantom{1} & {$14\pm1$}\phantom{1}\\
{  } & {$57$} & {$10\pm1$}\phantom{1} & {$3.4\pm0.4$} & {$< 10$ }\phantom{10} & {$15.2\pm0.4$}\phantom{1} & {$35\pm1$}\phantom{1} & {$46\pm1$}\phantom{1} & {$18\pm1$}\phantom{1} & {$14.9\pm0.9$}\phantom{1}\\
{  } & {$65$} & \phantom{1}{$< 10$} & {$< 4$} & {$< 10$ }\phantom{10} & {$11.7\pm0.3$}\phantom{1} & {$37\pm1$}\phantom{1} & {$35\pm1$}\phantom{1} & {$16\pm1$}\phantom{1} & {$16\pm1$}\phantom{1}\\
{  } & {$73$} & \phantom{1}{$< 10$} & {$< 4$} & {$< 10$ }\phantom{10} & {$9.2\pm0.3$} & {$24.1\pm0.9$}\phantom{1} & {$27.7\pm0.8$}\phantom{1} & {$9.8\pm0.8$} & {$8.8\pm0.7$}\\
{  } & {$81$} & \phantom{1}{$< 10$} & {$< 3$} & {$< 10$ }\phantom{10} & {$7.0\pm0.4$} & {$18\pm1$}\phantom{1} & {$21\pm1$}\phantom{1} & {$9\pm1$} & {$8\pm1$}\\
{  } & {Unres} & {$1100\pm30$}\phantom{2} & {$370\pm10$}\phantom{1} & {$< 100$ }\phantom{1} & {$1100\pm70$ } \phantom{1}& {$3500\pm200$}\phantom{1} & {$3300\pm200$}\phantom{1} & {$540\pm20$}\phantom{1} & {$530\pm20$}\phantom{1}\\
{  } & {Integ} & \phantom{1}{$3000\pm1000$} & {$1100\pm400$}\phantom{1} & {$< 1000$} & {$3300\pm700$}\phantom{1} & {$4000\pm2000$} & {$9900\pm2000$} & {$4100\pm900$}\phantom{1} & {$4100\pm900$}\phantom{1}\\
\enddata
\tablenotetext{a}{Distance with respect to nuclear row in 2--D spectrum, while ``Unres'' and ``Integ'' refer to the unresolved nuclear source and integrated spectra, respectively.}
\tablenotetext{b}{Flux is presented in units of $10^{-17}$~erg~s\textsuperscript{$-1$}~cm\textsuperscript{$-2$}.}
\tablenotetext{c}{We were not able to measure the line strengths in the resolved emission near the nucleus due to the overwhelming strength of the unresolved nuclear source. See Section~\ref{ssec:2dresfit} for details.}
\end{deluxetable*}

%% file: n1052_ext_corr_v1.tex
\begin{deluxetable}{cccc}
  \tablecaption{NGC~1052 $A_{\textrm{V}}$ values and Intrinsic H$\alpha$/H$\beta$ ratios\label{table: 1052_ext_corr}}
\tabletypesize{\scriptsize}
\tablewidth{0pt}
\tablecolumns{4}
\tablehead{
\colhead{Distance (pc)\tablenotemark{a}} & \colhead{$A_{\textrm{V}}$} & {[H$\alpha$/H$\beta]_{int}$\tablenotemark{b}} & {Mechanism}}
\startdata
{$-91$}        & {0.000} & {3.01} & {Shock}\\
{$-82$}        & {0.000} & {3.01} & {Shock}\\
{$-73$}        & {0.000} & {3.01} & {Shock}\\
{$-65$}        & {0.000} & {3.01} & {Shock}\\
{$-56$}        & {0.000} & {3.01} & {Shock}\\
{$-47$}        & {0.122} & {3.01} & {Shock}\\
{$-39$}        & {0.000} & {3.01} & {Shock}\\
{$-30$}        & {0.479} & {3.01} & {Shock}\\
{$-22$}        & {1.700} & {3.01} & {Shock}\\
{$-13$}        & {2.253} & {3.10} & {AGN--NLR}\\
{~~~0}         & {4.025} & {3.10} & {AGN--NLR}\\
{$+09$}        & {3.034} & {3.10} & {AGN--NLR}\\
{$+17$}        & {0.177} & {3.01} & {Shock}\\
{$+26$}        & {0.958} & {3.01} & {Shock}\\
{$+35$}        & {0.887} & {3.01} & {Shock}\\
{$+43$}        & {0.000} & {3.01} & {Shock}\\
{$+52$}        & {0.000} & {3.01} & {Shock}\\
{$+60$}        & {0.190} & {3.01} & {Shock}\\
{$+69$}        & {0.000} & {3.01} & {Shock}\\
{$+78$}        & {0.000} & {3.01} & {Shock}\\
{$+86$}        & {0.000} & {3.01} & {Shock}\\
{Unresolved} & {0.575} & {3.10} & {AGN--NLR}
\enddata
\tablenotetext{a}{Distance with respect to nuclear row in 2--D spectrum.}
\tablenotetext{b}{The adopted intrinsic value of the H$\alpha$/H$\beta$ ratio. The value for an AGN--NLR is from \citet[][chapter 11]{Osterbrock06}. The value for shocks is an average value from the \citet{Allen2008} shock models.}
\end{deluxetable}

%% file: n4278_ext_corr.tex
\begin{deluxetable}{rccc}
\tablecaption{NGC~4278 $A_{\textrm{V}}$ values and Intrinsic H$\alpha$/H$\beta$ ratios\label{table: 4278_ext_corr}}
\tabletypesize{\scriptsize}
\tablewidth{0pt}
\tablecolumns{4}
\tablehead{
\colhead{Distance (pc)\tablenotemark{a}} & \colhead{$A_{\textrm{V}}$} & \colhead{[H$\alpha$/H$\beta]_{int}$\tablenotemark{b}} & \colhead{Mechanism}}
\startdata
{$-72$}        & {0.000} & {3.01} & {Shock}\\
{$-65$}        & {0.310} & {3.01} & {Shock}\\
{$-58$}        & {0.159} & {3.01} & {Shock}\\
{$-51$}        & {0.209} & {3.01} & {Shock}\\
{$-43$}        & {0.359} & {3.01} & {Shock}\\
{$-36$}        & {1.463} & {3.01} & {Shock}\\
{$-29$}        & {0.012} & {3.01} & {Shock}\\
{$-22$}        & {0.378} & {3.01} & {Shock}\\
{$-15$}        & {1.192} & {3.01} & {Shock}\\
{$-07$}        & {1.111} & {3.01} & {Shock}\\
{00}         & {0.807} & {3.01} & {Shock}\\
{$+11$}        & {3.009} & {3.01} & {Shock}\\
{$+18$}        & {1.502} & {3.01} & {Shock}\\
{$+25$}        & {1.587} & {3.01} & {Shock}\\
{$+33$}        & {1.496} & {3.01} & {Shock}\\
{$+40$}        & {0.000} & {3.01} & {Shock}\\
{$+47$}        & {0.329} & {3.01} & {Shock}\\
{$+54$}        & {0.000} & {3.01} & {Shock}\\
{$+61$}        & {0.000} & {3.01} & {Shock}\\
{$+69$}        & {0.000} & {3.01} & {Shock}\\
{$+76$}        & {0.000} & {3.01} & {Shock}
\enddata
\tablenotetext{a}{Distance with respect to nuclear row in 2--D spectrum.}
\tablenotetext{b}{The adopted intrinsic value of the H$\alpha$/H$\beta$ ratio. This is an average value from the \citet{Allen2008} shock models.}
\end{deluxetable}

%% file: n4579_ext_corr.tex
\begin{deluxetable}{rccc}
\tablecaption{NGC~4579 $A_{\textrm{V}}$ values and Intrinsic H$\alpha$/H$\beta$ ratios\label{table: 4579_ext_corr}}
\tabletypesize{\scriptsize}
\tablewidth{0pt}
\tablecolumns{4}
\tablehead{
\colhead{Distance (pc)\tablenotemark{a}} & \colhead{$A_{\textrm{V}}$} & \colhead{[H$\alpha$/H$\beta]_{int}$\tablenotemark{b}} & \colhead{Mechanism}}
\startdata
{$-85$}        & {0.000} & {2.86} & {\ion{H}{2}}\\
{$-77$}        & {0.000} & {2.86} & {\ion{H}{2}}\\
{$-69$}        & {0.000} & {2.86} & {\ion{H}{2}}\\
{$-61$}        & {0.000} & {2.86} & {\ion{H}{2}}\\
{$-53$}        & {0.000} & {2.86} & {\ion{H}{2}}\\
{$-45$}        & {0.000} & {2.86} & {\ion{H}{2}}\\
{$-37$}        & {0.000} & {2.86} & {\ion{H}{2}}\\
{$-28$}        & {0.000} & {2.86} & {\ion{H}{2}}\\
{$-20$}        & {0.993} & {2.86} & {\ion{H}{2}}\\
{$-12$}        & {0.000} & {3.10} & {AGN--NLR}\\
{00}         & {0.000} & {3.10} & {AGN--NLR}\\
{$+08$}        & {0.000} & {3.10} & {AGN--NLR}\\
{$+16$}        & {0.000} & {3.10} & {AGN--NLR}\\
{$+24$}        & {0.000} & {2.86} & {\ion{H}{2}}\\
{$+33$}        & {1.749} & {2.86} & {\ion{H}{2}}\\
{$+41$}        & {1.018} & {2.86} & {\ion{H}{2}}\\
{$+49$}        & {0.596} & {2.86} & {\ion{H}{2}}\\
{$+57$}        & {0.024} & {2.86} & {\ion{H}{2}}\\
{$+65$}        & {0.209} & {2.86} & {\ion{H}{2}}\\
{$+73$}        & {0.256} & {2.86} & {\ion{H}{2}}\\
{$+81$}        & {0.073} & {2.86} & {\ion{H}{2}}\\
{Unresolved} & {3.012} & {3.10} & {AGN--NLR}
\enddata
\tablenotetext{a}{Distance with respect to nuclear row in 2--D spectrum.}
\tablenotetext{b}{The adopted intrinsic value of the H$\alpha$/H$\beta$ ratio. The value for an AGN--NLR is from \citet[][chapter 11]{Osterbrock06}. The value for \ion{H}{2} is the theoretical value for Case B recombination \citep{osterbrock1989}.}
\end{deluxetable}

%% file: liner_20180504_arxiv.bbl
\begin{thebibliography}{112}
\expandafter\ifx\csname natexlab\endcsname\relax\def\natexlab#1{#1}\fi

\bibitem[{{Allen} {et~al.}(2008){Allen}, {Groves}, {Dopita}, {Sutherland}, \&
  {Kewley}}]{Allen2008}
{Allen}, M.~G., {Groves}, B.~A., {Dopita}, M.~A., {Sutherland}, R.~S., \&
  {Kewley}, L.~J. 2008, \apjs, 178, 20

\bibitem[{{Baldwin} {et~al.}(1981){Baldwin}, {Phillips}, \&
  {Terlevich}}]{BPT81}
{Baldwin}, J.~A., {Phillips}, M.~M., \& {Terlevich}, R. 1981, \pasp, 93, 5

\bibitem[{{Barth} {et~al.}(1999){Barth}, {Filippenko}, \& {Moran}}]{Barth1999}
{Barth}, A.~J., {Filippenko}, A.~V., \& {Moran}, E.~C. 1999, \apjl, 515, L61

\bibitem[{{Barth} {et~al.}(1996){Barth}, {Reichert}, {Filippenko}, {Ho},
  {Shields}, {Mushotzky}, \& {Puchnarewicz}}]{Barth1996}
{Barth}, A.~J., {Reichert}, G.~A., {Filippenko}, A.~V., {et~al.} 1996, \aj,
  112, 1829

\bibitem[{{Barth} \& {Shields}(2000)}]{Barth2000}
{Barth}, A.~J., \& {Shields}, J.~C. 2000, \pasp, 112, 753

\bibitem[{{Belfiore} {et~al.}(2016){Belfiore}, {Maiolino}, {Maraston},
  {Emsellem}, {Bershady}, {Masters}, {Yan}, {Bizyaev}, {Boquien}, {Brownstein},
  {Bundy}, {Drory}, {Heckman}, {Law}, {Roman-Lopes}, {Pan}, {Stanghellini},
  {Thomas}, {Weijmans}, \& {Westfall}}]{Belfiore2016}
{Belfiore}, F., {Maiolino}, R., {Maraston}, C., {et~al.} 2016, \mnras, 461,
  3111

\bibitem[{{Bertelli} {et~al.}(1994){Bertelli}, {Bressan}, {Chiosi}, {Fagotto},
  \& {Nasi}}]{padova94}
{Bertelli}, G., {Bressan}, A., {Chiosi}, C., {Fagotto}, F., \& {Nasi}, E. 1994,
  \aaps, 106

\bibitem[{{Binette} {et~al.}(1994){Binette}, {Magris}, {Stasi{\'n}ska}, \&
  {Bruzual}}]{Binette94}
{Binette}, L., {Magris}, C.~G., {Stasi{\'n}ska}, G., \& {Bruzual}, A.~G. 1994,
  \aap, 292, 13

\bibitem[{{Binette} {et~al.}(1996){Binette}, {Wilson}, \&
  {Storchi-Bergmann}}]{Binette1996}
{Binette}, L., {Wilson}, A.~S., \& {Storchi-Bergmann}, T. 1996, \aap, 312, 365

\bibitem[{{Blandford} \& {Begelman}(1999)}]{Blanford99}
{Blandford}, R.~D., \& {Begelman}, M.~C. 1999, \mnras, 303, L1

\bibitem[{{Boroson} \& {Green}(1992)}]{Boroson1992}
{Boroson}, T.~A., \& {Green}, R.~F. 1992, \apjs, 80, 109

\bibitem[{{Bowers}(1997)}]{bowers97}
{Bowers}, C.~D. 1997, in The 1997 HST Calibration Workshop, ed. S.~Casertano,
  Space Telescope Science Institute, electronic version at
  http://www.stsci.edu/hst/stis/documents/calworkshop/1997

\bibitem[{{Brenneman} {et~al.}(2009){Brenneman}, {Weaver}, {Kadler}, {Tueller},
  {Marscher}, {Ros}, {Zensus}, {Kovalev}, {Aller}, {Aller}, {Irwin}, {Kerp}, \&
  {Kaufmann}}]{Brenneman2009}
{Brenneman}, L.~W., {Weaver}, K.~A., {Kadler}, M., {et~al.} 2009, \apj, 698,
  528

\bibitem[{{Bruzual} \& {Charlot}(2003)}]{Bruzual2003}
{Bruzual}, G., \& {Charlot}, S. 2003, \mnras, 344, 1000

\bibitem[{{Bujarrabal} {et~al.}(1998){Bujarrabal}, {Alcolea}, {Sahai},
  {Zamorano}, \& {Zijlstra}}]{Bujarabal94}
{Bujarrabal}, V., {Alcolea}, J., {Sahai}, R., {Zamorano}, J., \& {Zijlstra},
  A.~A. 1998, \aap, 331, 361

\bibitem[{{Capetti} {et~al.}(1997){Capetti}, {Axon}, \&
  {Macchetto}}]{Capetti1997}
{Capetti}, A., {Axon}, D.~J., \& {Macchetto}, F.~D. 1997, \apj, 487, 560

\bibitem[{{Cappellari} \& {Emsellem}(2004)}]{Cappellari2004}
{Cappellari}, M., \& {Emsellem}, E. 2004, \pasp, 116, 138

\bibitem[{{Cardelli} {et~al.}(1989){Cardelli}, {Clayton}, \&
  {Mathis}}]{Cardelli89}
{Cardelli}, J.~A., {Clayton}, G.~C., \& {Mathis}, J.~S. 1989, \apj, 345, 245

\bibitem[{{Cecil} {et~al.}(1995){Cecil}, {Morse}, \& {Veilleux}}]{Cecil1995}
{Cecil}, G., {Morse}, J.~A., \& {Veilleux}, S. 1995, \apj, 452, 613

\bibitem[{{Cid Fernandes} {et~al.}(2004){Cid Fernandes}, {Gonz{\'a}lez
  Delgado}, {Schmitt}, {Storchi-Bergmann}, {Martins}, {P{\'e}rez}, {Heckman},
  {Leitherer}, \& {Schaerer}}]{Cid04}
{Cid Fernandes}, R., {Gonz{\'a}lez Delgado}, R.~M., {Schmitt}, H., {et~al.}
  2004, \apj, 605, 105

\bibitem[{{Claussen} {et~al.}(1998){Claussen}, {Diamond}, {Braatz}, {Wilson},
  \& {Henkel}}]{claussen1998}
{Claussen}, M.~J., {Diamond}, P.~J., {Braatz}, J.~A., {Wilson}, A.~S., \&
  {Henkel}, C. 1998, \apjl, 500, L129

\bibitem[{{Contini}(1997)}]{contini97}
{Contini}, M. 1997, \aap, 323, 71

\bibitem[{{Contini} \& {Viegas}(2001)}]{contini01}
{Contini}, M., \& {Viegas}, S.~M. 2001, \apjs, 132, 211

\bibitem[{{Dopita}(2002)}]{Dopita2002a}
{Dopita}, M.~A. 2002, in Revista Mexicana de Astronomia y Astrofisica, vol.~27,
  Vol.~13, Revista Mexicana de Astronomia y Astrofisica Conference Series, ed.
  W.~J. {Henney}, W.~{Steffen}, L.~{Binette}, \& A.~{Raga}, 177--182

\bibitem[{{Dopita} {et~al.}(2002){Dopita}, {Groves}, {Sutherland}, {Binette},
  \& {Cecil}}]{Dopita2002b}
{Dopita}, M.~A., {Groves}, B.~A., {Sutherland}, R.~S., {Binette}, L., \&
  {Cecil}, G. 2002, \apj, 572, 753

\bibitem[{{Dopita} {et~al.}(1997){Dopita}, {Koratkar}, {Allen}, {Tsvetanov},
  {Ford}, {Bicknell}, \& {Sutherland}}]{Dopita1997}
{Dopita}, M.~A., {Koratkar}, A.~P., {Allen}, M.~G., {et~al.} 1997, \apj, 490,
  202

\bibitem[{{Dopita} {et~al.}(1996){Dopita}, {Koratkar}, {Evans}, {Allen},
  {Bicknell}, {Sutherland}, {Hawley}, \& {Sadler}}]{Dopita1996}
{Dopita}, M.~A., {Koratkar}, A.~P., {Evans}, I.~N., {et~al.} 1996, in
  Astronomical Society of the Pacific Conference Series, Vol. 103, The Physics
  of Liners in View of Recent Observations, ed. M.~{Eracleous}, A.~{Koratkar},
  C.~{Leitherer}, \& L.~{Ho}, 44

\bibitem[{{Dopita} \& {Sutherland}(1995)}]{Dopita1995}
{Dopita}, M.~A., \& {Sutherland}, R.~S. 1995, \apj, 455, 468

\bibitem[{{Dopita} {et~al.}(2006){Dopita}, {Fischera}, {Sutherland}, {Kewley},
  {Leitherer}, {Tuffs}, {Popescu}, {van Breugel}, \& {Groves}}]{Dopita2006}
{Dopita}, M.~A., {Fischera}, J., {Sutherland}, R.~S., {et~al.} 2006, \apjs,
  167, 177

\bibitem[{{Dopita} {et~al.}(2015){Dopita}, {Ho}, {Dressel}, {Sutherland},
  {Kewley}, {Davies}, {Hampton}, {Shastri}, {Kharb}, {Jose}, {Bhatt}, {Ramya},
  {Scharw{\"a}chter}, {Jin}, {Banfield}, {Zaw}, {James}, {Juneau}, \&
  {Srivastava}}]{Dopita2015}
{Dopita}, M.~A., {Ho}, I.-T., {Dressel}, L.~L., {et~al.} 2015, \apj, 801, 42

\bibitem[{{Dudik} {et~al.}(2005){Dudik}, {Satyapal}, {Gliozzi}, \&
  {Sambruna}}]{Dudik2005}
{Dudik}, R.~P., {Satyapal}, S., {Gliozzi}, M., \& {Sambruna}, R.~M. 2005, \apj,
  620, 113

\bibitem[{{Dudik} {et~al.}(2009){Dudik}, {Satyapal}, \& {Marcu}}]{Dudik2009}
{Dudik}, R.~P., {Satyapal}, S., \& {Marcu}, D. 2009, \apj, 691, 1501

\bibitem[{{Eracleous} {et~al.}(2010{\natexlab{a}}){Eracleous}, {Hwang}, \&
  {Flohic}}]{Eracleous2010a}
{Eracleous}, M., {Hwang}, J.~A., \& {Flohic}, H.~M.~L.~G. 2010{\natexlab{a}},
  \apj, 711, 796

\bibitem[{{Eracleous} {et~al.}(2010{\natexlab{b}}){Eracleous}, {Hwang}, \&
  {Flohic}}]{Eracleous2010b}
---. 2010{\natexlab{b}}, \apjs, 187, 135

\bibitem[{{Eracleous} {et~al.}(2002){Eracleous}, {Shields}, {Chartas}, \&
  {Moran}}]{Eracleous2002}
{Eracleous}, M., {Shields}, J.~C., {Chartas}, G., \& {Moran}, E.~C. 2002, \apj,
  565, 108

\bibitem[{{Falcke} {et~al.}(2000){Falcke}, {Nagar}, {Wilson}, \&
  {Ulvestad}}]{Falcke2000}
{Falcke}, H., {Nagar}, N.~M., {Wilson}, A.~S., \& {Ulvestad}, J.~S. 2000, \apj,
  542, 197

\bibitem[{{Falcke} {et~al.}(1998){Falcke}, {Wilson}, \& {Simpson}}]{Falcke1998}
{Falcke}, H., {Wilson}, A.~S., \& {Simpson}, C. 1998, \apj, 502, 199

\bibitem[{{Falco} {et~al.}(1999){Falco}, {Kurtz}, {Geller}, {Huchra}, {Peters},
  {Berlind}, {Mink}, {Tokarz}, \& {Elwell}}]{UZC1999}
{Falco}, E.~E., {Kurtz}, M.~J., {Geller}, M.~J., {et~al.} 1999, \pasp, 111, 438

\bibitem[{{Ferland} \& {Netzer}(1983)}]{Ferland1983}
{Ferland}, G.~J., \& {Netzer}, H. 1983, \apj, 264, 105

\bibitem[{{Ferruit} {et~al.}(1999){Ferruit}, {Wilson}, {Whittle}, {Simpson},
  {Mulchaey}, \& {Ferland}}]{Ferruit1999}
{Ferruit}, P., {Wilson}, A.~S., {Whittle}, M., {et~al.} 1999, \apj, 523, 147

\bibitem[{{Filho} {et~al.}(2002){Filho}, {Barthel}, \& {Ho}}]{Filho02}
{Filho}, M.~E., {Barthel}, P.~D., \& {Ho}, L.~C. 2002, \aap, 385, 425

\bibitem[{{Filho} {et~al.}(2006){Filho}, {Barthel}, \& {Ho}}]{Filho06}
---. 2006, \aap, 451, 71

\bibitem[{{Filippenko}(1996)}]{Filippenko1996}
{Filippenko}, A.~V. 1996, in Astronomical Society of the Pacific Conference
  Series, Vol. 103, The Physics of Liners in View of Recent Observations, ed.
  M.~{Eracleous}, A.~{Koratkar}, C.~{Leitherer}, \& L.~{Ho}, 17

\bibitem[{{Filippenko} \& {Terlevich}(1992)}]{Filippenko1992}
{Filippenko}, A.~V., \& {Terlevich}, R. 1992, \apjl, 397, L79

\bibitem[{{Flohic} {et~al.}(2006){Flohic}, {Eracleous}, {Chartas}, {Shields},
  \& {Moran}}]{Flohic2006}
{Flohic}, H.~M.~L.~G., {Eracleous}, M., {Chartas}, G., {Shields}, J.~C., \&
  {Moran}, E.~C. 2006, \apj, 647, 140

\bibitem[{{Fosbury} {et~al.}(1978){Fosbury}, {Mebold}, {Goss}, \&
  {Dopita}}]{Fosbury78}
{Fosbury}, R.~A.~E., {Mebold}, U., {Goss}, W.~M., \& {Dopita}, M.~A. 1978,
  \mnras, 183, 549

\bibitem[{{Frank} \& {Mellema}(1994)}]{Frank94}
{Frank}, A., \& {Mellema}, G. 1994, \aap, 289, 937

\bibitem[{{Gabel} {et~al.}(2000){Gabel}, {Bruhweiler}, {Crenshaw}, {Kraemer},
  \& {Miskey}}]{Gabel2000}
{Gabel}, J.~R., {Bruhweiler}, F.~C., {Crenshaw}, D.~M., {Kraemer}, S.~B., \&
  {Miskey}, C.~L. 2000, \apj, 532, 883

\bibitem[{{Girardi} {et~al.}(2000){Girardi}, {Bressan}, {Bertelli}, \&
  {Chiosi}}]{Girardi00}
{Girardi}, L., {Bressan}, A., {Bertelli}, G., \& {Chiosi}, C. 2000, \aaps, 141,
  371

\bibitem[{{Giroletti} {et~al.}(2005){Giroletti}, {Taylor}, \&
  {Giovannini}}]{Giroletti2005}
{Giroletti}, M., {Taylor}, G.~B., \& {Giovannini}, G. 2005, \apj, 622, 178

\bibitem[{{Gonz{\'a}lez Delgado} {et~al.}(2004){Gonz{\'a}lez Delgado}, {Cid
  Fernandes}, {P{\'e}rez}, {Martins}, {Storchi-Bergmann}, {Schmitt}, {Heckman},
  \& {Leitherer}}]{Gonzalez04}
{Gonz{\'a}lez Delgado}, R.~M., {Cid Fernandes}, R., {P{\'e}rez}, E., {et~al.}
  2004, \apj, 605, 127

\bibitem[{{Gonzalez Delgado} \& {Perez}(1996)}]{Gonzalez1996}
{Gonzalez Delgado}, R.~M., \& {Perez}, E. 1996, \mnras, 281, 1105

\bibitem[{{Gonz{\'a}lez Delgado} {et~al.}(2008){Gonz{\'a}lez Delgado},
  {P{\'e}rez}, {Cid Fernandes}, \& {Schmitt}}]{Gonzalez08}
{Gonz{\'a}lez Delgado}, R.~M., {P{\'e}rez}, E., {Cid Fernandes}, R., \&
  {Schmitt}, H. 2008, \aj, 135, 747

\bibitem[{{Gonz{\'a}lez-Mart{\'{\i}}n}
  {et~al.}(2009){Gonz{\'a}lez-Mart{\'{\i}}n}, {Masegosa}, {M{\'a}rquez},
  {Guainazzi}, \& {Jim{\'e}nez-Bail{\'o}n}}]{Gonzalez2009}
{Gonz{\'a}lez-Mart{\'{\i}}n}, O., {Masegosa}, J., {M{\'a}rquez}, I.,
  {Guainazzi}, M., \& {Jim{\'e}nez-Bail{\'o}n}, E. 2009, \aap, 506, 1107

\bibitem[{{Graves} {et~al.}(2007){Graves}, {Faber}, {Schiavon}, \&
  {Yan}}]{Graves07}
{Graves}, G.~J., {Faber}, S.~M., {Schiavon}, R.~P., \& {Yan}, R. 2007, \apj,
  671, 243

\bibitem[{{Groves} {et~al.}(2004){Groves}, {Dopita}, \&
  {Sutherland}}]{Groves2004}
{Groves}, B.~A., {Dopita}, M.~A., \& {Sutherland}, R.~S. 2004, \apjs, 153, 9

\bibitem[{{Halpern} \& {Steiner}(1983)}]{Halpern1983}
{Halpern}, J.~P., \& {Steiner}, J.~E. 1983, \apjl, 269, L37

\bibitem[{{Heckman}(1980)}]{Heckman1980}
{Heckman}, T.~M. 1980, \aap, 87, 152

\bibitem[{{Helmboldt} {et~al.}(2007){Helmboldt}, {Taylor}, {Tremblay},
  {Fassnacht}, {Walker}, {Myers}, {Sjouwerman}, {Pearson}, {Readhead},
  {Weintraub}, {Gehrels}, {Romani}, {Healey}, {Michelson}, {Blandford}, \&
  {Cotter}}]{Helboldt07}
{Helmboldt}, J.~F., {Taylor}, G.~B., {Tremblay}, S., {et~al.} 2007, \apj, 658,
  203

\bibitem[{{Ho}(2008)}]{Ho2008}
{Ho}, L.~C. 2008, \araa, 46, 475

\bibitem[{{Ho} {et~al.}(1993){Ho}, {Filippenko}, \& {Sargent}}]{Ho1993}
{Ho}, L.~C., {Filippenko}, A.~V., \& {Sargent}, W.~L.~W. 1993, \apj, 417, 63

\bibitem[{{Ho} {et~al.}(1997{\natexlab{a}}){Ho}, {Filippenko}, \&
  {Sargent}}]{HoFS97III}
---. 1997{\natexlab{a}}, \apjs, 112, 315

\bibitem[{{Ho} {et~al.}(1997{\natexlab{b}}){Ho}, {Filippenko}, \&
  {Sargent}}]{HoFS97V}
---. 1997{\natexlab{b}}, \apj, 487, 568

\bibitem[{{Ho} {et~al.}(2003){Ho}, {Filippenko}, \& {Sargent}}]{HoFS03}
---. 2003, \apj, 583, 159

\bibitem[{{Jensen} {et~al.}(2003){Jensen}, {Tonry}, {Barris}, {Thompson},
  {Liu}, {Rieke}, {Ajhar}, \& {Blakeslee}}]{Jensen03}
{Jensen}, J.~B., {Tonry}, J.~L., {Barris}, B.~J., {et~al.} 2003, \apj, 583, 712

\bibitem[{{Jones} {et~al.}(1984){Jones}, {Wrobel}, \& {Shaffer}}]{jones84}
{Jones}, D.~L., {Wrobel}, J.~M., \& {Shaffer}, D.~B. 1984, \apj, 276, 480

\bibitem[{{Kauffmann} {et~al.}(2003){Kauffmann}, {Heckman}, {Tremonti},
  {Brinchmann}, {Charlot}, {White}, {Ridgway}, {Brinkmann}, {Fukugita}, {Hall},
  {Ivezi{\'c}}, {Richards}, \& {Schneider}}]{Kauffmann03}
{Kauffmann}, G., {Heckman}, T.~M., {Tremonti}, C., {et~al.} 2003, \mnras, 346,
  1055

\bibitem[{{Kellermann} {et~al.}(1998){Kellermann}, {Vermeulen}, {Zensus}, \&
  {Cohen}}]{Kellermann98}
{Kellermann}, K.~I., {Vermeulen}, R.~C., {Zensus}, J.~A., \& {Cohen}, M.~H.
  1998, \aj, 115, 1295

\bibitem[{{Kewley} {et~al.}(2001){Kewley}, {Dopita}, {Sutherland}, {Heisler},
  \& {Trevena}}]{Kewley01}
{Kewley}, L.~J., {Dopita}, M.~A., {Sutherland}, R.~S., {Heisler}, C.~A., \&
  {Trevena}, J. 2001, \apj, 556, 121

\bibitem[{{Kewley} {et~al.}(2006){Kewley}, {Groves}, {Kauffmann}, \&
  {Heckman}}]{Kewley2006}
{Kewley}, L.~J., {Groves}, B., {Kauffmann}, G., \& {Heckman}, T. 2006, \mnras,
  372, 961

\bibitem[{{Kriss}(1994)}]{Kriss94}
{Kriss}, G. 1994, in Astronomical Society of the Pacific Conference Series,
  Vol.~61, Astronomical Data Analysis Software and Systems III, ed. D.~R.
  {Crabtree}, R.~J. {Hanisch}, \& J.~{Barnes}, 437

\bibitem[{{Kurpiewski} \& {Jaroszy{\'n}ski}(1999)}]{Kurp1999}
{Kurpiewski}, A., \& {Jaroszy{\'n}ski}, M. 1999, \aap, 346, 713

\bibitem[{{Laor}(1998)}]{Laor1998}
{Laor}, A. 1998, \apjl, 496, L71

\bibitem[{{Lauer} {et~al.}(1995){Lauer}, {Ajhar}, {Byun}, {Dressler}, {Faber},
  {Grillmair}, {Kormendy}, {Richstone}, \& {Tremaine}}]{Lauer95}
{Lauer}, T.~R., {Ajhar}, E.~A., {Byun}, Y.-I., {et~al.} 1995, \aj, 110, 2622

\bibitem[{{Luridiana} {et~al.}(2015){Luridiana}, {Morisset}, \&
  {Shaw}}]{luridian2015}
{Luridiana}, V., {Morisset}, C., \& {Shaw}, R.~A. 2015, \aap, 573, A42

\bibitem[{{Maoz}(2007)}]{Maoz2007}
{Maoz}, D. 2007, \mnras, 377, 1696

\bibitem[{{Martins} {et~al.}(2004){Martins}, {Leitherer}, {Cid Fernandes},
  {Gonz{\'a}lez Delgado}, {Schmitt}, {Storchi-Bergmann}, \&
  {Heckman}}]{Martins04}
{Martins}, L.~P., {Leitherer}, C., {Cid Fernandes}, R., {et~al.} 2004, in IAU
  Symposium, Vol. 222, The Interplay Among Black Holes, Stars and ISM in
  Galactic Nuclei, ed. T.~{Storchi-Bergmann}, L.~C. {Ho}, \& H.~R. {Schmitt},
  337--338

\bibitem[{{Masegosa} {et~al.}(2011){Masegosa}, {M{\'a}rquez}, {Ramirez}, \&
  {Gonz{\'a}lez-Mart{\'{\i}}n}}]{Masegosa11}
{Masegosa}, J., {M{\'a}rquez}, I., {Ramirez}, A., \&
  {Gonz{\'a}lez-Mart{\'{\i}}n}, O. 2011, \aap, 527, A23

\bibitem[{{Nagao} {et~al.}(2002){Nagao}, {Murayama}, {Shioya}, \&
  {Taniguchi}}]{Nagao02}
{Nagao}, T., {Murayama}, T., {Shioya}, Y., \& {Taniguchi}, Y. 2002, \apj, 567,
  73

\bibitem[{{Nagar} {et~al.}(2005){Nagar}, {Falcke}, \& {Wilson}}]{Nagar2005}
{Nagar}, N.~M., {Falcke}, H., \& {Wilson}, A.~S. 2005, \aap, 435, 521

\bibitem[{{Narayan} \& {Yi}(1995)}]{Narayan95}
{Narayan}, R., \& {Yi}, I. 1995, \apj, 452, 710

\bibitem[{Nemmen {et~al.}(2014)Nemmen, Storchi-Bergmann, \&
  Eracleous}]{Nemmen2014}
Nemmen, R.~S., Storchi-Bergmann, T., \& Eracleous, M. 2014, \mnras, 438, 2804

\bibitem[{{Osterbrock}(1989)}]{osterbrock1989}
{Osterbrock}, D.~E. 1989, {Astrophysics of gaseous nebulae and active galactic
  nuclei}

\bibitem[{{Osterbrock} \& {Ferland}(2006)}]{Osterbrock06}
{Osterbrock}, D.~E., \& {Ferland}, G.~J. 2006, {Astrophysics of gaseous nebulae
  and active galactic nuclei}

\bibitem[{{Papaderos} {et~al.}(2013){Papaderos}, {Gomes}, {V{\'{\i}}lchez},
  {Kehrig}, {Lehnert}, {Ziegler}, {S{\'a}nchez}, {Husemann}, {Monreal-Ibero},
  {Garc{\'{\i}}a-Benito}, {Bland-Hawthorn}, {Cortijo-Ferrero}, {de
  Lorenzo-C{\'a}ceres}, {del Olmo}, {Falc{\'o}n-Barroso}, {Galbany},
  {Iglesias-P{\'a}ramo}, {L{\'o}pez-S{\'a}nchez}, {Marquez}, {Moll{\'a}},
  {Mast}, {van de Ven}, \& {Wisotzki}}]{Papaderos2013}
{Papaderos}, P., {Gomes}, J.~M., {V{\'{\i}}lchez}, J.~M., {et~al.} 2013, \aap,
  555, L1

\bibitem[{{Phillips} {et~al.}(1986){Phillips}, {Jenkins}, {Dopita}, {Sadler},
  \& {Binette}}]{phill86}
{Phillips}, M.~M., {Jenkins}, C.~R., {Dopita}, M.~A., {Sadler}, E.~M., \&
  {Binette}, L. 1986, \aj, 91, 1062

\bibitem[{{Pogge} {et~al.}(2000){Pogge}, {Maoz}, {Ho}, \&
  {Eracleous}}]{Pogge00}
{Pogge}, R.~W., {Maoz}, D., {Ho}, L.~C., \& {Eracleous}, M. 2000, \apj, 532,
  323

\bibitem[{{Sabra} {et~al.}(2003){Sabra}, {Shields}, {Ho}, {Barth}, \&
  {Filippenko}}]{Sabra03}
{Sabra}, B.~M., {Shields}, J.~C., {Ho}, L.~C., {Barth}, A.~J., \& {Filippenko},
  A.~V. 2003, \apj, 584, 164

\bibitem[{{S{\'a}nchez-Bl{\'a}zquez} {et~al.}(2006){S{\'a}nchez-Bl{\'a}zquez},
  {Peletier}, {Jim{\'e}nez-Vicente}, {Cardiel}, {Cenarro},
  {Falc{\'o}n-Barroso}, {Gorgas}, {Selam}, \& {Vazdekis}}]{Sanchez2006}
{S{\'a}nchez-Bl{\'a}zquez}, P., {Peletier}, R.~F., {Jim{\'e}nez-Vicente}, J.,
  {et~al.} 2006, \mnras, 371, 703

\bibitem[{{Sarzi} {et~al.}(2010){Sarzi}, {Shields}, {Schawinski}, {Jeong},
  {Shapiro}, {Bacon}, {Bureau}, {Cappellari}, {Davies}, {de Zeeuw}, {Emsellem},
  {Falc{\'o}n-Barroso}, {Krajnovi{\'c}}, {Kuntschner}, {McDermid}, {Peletier},
  {van den Bosch}, {van de Ven}, \& {Yi}}]{Sarzi10}
{Sarzi}, M., {Shields}, J.~C., {Schawinski}, K., {et~al.} 2010, \mnras, 402,
  2187

\bibitem[{{Sevenster} \& {Chapman}(2001)}]{Sevenster01}
{Sevenster}, M.~N., \& {Chapman}, J.~M. 2001, \apjl, 546, L119

\bibitem[{{Shields}(1992)}]{Shields1992}
{Shields}, J.~C. 1992, \apjl, 399, L27

\bibitem[{{Shull} \& {McKee}(1979)}]{Shull79}
{Shull}, J.~M., \& {McKee}, C.~F. 1979, \apj, 227, 131

\bibitem[{{Sim{\~o}es Lopes} {et~al.}(2007){Sim{\~o}es Lopes},
  {Storchi-Bergmann}, {de F{\'a}tima Saraiva}, \& {Martini}}]{Simoeslopes07}
{Sim{\~o}es Lopes}, R.~D., {Storchi-Bergmann}, T., {de F{\'a}tima Saraiva}, M.,
  \& {Martini}, P. 2007, \apj, 655, 718

\bibitem[{{Singh} {et~al.}(2013){Singh}, {van de Ven}, {Jahnke}, {Lyubenova},
  {Falc{\'o}n-Barroso}, {Alves}, {Cid Fernandes}, {Galbany},
  {Garc{\'{\i}}a-Benito}, {Husemann}, {Kennicutt}, {Marino}, {M{\'a}rquez},
  {Masegosa}, {Mast}, {Pasquali}, {S{\'a}nchez}, {Walcher}, {Wild}, {Wisotzki},
  \& {Ziegler}}]{Singh2013}
{Singh}, R., {van de Ven}, G., {Jahnke}, K., {et~al.} 2013, \aap, 558, A43

\bibitem[{{Sugai} {et~al.}(2005){Sugai}, {Hattori}, {Kawai}, {Ozaki}, {Kosugi},
  {Ohtani}, {Hayashi}, {Ishigaki}, {Ishii}, {Sasaki}, {Takeyama}, {Yutani},
  {Usuda}, {Hayashi}, \& {Namikawa}}]{Sugai2005}
{Sugai}, H., {Hattori}, T., {Kawai}, A., {et~al.} 2005, \apj, 629, 131

\bibitem[{{Taniguchi} {et~al.}(2000){Taniguchi}, {Shioya}, \&
  {Murayama}}]{Taniguchi2000}
{Taniguchi}, Y., {Shioya}, Y., \& {Murayama}, T. 2000, \aj, 120, 1265

\bibitem[{{Terlevich} \& {Melnick}(1985)}]{Terlevich1985}
{Terlevich}, R., \& {Melnick}, J. 1985, \mnras, 213, 841

\bibitem[{{Tully}(1988)}]{Tully1988}
{Tully}, R.~B. 1988, {Nearby galaxies catalog}

\bibitem[{{Tully} {et~al.}(2013){Tully}, {Courtois}, {Dolphin}, {Fisher},
  {H{\'e}raudeau}, {Jacobs}, {Karachentsev}, {Makarov}, {Makarova},
  {Mitronova}, {Rizzi}, {Shaya}, {Sorce}, \& {Wu}}]{Tully2013}
{Tully}, R.~B., {Courtois}, H.~M., {Dolphin}, A.~E., {et~al.} 2013, \aj, 146,
  86

\bibitem[{{Ulvestad} \& {Ho}(2001)}]{Ulvestad2001}
{Ulvestad}, J.~S., \& {Ho}, L.~C. 2001, \apjl, 562, L133

\bibitem[{{Van de Steene} \& {van Hoof}(2003)}]{Vandesteene03}
{Van de Steene}, G.~C., \& {van Hoof}, P.~A.~M. 2003, \aap, 406, 773

\bibitem[{{van Dokkum}(2001)}]{Dokkum2001}
{van Dokkum}, P.~G. 2001, \pasp, 113, 1420

\bibitem[{{van Velzen} {et~al.}(2012){van Velzen}, {Falcke}, {Schellart},
  {Nierstenh{\"o}fer}, \& {Kampert}}]{van2012}
{van Velzen}, S., {Falcke}, H., {Schellart}, P., {Nierstenh{\"o}fer}, N., \&
  {Kampert}, K.-H. 2012, \aap, 544, A18

\bibitem[{{Vanden Berk} {et~al.}(2001){Vanden Berk}, {Richards}, {Bauer},
  {Strauss}, {Schneider}, {Heckman}, {York}, {Hall}, {Fan}, {Knapp},
  {Anderson}, {Annis}, {Bahcall}, {Bernardi}, {Briggs}, {Brinkmann}, {Brunner},
  {Burles}, {Carey}, {Castander}, {Connolly}, {Crocker}, {Csabai}, {Doi},
  {Finkbeiner}, {Friedman}, {Frieman}, {Fukugita}, {Gunn}, {Hennessy},
  {Ivezi{\'c}}, {Kent}, {Kunszt}, {Lamb}, {Leger}, {Long}, {Loveday}, {Lupton},
  {Meiksin}, {Merelli}, {Munn}, {Newberg}, {Newcomb}, {Nichol}, {Owen}, {Pier},
  {Pope}, {Rockosi}, {Schlegel}, {Siegmund}, {Smee}, {Snir}, {Stoughton},
  {Stubbs}, {SubbaRao}, {Szalay}, {Szokoly}, {Tremonti}, {Uomoto}, {Waddell},
  {Yanny}, \& {Zheng}}]{Vanden2001}
{Vanden Berk}, D.~E., {Richards}, G.~T., {Bauer}, A., {et~al.} 2001, \aj, 122,
  549

\bibitem[{Veilleux \& Osterbrock(1987)}]{veilleux87}
Veilleux, S., \& Osterbrock, D.~E. 1987, \apjs, 63, 295

\bibitem[{{Vestergaard} \& {Wilkes}(2001)}]{Vestergaard2000}
{Vestergaard}, M., \& {Wilkes}, B.~J. 2001, \apjs, 134, 1

\bibitem[{{Walsh} {et~al.}(2008){Walsh}, {Barth}, {Ho}, {Filippenko}, {Rix},
  {Shields}, {Sarzi}, \& {Sargent}}]{Walsh2008}
{Walsh}, J.~L., {Barth}, A.~J., {Ho}, L.~C., {et~al.} 2008, \aj, 136, 1677

\bibitem[{{Wilson} \& {Raymond}(1999)}]{Wilson99}
{Wilson}, A.~S., \& {Raymond}, J.~C. 1999, \apjl, 513, L115

\bibitem[{{Wrobel}(1984)}]{Wrobel84}
{Wrobel}, J.~M. 1984, \apj, 284, 531

\bibitem[{{Yan} \& {Blanton}(2012)}]{Yan2012}
{Yan}, R., \& {Blanton}, M.~R. 2012, \apj, 747, 61

\bibitem[{{Yan} {et~al.}(2006){Yan}, {Newman}, {Faber}, {Konidaris}, {Koo}, \&
  {Davis}}]{Yan06}
{Yan}, R., {Newman}, J.~A., {Faber}, S.~M., {et~al.} 2006, \apj, 648, 281

\end{thebibliography}
